\documentclass[aps,prd, showpacs]{revtex4}

\usepackage{graphicx} 
\usepackage[scriptsize,hang]{caption}
\usepackage[scriptsize,hang]{subfigure}
\usepackage{pstricks,pstricks-add}
\usepackage{pst-plot}
\usepackage{color}
\usepackage{placeins}

\usepackage{amsmath,amsfonts,bbm,dsfont,mathrsfs}

\newcommand{\half}{\frac{1}{2}}
\newcommand{\dd}{{\mathrm{d}}}

\begin{document}

\title{Geodesic motion in the five-dimensional Myers-Perry-AdS spacetime}

\author{Saskia Grunau}
\email[]{saskia.grunau@uni-oldenburg.de}
\author{Hendrik Neumann}
\email[]{hendrik.neumann@uni-oldenburg.de}
\author{Stephan Reimers}
\email[]{stephan.reimers@uni-oldenburg.de}
\affiliation{Institut f\"ur Physik, Universit\"at Oldenburg, D--26111 Oldenburg, Germany}

\date{\today}

\begin{abstract}
In this article we study the geodesic motion of test particles and light in the five-dimensional Myers-Perry-anti de Sitter spacetime. We derive the equations of motion and present their solutions in terms of the Weierstra{\ss} $\wp$, $\sigma$, and $\zeta$ functions. With the help of parametric diagrams and effective potentials we analyze the geodesic motion and give a list of all possible orbit types for timelike, null and spacelike geodesics. We plot examples of the orbits and take a look at the photon region in five dimensions. Furthermore we study spacelike geodesics and their relation to the AdS/CFT correspondence.
\end{abstract}

\pacs{04.20.Jb, 02.30.Hq, 04.50.Gh}

\maketitle

\section{Introduction}

In 1997, quantum field theory was connected to gravity by famous anti-de Sitter/ conformal field theory (AdS/CFT) correspondence \cite{Maldacena:1997re}. In particular, compactifications of string theory on anti-de Sitter space were related to a conformal field theory. This made asymptotically anti-de Sitter black holes an interesting phenomenon to study.

The first rotating AdS black hole in four dimensions was described by Carter \cite{Carter:1968ks}, shortly after Kerr proposed his asymptotically flat rotating black hole solution \cite{Kerr:1963ud}. The higher-dimensional generalization of the Kerr black hole was found by Myers and Perry \cite{Myers:1986un}. The $d$-dimensional Myers-Perry black hole possesses $\lfloor(d-1)/{2}\rfloor$ rotation parameters associated with $\lfloor(d-1)/{2}\rfloor$ independent planes of rotation.
Hawking et al. \cite{Hawking:1998kw} found a five-dimensional rotating AdS black hole with two rotation parameters. This solution was extended to higher dimensions \cite{Gibbons:2004js, Gibbons:2004uw} and moreover the NUT (Newman-Unti-Tamburino) parameter was included \cite{Chen:2006xh}.

The study of the orbits of test particles and light in a spacetime is a powerful method to explore black holes and to test different models and theories. Observable quantities like the periastron shift of bound orbits, the light deflection angle of escape orbits or the shadow of a black hole can be compared to observations. Geodesics provide insight to the structure of the spacetime and information on the black hole. In particular, geodesics can be related to two-point correlators in AdS/CFT \cite{Balasubramanian:1999zv}.

The Hamilton-Jacobi formalism has proved to be very efficient in deriving the equations of motion. Carter \cite{Carter:1968ks} showed that the Hamilton-Jacobi equation for test particles in the Kerr spacetime separates. The equations of motion in the four-dimensional Kerr spacetime can be solved analytically in terms of elliptic functions. However, when the cosmological constant is added to the four-dimensional Kerr spacetime hyperelliptic functions are required for the analytical solution 
\cite{Kraniotis:2005zm} -- \cite{Hackmann:2010zz}. 

In \cite{Frolov:2003en} the equations of motion in the five-dimensional Myers-Perry spacetime were given and different types of orbits were analyzed. In higher dimensions the separability of rotating (A)dS spacetimes was shown in \cite{Vasudevan:2004ca}--\cite{Frolov:2008jr}. The separability of the  Myers-Perry spacetime and its charged version in arbitrary dimensions was shown in \cite{Chervonyi:2015ima} by constructing Killing-Yano tensors. A method to solve the equations of motion in the higher-dimensional Myers-Perry spacetime with a single rotation parameter was presented in \cite{Enolski:2010if}. There hyperelliptic functions were involved. However, in the five-dimensional Myers-Perry spacetime with two rotation parameters it is possible to solve the geodesic equations analytically in terms of elliptic functions \cite{Kagramanova:2012hw, Diemer:2014lba}. 

Although the four-dimensional Kerr-Newman solution of Einstein's field equation could not be generalized to higher dimensions yet, solutions of the Einstein-Maxwell-Chern-Simons field equations in the five-dimensional minimal gauged supergravity have been found \cite{Chong:2005hr, Gauntlett:2002nw}. This solution is determined by the mass, two angular momenta, an electric charge and a (negative) cosmological constant. The geodesic equations of this spacetime have been solved analytically in \cite{Reimers:2016czc} in the case of a vanishing cosmological constant. For a nonvanishing cosmological constant, but a vanishing electrical charge, the solution reduces to the Myers-Perry-AdS spacetime. Delsate et al. investigated the geodesic motion in the five-dimensional Myers-Perry-AdS spacetime with two equal angular momenta \cite{Delsate:2015ina}. The geodesics in another spacetime in minimal supergravity were studied in \cite{Gibbons:1999uv} and later in \cite{Diemer:2013fza}. There a charged rotating extremal black hole, the Breckenridge-Myers-Peet-Vafa solution, was considered. Charged geodesics in the same spacetime were studied in \cite{Herdeiro:2000ap}.
\\

In the present article we study the geodesic motion of test particles and light in the general five-dimensional Myers-Perry-AdS spacetime with two independent angular momenta and a negative cosmological constant. We start with a short discussion of the Myers-Perry-AdS spacetime and derive the equations of motion in Sec. \ref{sec:mp-spacetime}. In Sec. \ref{sec:classification} we analyze timelike and null geodesics using parametric diagrams and the effective potential to give a list of all possible types. We also study the photon region and the conditions for orbits ending in the singularity. Furthermore in Sec. \ref{sec:spacelike} we look at spacelike geodesics and comment on the application of these geodesics in AdS/CFT.
The equations of motion are solved analytically in terms of the Weierstra{\ss} $\wp$, $\sigma$, and $\zeta$ functions in Sec. \ref{sec:solutions}. Examples of the orbits are shown in section \ref{sec:orbits}. In Sec. \ref{sec:conclusion} we conclude and give an outlook on possible future work.

\section{The Myers-Perry-anti de Sitter spacetime}
\label{sec:mp-spacetime}

\begin{figure}[h]
	\centering
	\subfigure[$a=0.5$ and $b=0$]{
		\includegraphics[width=0.31\textwidth]{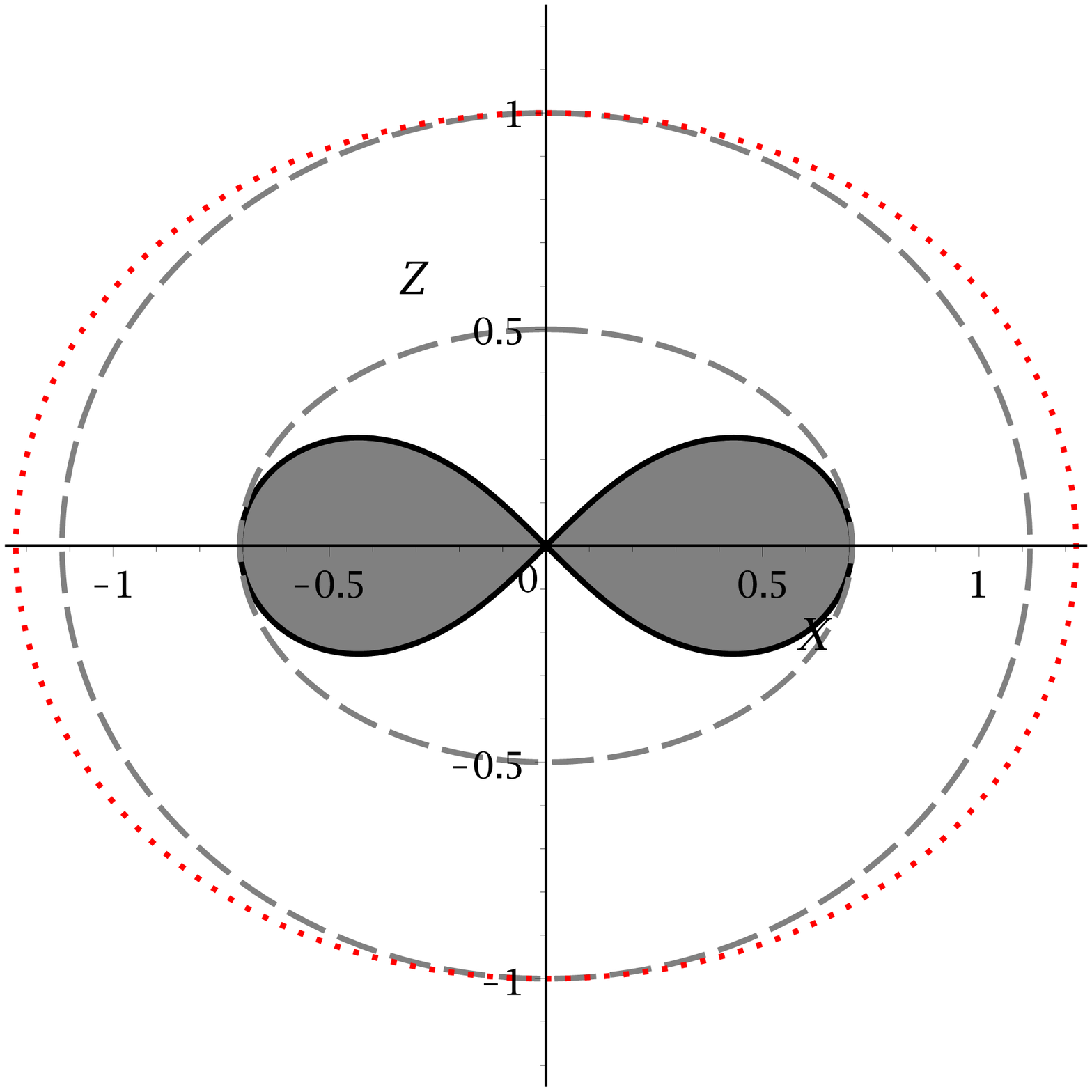}
	}
	\subfigure[$a=0.5$ and $b=0.1$]{
		\includegraphics[width=0.31\textwidth]{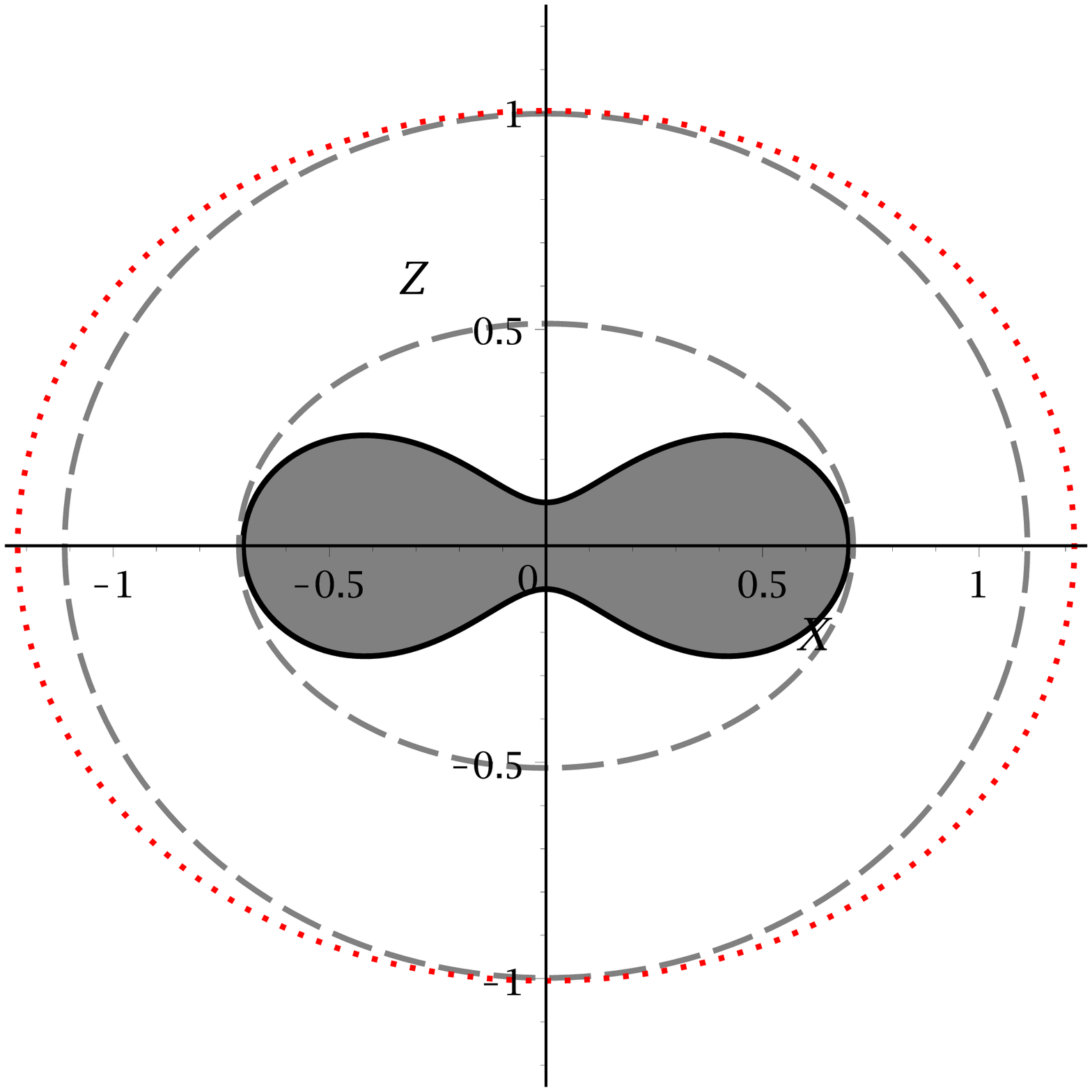}
	}
	\subfigure[$a=0.5$ and $b=0.4$]{
		\includegraphics[width=0.31\textwidth]{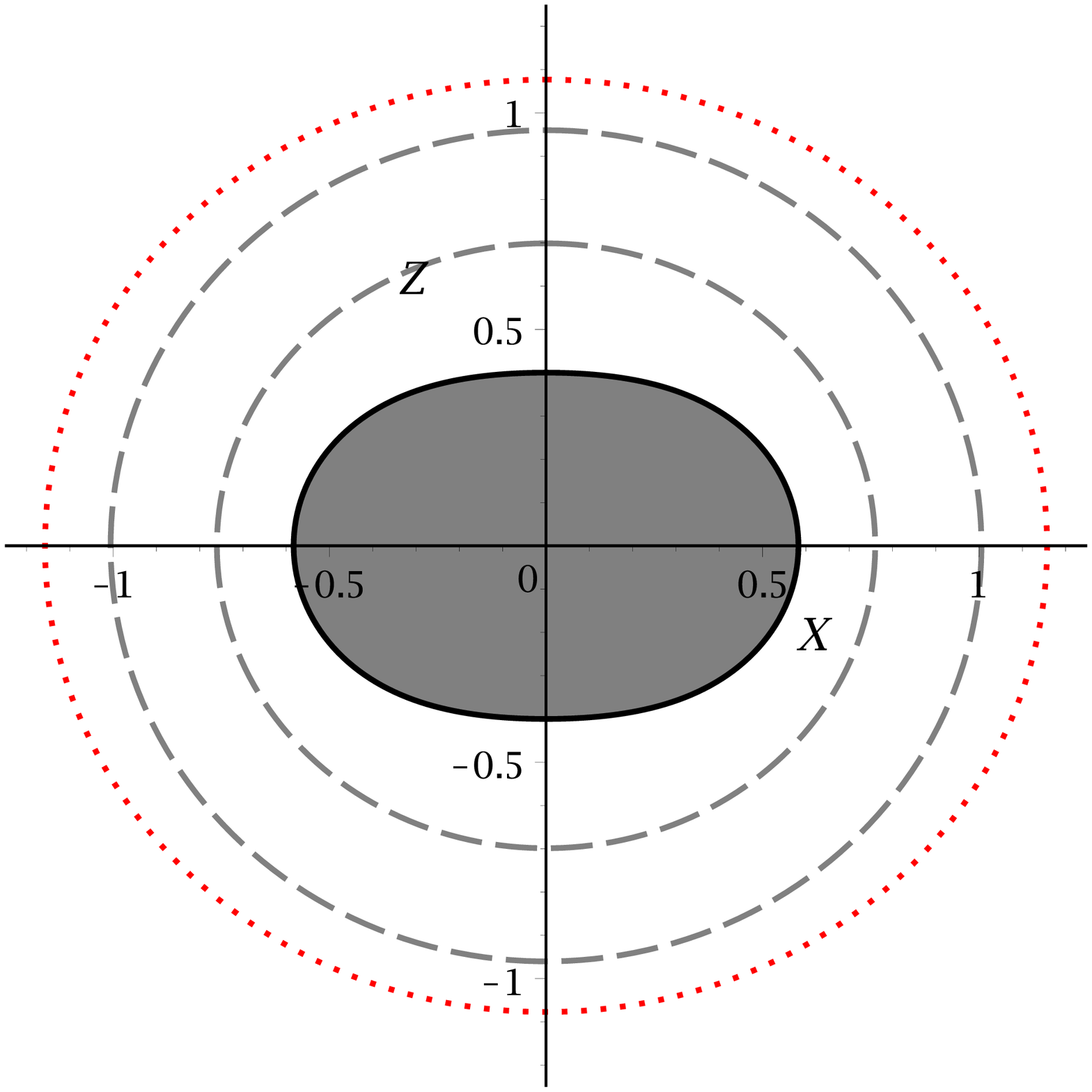}
	}
	\caption{Plots of the singularity given by $\rho^2=0$ (grey structure with black boundary) and the horizons given by $\Delta_x=0$ (grey dashed lines) with $M=0.5$, $g=0.01$, and various rotation parameters. The red dotted line is the static limit, the boundary of the ergoregion. The Cartesian coordinates from Eq. \eqref{eqn:coordinates} are used in the plane $\phi=\psi=0$.}
	\label{pic:singularity}
\end{figure}

In a coordinate frame which is nonrotating at infinity, the Myers-Perry-AdS spacetime is given by (compare \cite{Chong:2005hr} with $q=0$)
\begin{align}
	\dd s^2 = &-\frac{\Delta_\theta (1+g^2r^2)\dd t^2}{\Xi_a\Xi_b} + \frac{2M}{\rho^2} \left( \frac{\Delta_\theta\dd t}{\Xi_a\Xi_b} - a \sin^2\theta \, \frac{\dd\phi}{\Xi_a} - b \cos^2\theta \, \frac{\dd\psi}{\Xi_b}  \right)^2  \nonumber\\ 
	& + \frac{r^2+a^2}{\Xi_a}\sin^2\theta \, \dd\phi^2 + \frac{r^2+b^2}{\Xi_b}\cos^2\theta \, \dd\psi^2  + \frac{\rho^2 \dd \theta^2}{\Delta_\theta}  + \frac{\rho^2 r^2 \dd r^2}{\Delta_r}\, ,
	\label{eqn:mp-ads-metric}
\end{align}
The metric is given in asymptotically static Boyer-Lindquist-like coordinates. It is characterized by its mass related to the parameter $M$, two independent rotation parameters $a$, $b$ and the (negative) cosmological constant which is represented by $g^2=-\frac{\Lambda}{4}$. In the following we assume without loss of generality that $a\geq b$. The metric functions are
\begin{align}
	\Xi_a &= 1-a^2g^2 \, , \nonumber\\
	\Xi_b &= 1-b^2g^2 \, , \nonumber\\
	\Delta_\theta &= 1- a^2g^2\cos^2\theta - b^2g^2\sin^2\theta \, , \nonumber\\
	\Delta_r &= (r^2+a^2)(r^2+b^2)(1+g^2r^2)-2Mr^2 \, , \nonumber\\
	\rho^2 &=  r^2+a^2\cos^2\theta +b^2\sin^2\theta \, .
\end{align}
Myers and Perry introduced a new radial coordinate $x =r^2$, so that the whole spacetime is covered \cite{Myers:1986un, Gibbons:2009um}. Applying this to the Myers-Perry-AdS metric \eqref{eqn:mp-ads-metric} yields
\begin{align}
	\dd s^2 = &-\frac{\Delta_\theta (1+g^2x)\dd t^2}{\Xi_a\Xi_b} + \frac{2M}{\rho^2} \left( \frac{\Delta_\theta\dd t}{\Xi_a\Xi_b} - a \sin^2\theta \, \frac{\dd\phi}{\Xi_a} - b \cos^2\theta \, \frac{\dd\psi}{\Xi_b}  \right)^2  \nonumber\\ 
	& + \frac{x+a^2}{\Xi_a}\sin^2\theta \, \dd\phi^2 + \frac{x+b^2}{\Xi_b}\cos^2\theta \, \dd\psi^2  + \frac{\rho^2 \dd \theta^2}{\Delta_\theta}  + \frac{\rho^2 \dd x^2}{4\Delta_x}\, ,
	\label{eqn:mp-ads-metric2}
\end{align}
with
\begin{align}
	\Delta_x &= (x+a^2)(x+b^2)(1+g^2x)-2Mx \, , \nonumber\\
	\rho^2 &=  x+a^2\cos^2\theta +b^2\sin^2\theta \, .
\end{align}
The Myers-Perry-AdS black hole has a singularity for $\rho^2=0$, which is the same condition as in the asymptotically flat case $g=0$. The singularity is a closed surface in the range $x\in[-a^2, -b^2]$ and cannot be traversed, even when one of the two rotations parameters is vanishing \cite{Diemer:2014lba}. We shift the coordinate $x$ by $+a^2$ in the usual transformation from Boyer-Lindquist to Cartesian coordinates
\begin{align}
	X &= \sqrt{x+2a^2}\, \sin\theta \cos\phi \, , \nonumber\\
	Y &= \sqrt{x+2a^2}\, \sin\theta \sin\phi \, , \nonumber\\
	Z &= \sqrt{x+b^2+a^2}\, \cos\theta \cos\psi \, , \nonumber\\
	W &= \sqrt{x+b^2+a^2}\, \cos\theta \sin\psi \, .
	\label{eqn:coordinates}
\end{align}
The coordinate ranges are $x\in[-a^2,\infty]$, $\theta\in[0,\frac{\pi}{2}]$, $\phi\in[0,2\pi]$ and $\psi\in[0,2\pi]$. Figure \ref{pic:singularity} shows the singularity for different rotation parameters in the $X$-$Z$ plane ($\phi=\psi=0$).

The horizons of the black hole are given by $\Delta_x=0$, where $\Delta_x$ is a third order polynomial in $x$. Applying the rule of Descartes we find that  $\Delta_x$ has a single negative and possibly two positive zeros if $2M\geq a^2+b^2+a^2b^2g^2$. There are one or three negative zeros if $2M< a^2+b^2+a^2b^2g^2$. For positive $M$ all negative zeros are smaller than $-a^2$, but for negative $M$ a negative zero can be greater than $-b^2$. However, at least in the case $g=0$ it was pointed out by Gibbons and Herdeiro \cite{Gibbons:2009um} that this negative zero for $M<0$ does not correspond to a regular horizon.

\begin{figure}[h]
	\centering
	\includegraphics[width=0.31\textwidth]{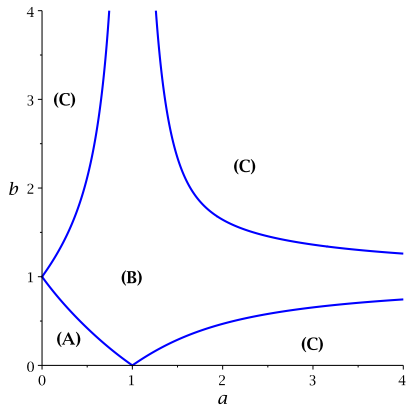}
	\caption{Parametric $a$-$b$-diagram ($M=0.5$, $g=1$) showing three regions with different numbers of zeros of $\Delta_x$. In region (A) $\Delta_x$ has a single negative and two positive zeros, therefore two horizons exist. There is a single negative zero in regions (B) and three negative zeros in region (C). Regions (B) and (C) correspond to a naked singularity.}
	\label{pic:horizon-parameter}
\end{figure}

The number of zeros changes if $\Delta_x$ has a double zero. Figure \ref{pic:horizon-parameter} shows a parametric $a$-$b$ diagram for positive $M$, plotted from the conditions  $\Delta_x=0$ and  $\frac{\dd \Delta_x}{\dd x}=0$. There are three regions with different numbers of zeros. Positive horizons are only present for small rotation parameters in region (A), here $\Delta_x$ has two positive zeros and a single negative zero which is smaller than $-a^2$. Regions (B) (one zero $<-a^2$) and (C) (three zeros $<-a^2$) correspond to a naked singularity.

The boundary of the ergoregion is defined by $g_{tt}=0$ and therefore
\begin{equation}
	2M\Delta_\theta-\rho^2 (1+g^2x)=0 \, ,
\end{equation}
which has two solutions
\begin{equation}
	x_{\rm ergo}^\pm = \frac{1}{2g^2} \left( \Delta_\theta -2 \pm \sqrt{ \Delta_\theta ( \Delta_\theta +8Mg^2)} \right)\, .
\end{equation}
The solution $x_{\rm ergo}^-$ is smaller than $-a^2$ (if the parameters are chosen such that horizons exist) and lies not within the allowed range of $x$. The ergoregion is the space between the event horizon and $x_{\rm ergo}^+$, the so called static limit (see Fig. \ref{pic:singularity}).\\

Equations of motion for test particles in the Myers-Perry-AdS spactime in a nonrotating coordinate frame \eqref{eqn:mp-ads-metric} can be derived with the Hamilton-Jacobi formalism. To solve the Hamilton-Jacobi equation
\begin{equation}
	\frac{\partial S}{\partial \lambda} + \half g^{\mu\nu} \frac{\partial S}{\partial x^\mu}\frac{\partial S}{\partial x^\nu}=0
\label{eqn:hjd}
\end{equation}
we use the following ansatz for the action $S$:
\begin{equation}
	S= \half\delta\lambda - Et + L\phi + J\psi + S_r(r)+S_\theta (\theta) \, ,
\end{equation}
where $E$ is the energy of the test particle and $L$, $J$ are the two angular momenta. $\delta$ represents the mass of the test particle. We choose $\delta=1$ for timelike geodesics,$\delta=0$ for null geodesics, and $\delta=-1$ for spacelike geodesics. Along the geodesics $\lambda$ is an affine parameter. The Hamilton-Jacobi equation \eqref{eqn:hjd} separates with the help of the Carter \cite{Carter:1968rr} constant $K$ and yields five differential equations of motion
\begin{align}
	\left(\frac{\dd x}{\dd \gamma}\right) ^2 =& X(x) \, , \label{eqn:x-equation}\\
	\left(\frac{\dd \theta}{\dd \gamma}\right) ^2 =& \Theta (\theta)  \, , \label{eqn:theta-equation}\\
	\left(\frac{\dd \phi}{\dd \gamma}\right) =& \frac{1}{\Delta_x} \left\lbrace L\Xi_a (1+g^2x)(b^2+x)(b^2-a^2) + 2M \left[ Ea(b^2+x) -L(b^2+a^2g^2x) -Jab(1+g^2x) \right] \right\rbrace \nonumber\\ 
	&+ \frac{L\Xi_a}{\sin^2\theta}\, , \label{eqn:phi-equation}\\
	\left(\frac{\dd \psi}{\dd \gamma}\right) =& \frac{1}{\Delta_x} \left\lbrace J\Xi_b (1+g^2x)(a^2+x)(a^2-b^2) + 2M \left[ Eb(a^2+x) -J(a^2+b^2g^2x) -Lab(1+g^2x) \right] \right\rbrace \nonumber\\ 
	&+ \frac{J\Xi_b}{\cos^2\theta} \, , \label{eqn:psi-equation}\\
	\left(\frac{\dd t}{\dd \gamma}\right) =& \frac{1}{\Delta_x}\left\lbrace E\left[ 2M( (a^2+b^2)x+a^2b^2) +(x+a^2)(x+b^2)(x-(a^2+b^2)gx - a^2b^2g^2 ) \right] \right. \nonumber\\
	&\left. -2M\left[ aL(b^2+x)+Jb(a^2+x) \right] \right\rbrace +\frac{E}{\Delta_\theta}\left(\Xi_a a^2\cos^2\theta+\Xi_b b^2\sin^2\theta \right) \, , \label{eqn:t-equation}
\end{align}
with the polynomial $X(x)$ of order four and the function $\Theta(\theta)$
\begin{align}
	X(x)  =& -4\left\lbrace \Delta_x(K+\delta x) + E^2\left[ -2M( (a^2+b^2)x+a^2b^2) -(x+a^2)(x+b^2)(x-(a^2+b^2)g^2x - a^2b^2g^2 ) \right] \right. \nonumber\\
	& + L^2 \left[ \Xi_a(1+g^2x)(b^2+x)(b^2-a^2)-2M(a^2g^2x+b^2) \right] \nonumber\\
	& + J^2 \left[ \Xi_b(1+g^2x)(a^2+x)(a^2-b^2)-2M(b^2g^2x+a^2) \right]\nonumber\\
	& \left.+ 4M \left[ ELa(b^2+x) +EJb(a^2+x) -LJab(1+g^2x) \right] \right\rbrace \, , \label{eqn:Xfunc}\\
	\Theta (\theta) = & K\Delta_\theta   +E^2(\Xi_a a^2\cos^2\theta+\Xi_b b^2\sin^2\theta) - \Delta_\theta \left[ \delta (a^2\cos^2\theta+b^2\sin^2\theta) +\frac{L^2\Xi_a}{\sin^2\theta} +\frac{J^2\Xi_b}{\cos^2\theta} \right] \, . \label{eqn:Thetafunc}
\end{align}
We applied the Mino \cite{Mino:2003yg} time $\gamma$ as $\rho^2 \dd\gamma = \dd\lambda$ to remove the factor  $\rho^2$ from all equations. From now on we will use dimensionless quantities in the equations of motion which is achieved by setting
\begin{align}
	& x\rightarrow 2Mx\, , \ t\rightarrow\sqrt{2M}t\, ,\ \gamma \rightarrow \frac{\gamma}{\sqrt{2M}}\, , \ a\rightarrow\sqrt{2M}a\, , \ b\rightarrow\sqrt{2M}b\, , \ \\ 
	&L\rightarrow\sqrt{2M}L\, , \ J\rightarrow\sqrt{2M}J\, , \ K\rightarrow\sqrt{2M}K\, , 
  \ g\rightarrow \frac{g}{\sqrt{2M}}\,  .
\end{align}
This is equivalent to setting $M=\half$.

\section{Classification of timelike and null geodesics}
\label{sec:classification}

In this section we analyze timelike and null geodesics. Using parametric diagrams and effective potentials we determine the possible orbit types, which are characterized by the $x$ equation \eqref{eqn:x-equation} and the $\theta$ equation \eqref{eqn:theta-equation}. The following orbits can be found in the Myers-Perry-AdS spacetime:
\begin{enumerate}
	\item Bound orbits (BO) with the range $x\in[x_1, x_2]$ and 
	\begin{enumerate}
		\item either $x_1,x_2>x_+$ 
		\item or $x_1,x_2<x_-$.
	\end{enumerate}
	\item Many-world bound orbits (MBO) with the range $x\in[x_1, x_2]$, where $x_1\leq x_-$ and $x_2\geq x_+$. These geodesics emerge into another universe every time the horizons are crossed twice.
	\item Escape orbits (EO) with the range $x\in[x_1, \infty)$ and $x_1\geq x_+$.
	\item Two-world escape orbits (TWEO) with the range $x\in[x_1, \infty)$ and $x_1\leq x_-$.These geodesics emerge into another universe after the horizons are crossed twice.
	\item Terminating orbits (TO) with the range
	\begin{enumerate}
		\item $x\in[x_0, \infty)$
		\item or $x\in[x_0, x_1]$
	\end{enumerate}
	where $x_0$ is on the closed surface $\rho^2=0$. These orbits end in the singularity and exist in special cases only.
\end{enumerate}

\subsection{Timelike geodesics}

First we will study timelike geodesics describing particles moving in the spacetime of the Myers-Perry-AdS black hole.

\subsubsection{The $\theta$ motion}

The $\theta$ motion is described by Eq. \eqref{eqn:theta-equation}. The function $\Theta(\theta)$ [see Eq. \ref{eqn:Thetafunc}] has poles at $\theta=0$ (or $\theta=\pi$) and at $\theta=\frac{\pi}{2}$. Therefore, the geodesics cannot reach $\theta=0$ (or $\theta=\pi$) as long as $L\neq 0$ and  $\theta=\frac{\pi}{2}$ as long as $J\neq0$.
For simplicity we substitute $\nu=\cos\theta^2$ so that function $\Theta(\theta)$ becomes
\begin{align}
	\Theta (\nu) = & K\Delta_\nu   +E^2(\Xi_a a^2\nu+\Xi_b b^2(1-\nu)) - \Delta_\nu \left[ \delta (a^2\nu+b^2(1-\nu)) +\frac{L^2\Xi_a}{1-\nu} +\frac{J^2\Xi_b}{\nu} \right]
\label{eqn:nu-equation}
\end{align}
with $\Delta_\nu= 1-a^2g^2\nu-b^2g^2(1-\nu)$.\\

The turning points of the $\theta$ motion are the zeros of Eqs. \eqref{eqn:theta-equation} or \eqref{eqn:nu-equation} in the range $\theta\in[0,\frac{\pi}{2}]$ or $\nu\in [0,1]$. The number of zeros changes if double zeros appear. From this condition parametric diagrams can be drawn to obtain parameter regions with a different number of zeros. It appears that the function $\Theta$ has either two zeros or none in the allowed range. In Fig. \ref{pic:parameterplot-timelike} the $\theta$ equation has two zeros in the white region and none in the grey region. Therefore, geodesic motion is not possible in the grey region.\\

\begin{figure}[h]
	\centering
	\subfigure[$K=2$]{
		\includegraphics[width=0.31\textwidth]{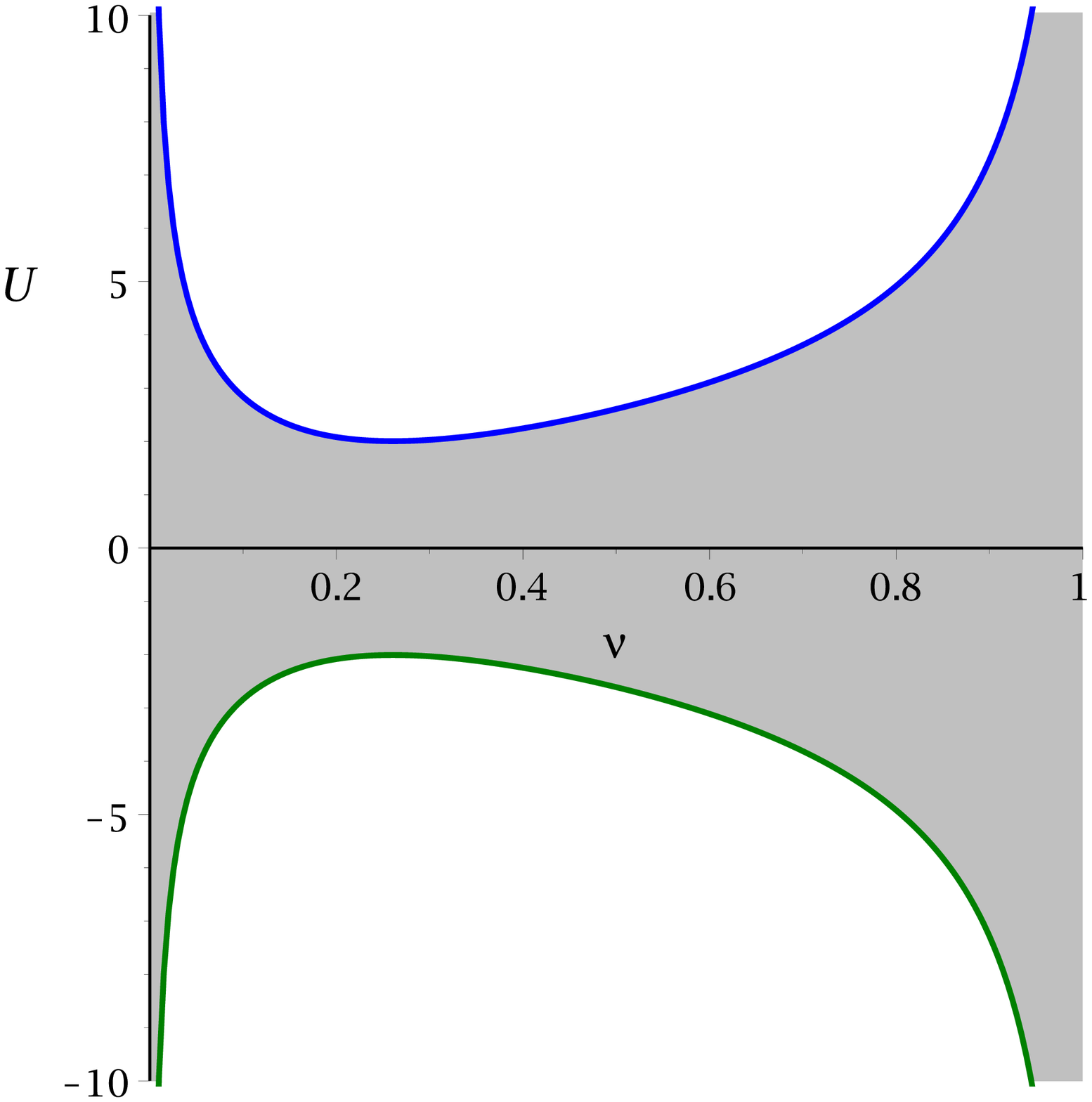}
	}
	\subfigure[$K=3.5$]{
		\includegraphics[width=0.31\textwidth]{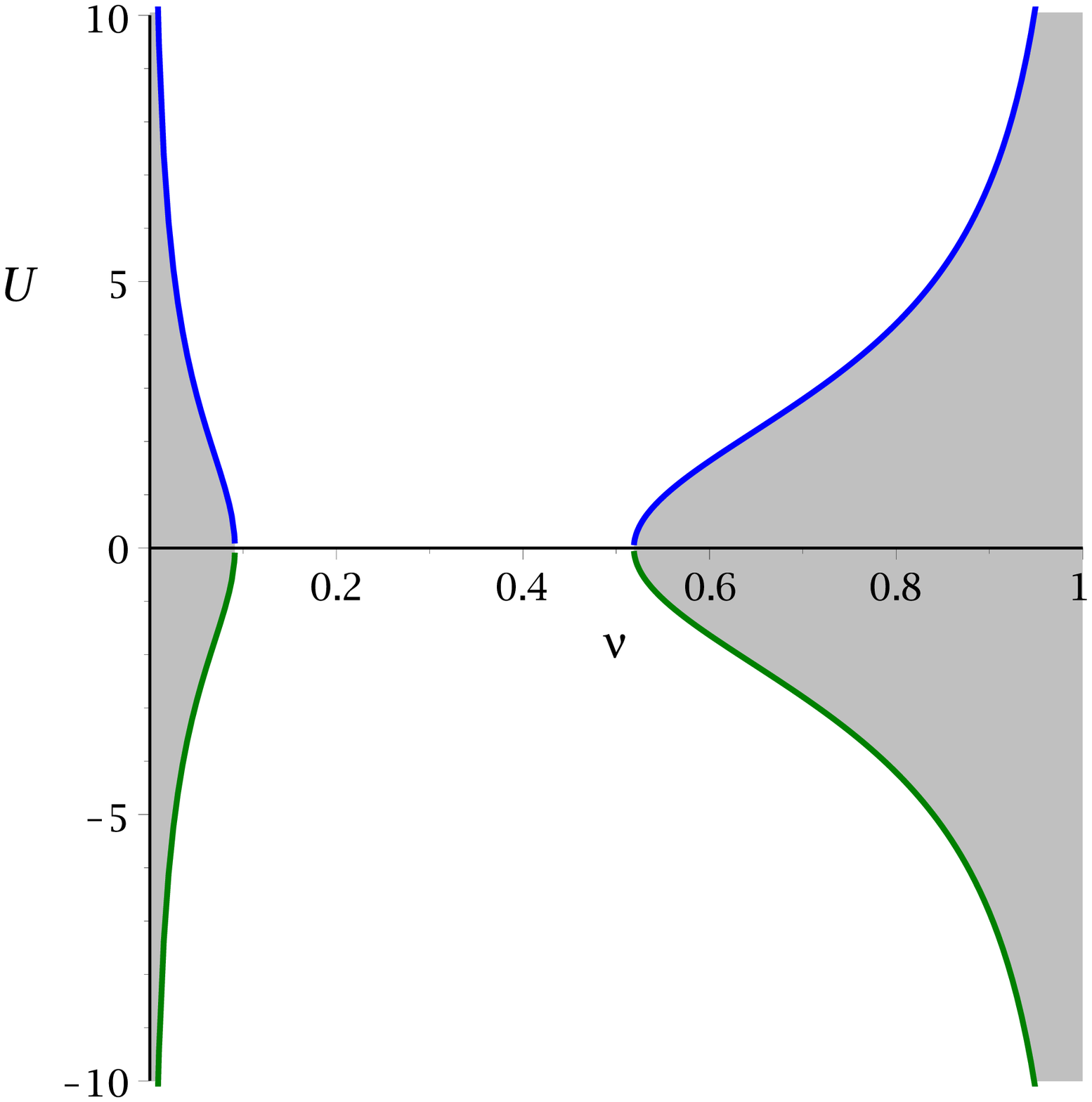}
	}
	\caption{Plots of the effective $\theta$-potential for $\delta=1$, $g=0.1$,$a=0.5$, $b=0.4$, $L=1.2$, $J=0.4$ and different $K$. The green and blue curves are the two parts of the effective potential. The grey areas are forbidden by the $\theta$ equation. Geodesic motion is possible in the white areas only.}
	\label{pic:theta-potential-timelike}
\end{figure}

Additionally we define an effective potential $U$ by $\Theta (\nu) = f(\nu)(E-U_-)(E-U_+)$ so that
\begin{equation}
 U_\pm = \pm \sqrt{  \frac{\Delta_\nu}{\Xi_a a^2\nu+\Xi_b b^2(1-\nu)} \left[  \delta (a^2\nu+b^2(1-\nu)) -K  +\frac{L^2\Xi_a}{1-\nu} +\frac{J^2\Xi_b}{\nu}  \right]   } \, .
\label{eqn:theta-potential}
\end{equation}
Figure \ref{pic:theta-potential-timelike} depicts two typical effective potentials for the $\theta$ motion. The green and blue curves are the two parts of the effective potential. In the grey area $\theta$ (or $\nu$) becomes imaginary and therefore motion is not allowed. A potential barrier prevents light and test particles from reaching $\nu=0$ and $\nu=1$.

In Fig. \ref{pic:theta-potential-timelike}(a) the function $\Theta (\nu)$ has two zeros except for some energy range where geodesic motion not possible, whereas in Fig. \ref{pic:theta-potential-timelike}(b) $\Theta (\nu)$ has two zeros for all energies.

\subsubsection{The $x$ motion}

\begin{figure}[h]
	\centering
	\subfigure[$\delta=1$, $g=0.1$, $a=0.55$, $b=0.4$, $L=0.6$, $J=0.5$]{
		\includegraphics[width=0.31\textwidth]{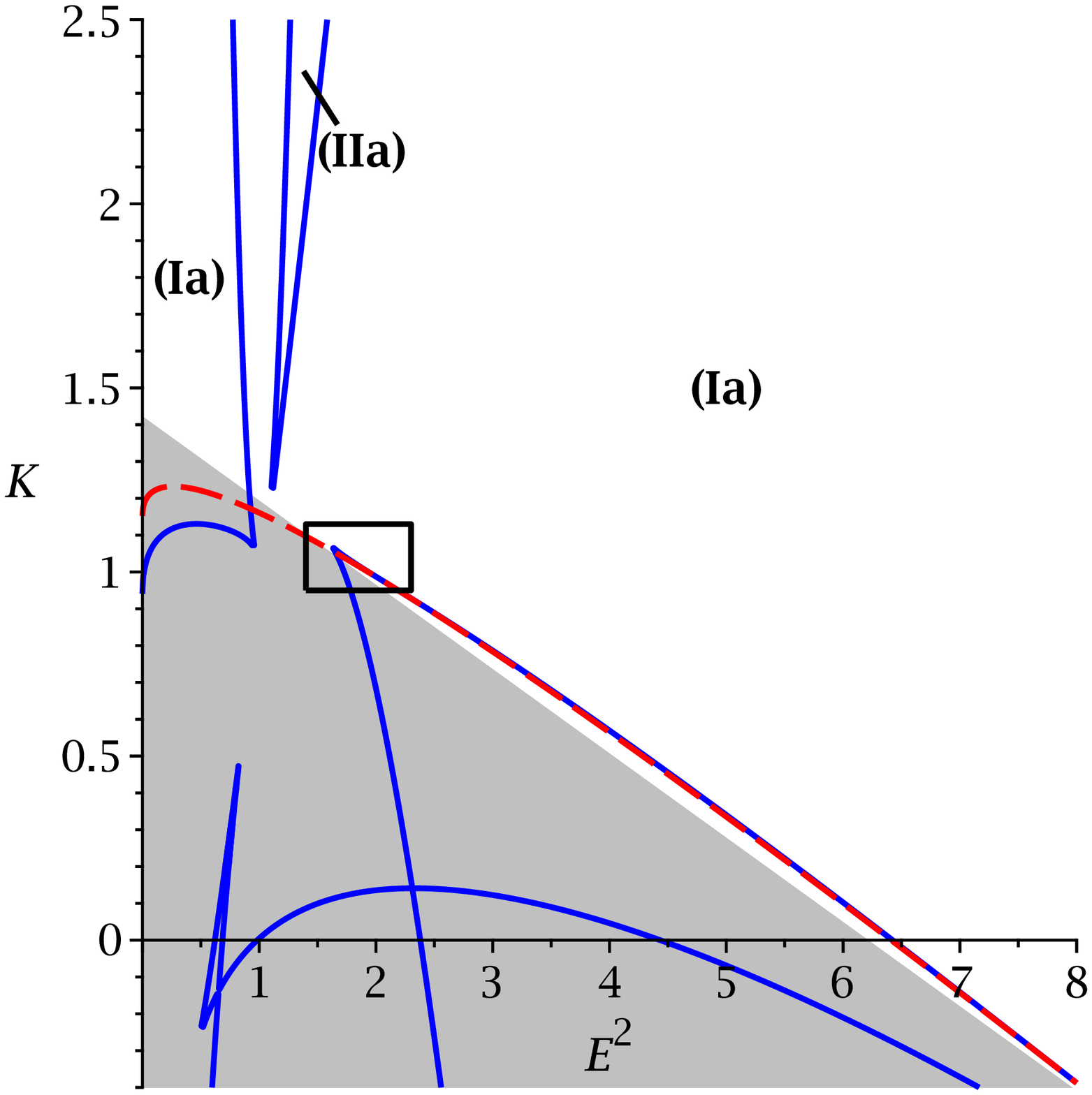}
	}
	\subfigure[Closeup of the box in Fig. \ref{pic:parameterplot-timelike}(a).]{
		\includegraphics[width=0.31\textwidth]{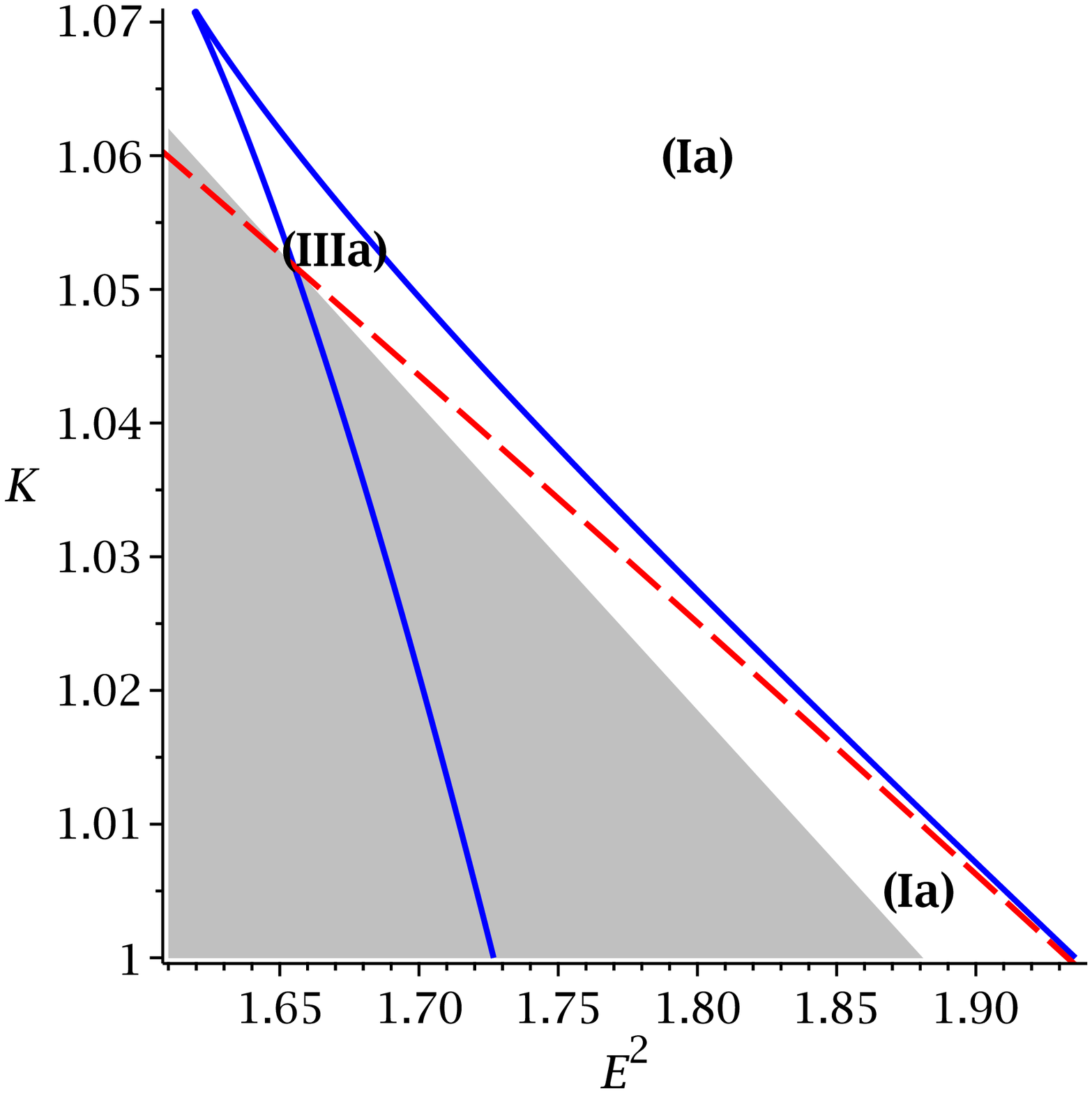}
	}
	\caption{Combined parametric $K$-$E^2$ diagrams of the $x$ equation and the $\theta$ equation for timelike geodesics. The blue curves correspond to double zeros of $X(x)$ and the red dashed curve corresponds to $X(x=-a^2)=0$. The curves separate regions with a different number of zeros. $X(x)$ has two zeros in region (Ia) and four zeros in regions (IIa) and (IIIa). The $\theta$ equation has two zeros in the white region and none in the grey region, so that geodesic motion in not possible in the grey region.}
	\label{pic:parameterplot-timelike}
\end{figure}

The radial motion is described by Eq. \eqref{eqn:x-equation} with the new coordinate $x=r^2$. The zeros of the polynomial $X(x)$ [see Eq. \eqref{eqn:Xfunc}] are the turning points of the geodesics and therefore the $x$ equation \eqref{eqn:x-equation} determines the possible orbit types. The number of zeros in the allowed coordinate range $x\in[-a^2,\infty]$ changes if double zeros occur, i.e., $X(x)=0$ and $\frac{\dd X(x)}{\dd x}=0$, or if a zero crosses $x=-a^2$ [wich leads to $X(-a^2)=0$]. From these conditions we plot parametric $K$-$E^2$ diagrams and also include the parametric diagrams for the $\theta$ equation. Figure \ref{pic:parameterplot-timelike} shows a typical example of a parametric plot in the Myers-Perry-AdS spacetime. The blue curves correspond to double zeros of $X(x)$ and the red dashed curve corresponds to $X(x=-a^2)=0$. The curves separate regions with a different number of zeros. $X(x)$ has two zeros in region (Ia) and four zeros in regions (IIa) and (IIIa). Although regions regions (IIa) and (IIIa) have the same amount of zeros, different orbit configurations occur, as we will see in the following. In both regions there are MBOs and BOs, but in region (IIa) the BOs are outside the black hole and in region (IIIa) the BOs are hidden behind the inner horizon. 

\begin{figure}[h]
	\centering
	\subfigure[$\delta=1$, $K=1.6$, $g=0.2$,  $a=0.55$, $b=0.4$, $L=0.5$, $J=0.7$]{
		\includegraphics[width=0.31\textwidth]{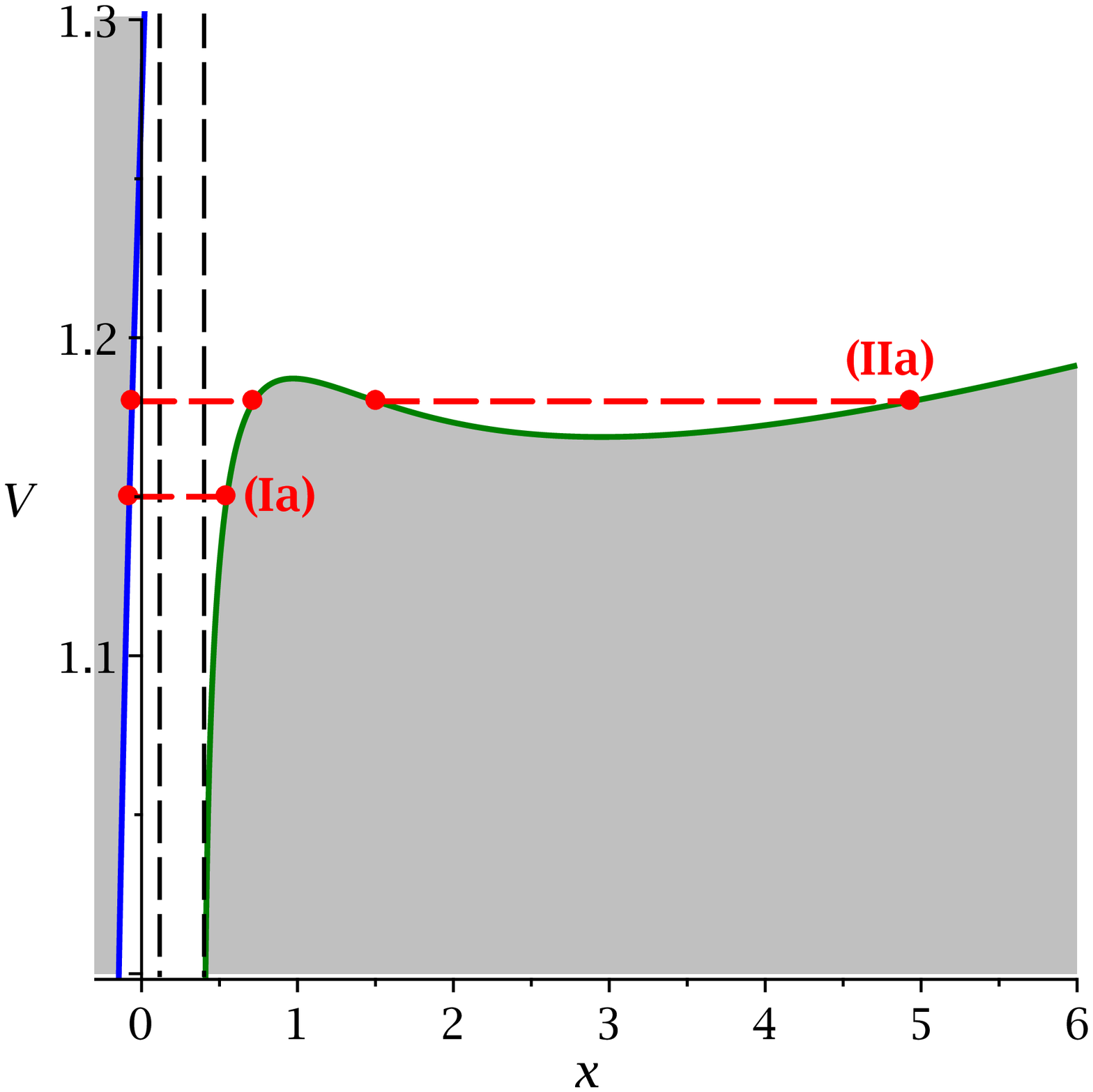}
	}
	\subfigure[$\delta=1$, $K=1.106$, $g=0.3$,  $a=0.55$, $b=0.4$, $L=0.65$, $J=0.5$]{
		\includegraphics[width=0.31\textwidth]{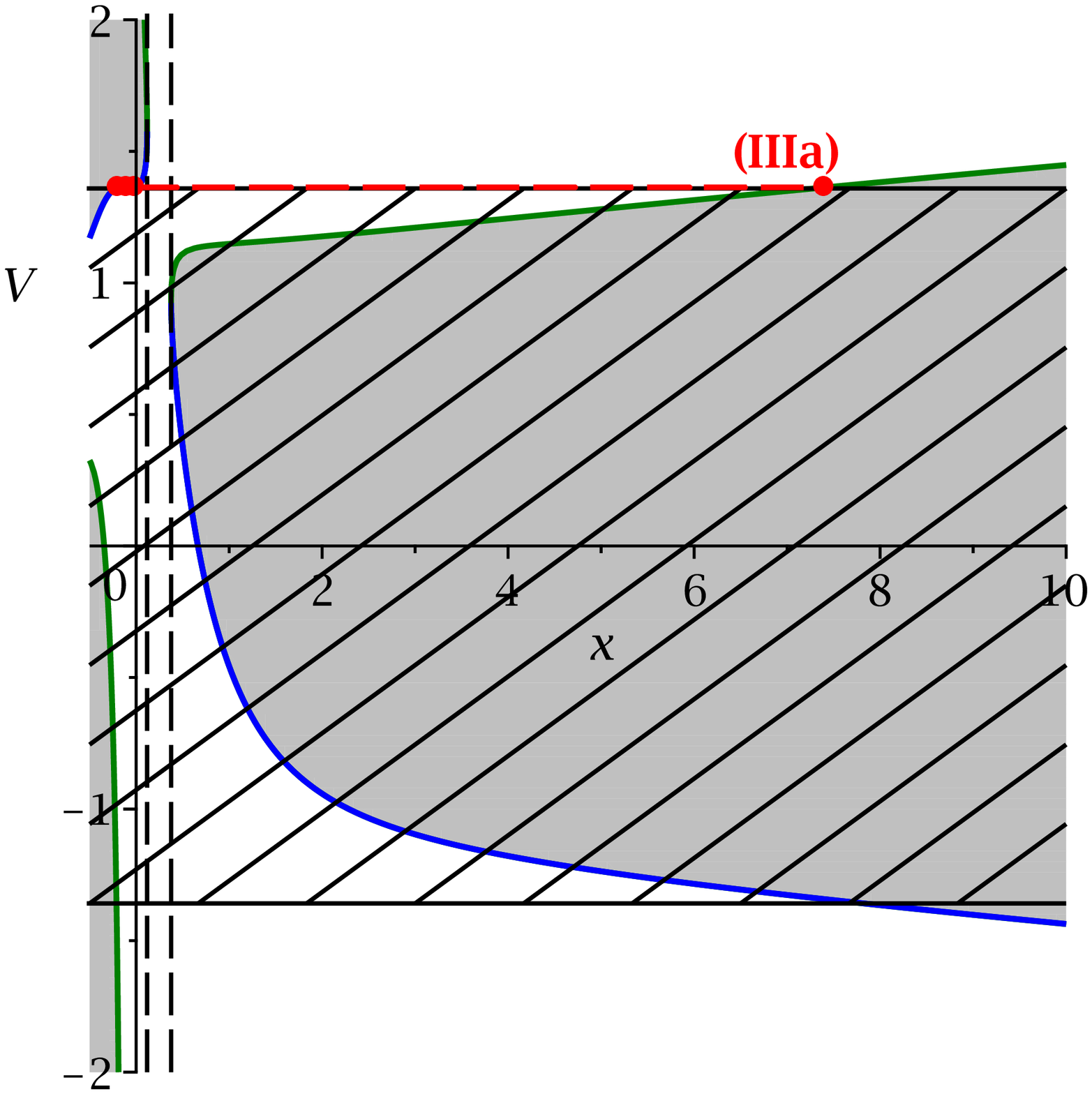}
	}
	\subfigure[Closeup of Fig. \ref{pic:potential-timelike}(b) ]{
		\includegraphics[width=0.31\textwidth]{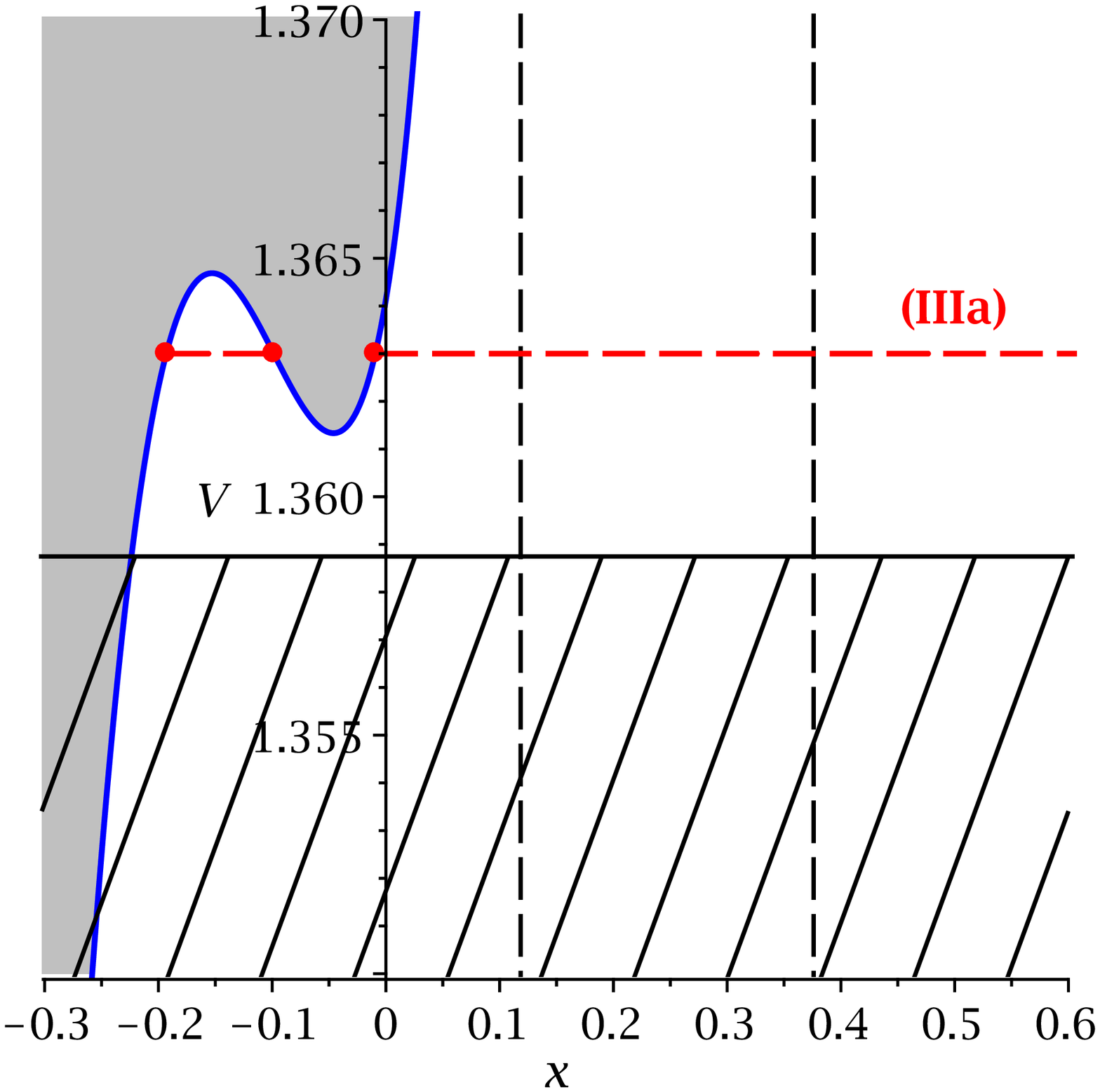}
	}
	\caption{Plots of the effective potential for timelike geodesics. The green and blue curves are the two parts of the effective potential. The grey areas are forbidden by the $x$ equation and the hatched areas are forbidden by the $\theta$ equation. Geodesic motion is possible in the white areas only. The vertical black dashed lines indicate the position of the horizons. The red dashed lines are example energies and the red dots mark the turning points. The red numbers refer to the orbit types of Table \ref{tab:orbits-timelike} and Fig. \ref{pic:parameterplot-timelike}}
	\label{pic:potential-timelike}
\end{figure}

Furthermore, we define an effective potential $V$ consisting of two parts $V_+$ and $V_-$
\begin{equation}
	\left( \frac{\dd x}{\dd\gamma} \right)^2 = X(x)= f(x) (E-V_+)(E-V-) \, .
\end{equation}
Then the effective potential for the $x$ motion is
\begin{equation}
	V_\pm =  \frac{-\beta\pm \sqrt{\beta^2-4\alpha\gamma}}{2\alpha} 
	\label{eq:radialpotential}
\end{equation}
with
\begin{align}
	\alpha =& \ 8M \left[ (a^2+b^2)x+a^2b^2 \right] +4 (x+a^2)(x+b^2)\left[ x-(a^2+b^2)g^2x - a^2b^2g^2 \right] \, , \nonumber \\
	\beta =& -16M  \left[ La(b^2+x) + Jb(a^2+x) \right] \, ,  \nonumber\\
	\gamma=& -4 L^2 \left[ \Xi_a(1+g^2x)(b^2+x)(b^2-a^2)-2M(a^2g^2x+b^2) \right]  
	    -4\Delta_x (K+\delta x) \nonumber \\
	  &  -4 J^2 \left[ \Xi_b(1+g^2x)(a^2+x)(a^2-b^2)-2M(b^2g^2x+a^2) \right]
	     +16M LJab(1+g^2x) \,  .
	\label{eq:radialpotential-coeff}
\end{align}
Figure \ref{pic:potential-timelike} shows examples of the effective potential for timelike geodesics. The green and blue curves are the two parts of the effective potential. The grey areas are forbidden by the $x$ equation and the hatched areas are forbidden by the $\theta$ equation. Geodesic motion is possible in the white areas only. The vertical black dashed lines indicate the position of the horizons. The red dashed lines are example energies for different orbits and the red dots mark the turning points. With the help of the effective potential we can now determine the orbit types in regions (Ia)-(IIIa) of the parametric diagrams (in the following we assume $x_i<x_{i+1}$):
\begin{enumerate}
	\item Region (Ia): The polynomial $X(x)$ has two zeros $x_1\leq x_-$ and $x_2\geq x_+$. $X(x)$ is positive for $x\in[x_1, x_2]$ and therefore MBOs crossing both horizons can be found in this region.
	\item Region (IIa): The polynomial $X(x)$ has four zeros $x_1\leq x_-$ and $x_2,x_3,x_4\geq x_+$.  $X(x)$ is positive for $x\in[x_1, x_2]$ and  $x\in[x_3, x_4]$. MBOs and BOs exist.
	\item Region (IIIa): The polynomial $X(x)$ has four zeros $x_1,x_2,x_3\leq x_-$ and $x_4\geq x_+$.  $X(x)$ is positive for $x\in[x_1, x_2]$ and  $x\in[x_3, x_4]$. There are BOs hidden behind the inner horizon and MBOs.
\end{enumerate}
An overview of the possible orbits types for timelike geodesics is shown in Table \ref{tab:orbits-timelike}.

\begin{table}[h!]
\begin{center}
\begin{tabular}{|cccc|}\hline
Region & Zeros & Range of $x$ & Orbit \\
\hline\hline
Ia & 2 &
\begin{pspicture}(0,-0.2)(6,0.2)
\psline[linewidth=0.5pt]{->}(0,0)(6,0)
\psline[linewidth=0.5pt](0,-0.2)(0,0.2)
\psline[linewidth=0.5pt,doubleline=true](2,-0.2)(2,0.2)
\psline[linewidth=0.5pt,doubleline=true](3.5,-0.2)(3.5,0.2)
\psline[linewidth=1.2pt]{*-*}(1.5,0)(4,0)
\end{pspicture}
  & MBO
\\ \hline
IIa & 4 &
\begin{pspicture}(0,-0.2)(6,0.2)
\psline[linewidth=0.5pt]{->}(0,0)(6,0)
\psline[linewidth=0.5pt](0,-0.2)(0,0.2)
\psline[linewidth=0.5pt,doubleline=true](2,-0.2)(2,0.2)
\psline[linewidth=0.5pt,doubleline=true](3.5,-0.2)(3.5,0.2)
\psline[linewidth=1.2pt]{*-*}(1.5,0)(4,0)
\psline[linewidth=1.2pt]{*-*}(4.5,0)(5.5,0)
\end{pspicture}
  & MBO, BO
\\ \hline
IIIa & 4 &
\begin{pspicture}(0,-0.2)(6,0.2)
\psline[linewidth=0.5pt]{->}(0,0)(6,0)
\psline[linewidth=0.5pt](0,-0.2)(0,0.2)
\psline[linewidth=0.5pt,doubleline=true](2,-0.2)(2,0.2)
\psline[linewidth=0.5pt,doubleline=true](3.5,-0.2)(3.5,0.2)
\psline[linewidth=1.2pt]{*-*}(1.5,0)(4,0)
\psline[linewidth=1.2pt]{*-*}(0.5,0)(1,0)
\end{pspicture}
  & BO, MBO
\\ \hline\hline
\end{tabular}
\caption{Types of orbits for timelike geodesics in the Myers-Perry-AdS spacetime. The range of the orbits is represented by thick lines. The turning points are marked by thick dots. The two vertical double lines indicate the position of the horizons and the single vertical line corresponds to the singularity.}
\label{tab:orbits-timelike}
\end{center}
\end{table}

\begin{figure}[h]
	\centering
	\subfigure[Parametric $a$-$b$-diagram]{
		\includegraphics[width=0.31\textwidth]{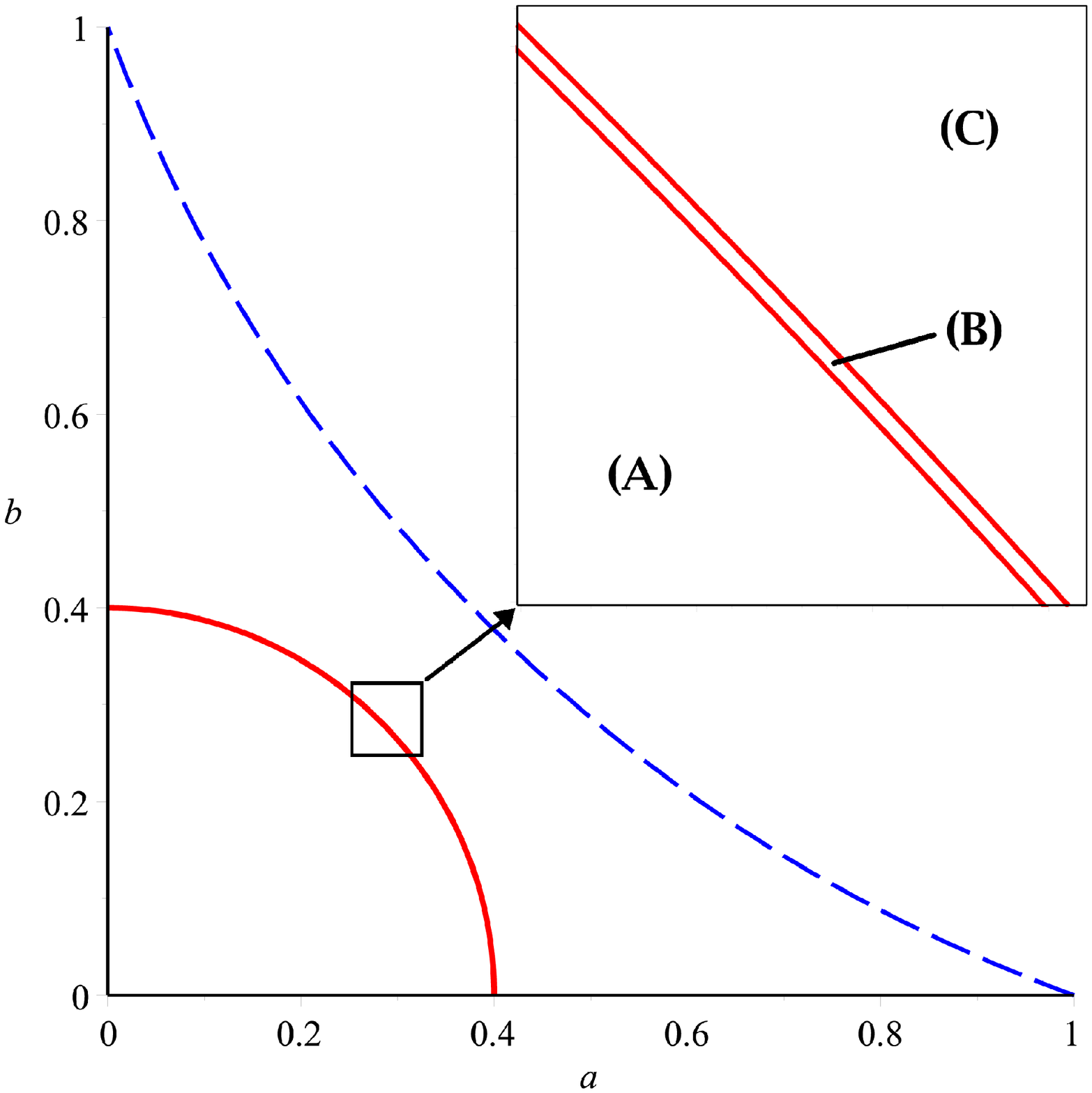}
	}
	\subfigure[Effective potential from region (B) with $a=0.3$, $b=.264$]{
		\includegraphics[width=0.31\textwidth]{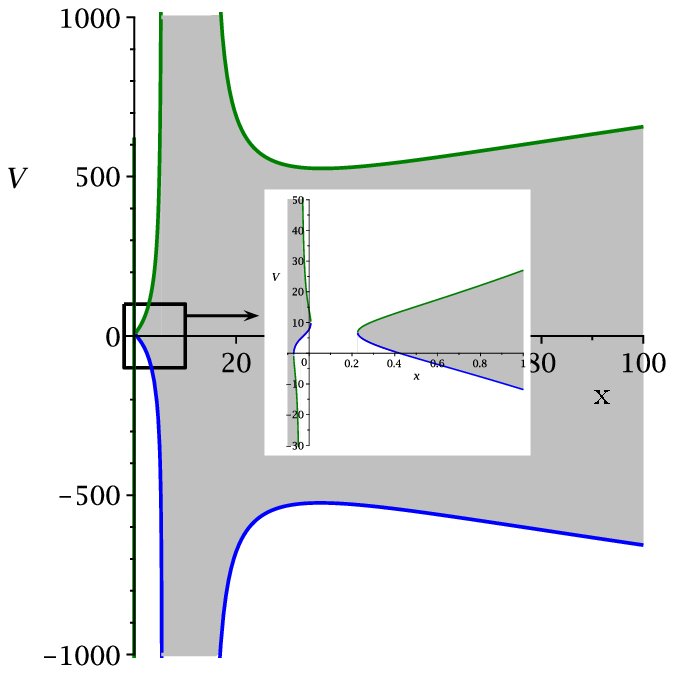}
	}
	\subfigure[Effective potential from region (C) with $a=0.3$, $b=.27$]{
		\includegraphics[width=0.31\textwidth]{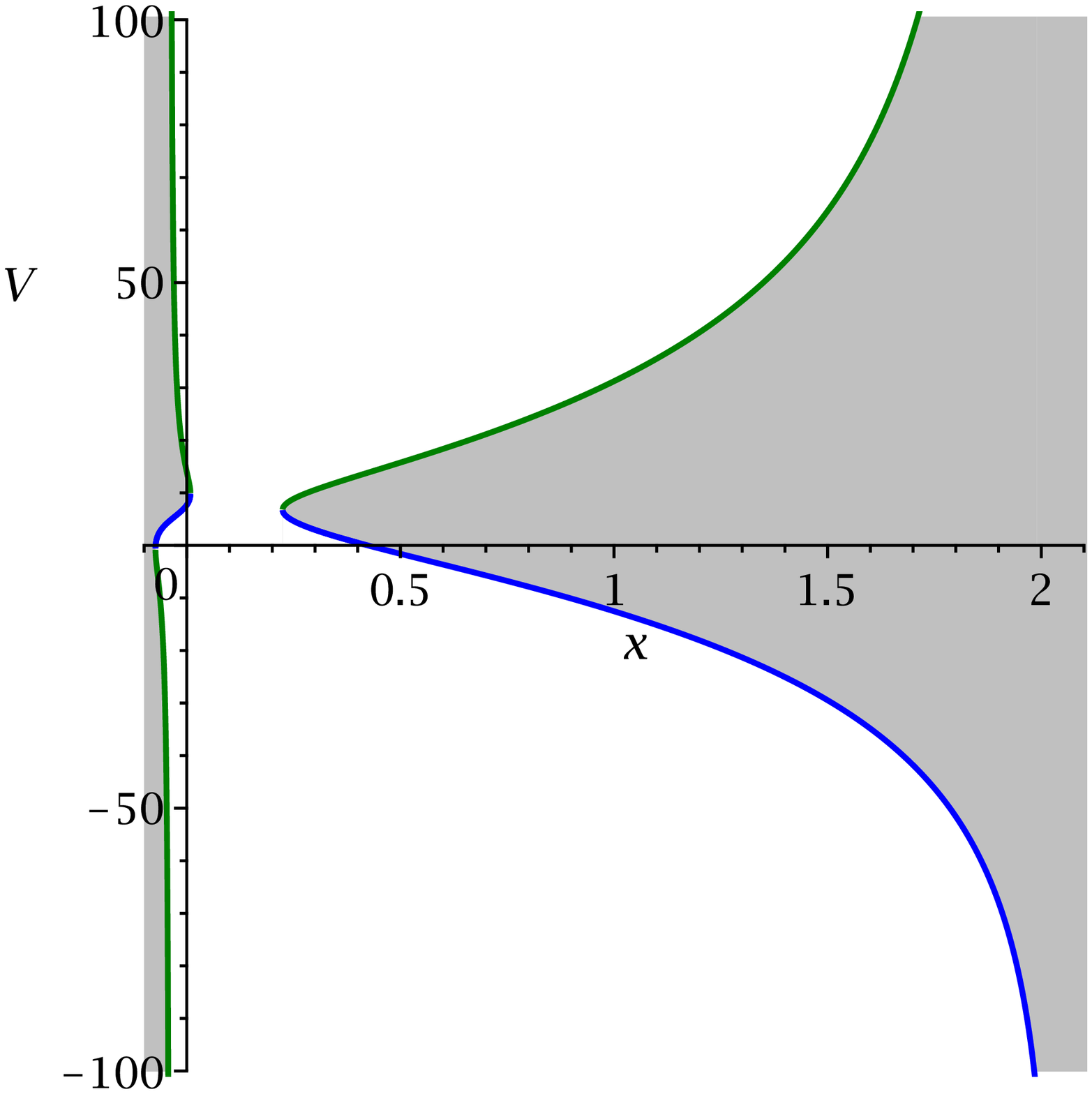}
	}
	\caption{Parametric $a$-$b$ diagram ($M=0.5$, $g=2.5$) showing three regions with different numbers of zeros of $\alpha$ and corresponding effective potentials from regions (B) and (C) with parameters $g=2.5$, $\delta = 1$, $L = 2$, $J = 1$, $K = 5$. In region (A) $\alpha$ has a single negative zero. In region (B) $\alpha$ has two positive zeros and a single negative zero. In region (C) $\alpha$ has a positive and a negative zero. The red curves in panel (a) separate the different regions and the blue dashed curve separates the black hole and the naked singularity. The blue and green curves in panel (b) and (c) are the effective potential $V$. In the grey region geodesic motion is not possible.}
	\label{pic:potdiverge}
\end{figure}

The asymptotic behavior of the effective potential is given by
\begin{equation}
\lim\limits_{x \rightarrow \infty} V_\pm = \pm \infty \, .
\end{equation}
Furthermore, from the Eqs. \eqref{eq:radialpotential} and \eqref{eq:radialpotential-coeff} it is obvious that the effective potential diverges for $x<\infty$ at the zeros of $\alpha$. To find the parameter regions in which the potential diverges for $x<\infty$ we plot parameteric $a$-$b$ diagrams, see Fig. \ref{pic:potdiverge}(a). The red curves in the diagram separate three regions with different numbers of zeros of $\alpha$. Below the blue dashed curve the black hole has two horizons and above the curve there is a naked singularity. In region (A) the potential diverges  for $x \rightarrow \infty$ and a negative $x$ value, since $\alpha$ has a single negative zero. The potential diverges for  $x \rightarrow \infty$ and additionally for two positive $x$ values in region (B), see Fig. \ref{pic:potdiverge}(b). In region (C) the potential diverges for a positive $x$ and a negative $x$; see Fig. \ref{pic:potdiverge}(c).

\FloatBarrier

\subsection{Null geodesics}

Null geodesics or lightlike geodesics describe the world lines of massless particles ($\delta =0$), such as photons. Their properties can be used in order to discuss spacetime observables like the light deflection, which is described by a lightlike escape orbit. Furthermore the shadow of the black hole can be calculated by means of null geodesics.

\subsubsection{The $\theta$ motion}
\begin{figure}[h]
	\centering
	\subfigure[$K=2$]{
		\includegraphics[width=0.31\textwidth]{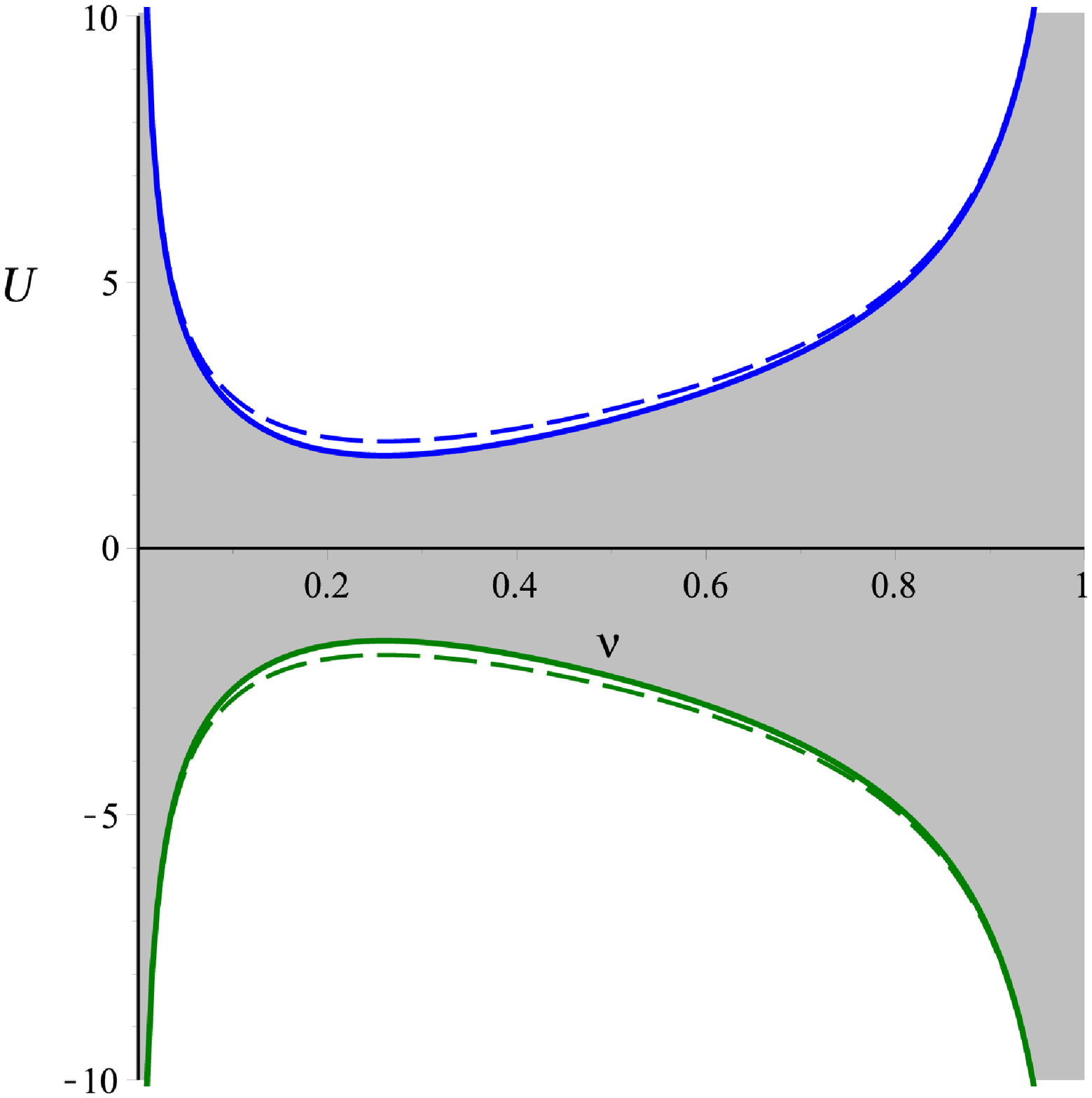}
	}
	\subfigure[$K=2.6$]{
		\includegraphics[width=0.31\textwidth]{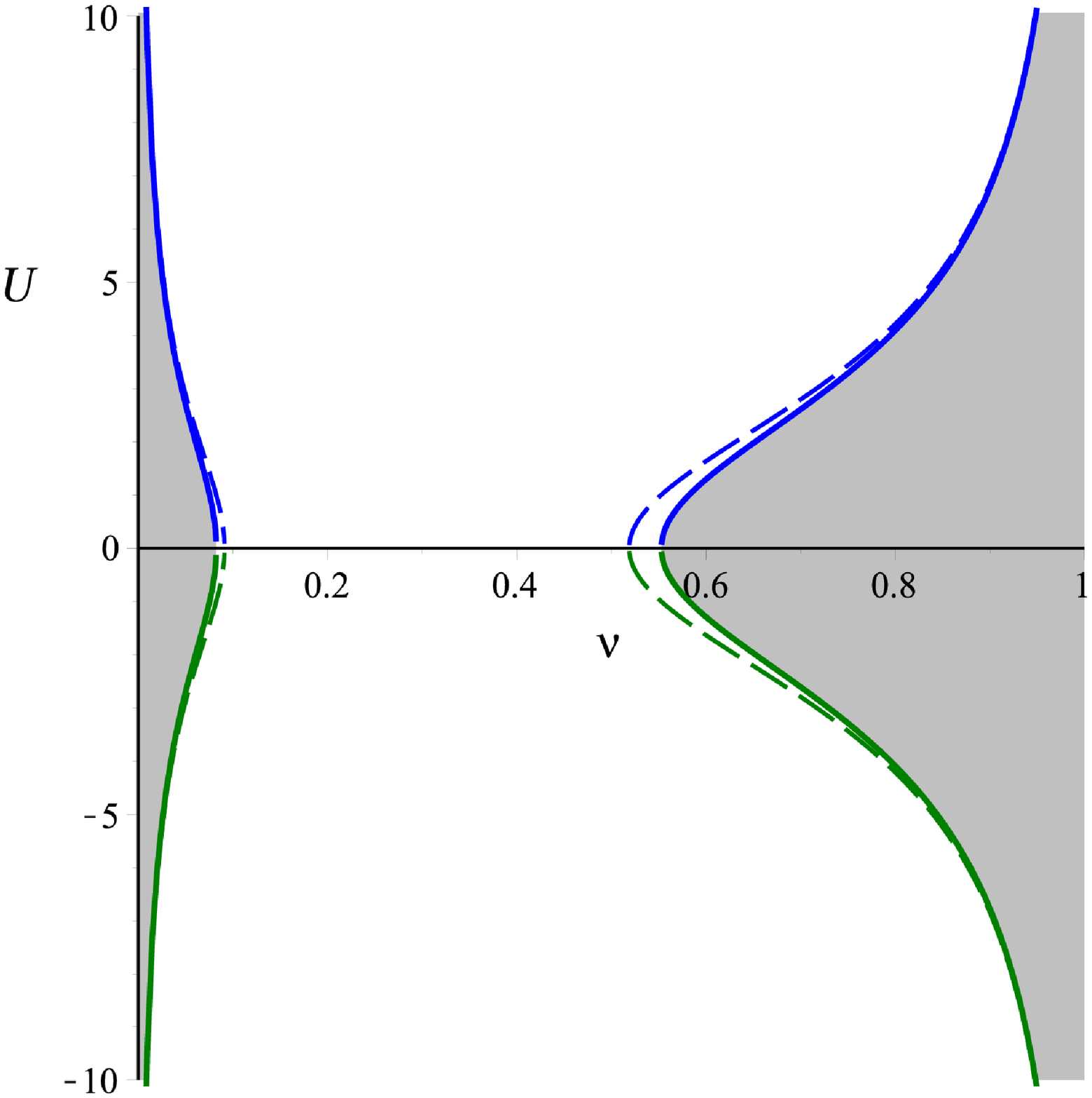}
	}
	\subfigure[$K=3.5$]{
		\includegraphics[width=0.31\textwidth]{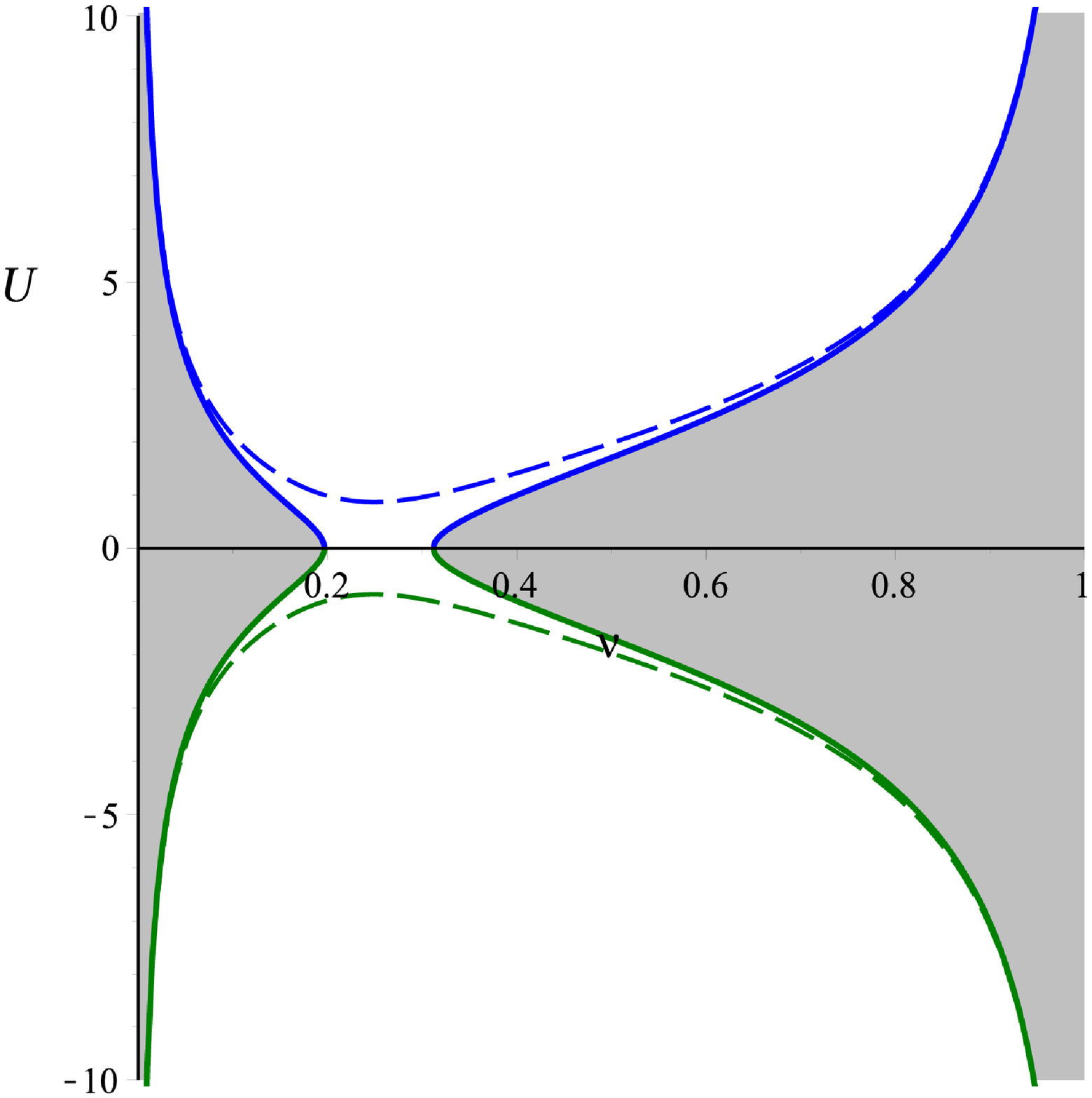}
	}
	\caption{Plots of the effective $\theta$ potential for $\delta=0$, $g=0.1$,$a=0.5$, $b=0.4$, $L=1.2$, $J=0.4$ and different $K$. The solid green and blue curves are the two parts of the lightlike effective potential and the dashed lines show the corresponding timelike effective potentials.The grey areas are forbidden by the $\theta$ equation. Geodesic motion is possible in the white areas only.}
	\label{pic:theta-potential-lightlike}
\end{figure}

We will start our analysis of null geodesics by studying the $\theta$ equation in its substituted form \eqref{eqn:nu-equation} for $\delta =0$. As we did in the case of timelike geodesics, we will define an effective potential
\begin{equation}
	U_\pm = \pm \sqrt{ \frac{\Delta_\nu}{\Xi_a a^2\nu+\Xi_b b^2(1-\nu)} \left[-K  +\frac{L^2\Xi_a}{1-\nu} +\frac{J^2\Xi_b}{\nu}  \right] } \,
\end{equation}
in order to investigate the possible set of zeros confining the $\theta$ motion. Figure \ref{pic:theta-potential-lightlike} depicts three typical effective potentials. 

We can see that the possible set of zeros of the $\theta$ equation is quite similar to the timelike case. Nevertheless, for certain values of the Carter constant $K$, it is possible to attain energy values in the lightlike case, which are not valid for timelike geodesics with the same set of parameters.

\FloatBarrier
\subsubsection{The $x$ motion}	

\begin{figure}[h]
	\centering
\includegraphics[width=0.31\textwidth]{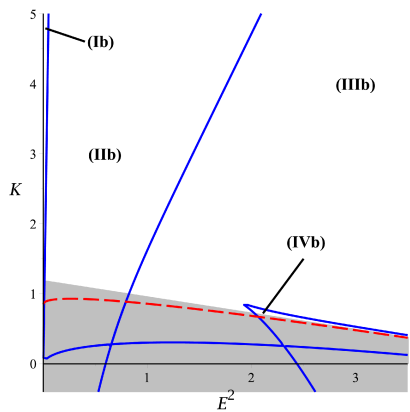}
	\caption{Combined parametric $K$-$E^2$ diagrams of the $x$ equation and the $\theta$ equation for lightlike geodesics with $\delta=0$, $g=0.1$, $a=0.55$, $b=0.4$, $L=0.6$, $J=0.5$. The blue curves correspond to double zeros of $X(x)$ and the red dashed curve corresponds to $X(x=-a^2)=0$. The curves separate regions with a different number of zeros. $X(x)$ has one real zero in region (Ib), three real zeros in region (IIb) and (IVb) and two real zeros in region (IIIb). The $\theta$ equation has two zeros in the white region and none in the grey region, so that geodesic motion in not possible in the grey region.}
	\label{pic:parameterplot-lightlike}
\end{figure}

In order to classify the possible set of orbit types for null geodesics, we will investigate the radial equation \eqref{eqn:x-equation} by means of parametric diagrams and effective potentials as we did for timelike geodesics. The combined parametric $K$-$E^2$ diagram for the lightlike $\theta$  and $r$ motion is shown in Fig. \ref{pic:parameterplot-lightlike}.

In contrast to the timelike orbits of the Myers-Perry-AdS spacetime, there is one more region in the case of lightlike orbits.

\begin{enumerate}
\item Region (Ib): The polynomial $X(x)$ has as single zero $x_1\leq x_-$. $X(x)$ is positive for $x > x_1$ and therefore TWEOs crossing both horizons can be found in this region.
\item Region (IIb): The polynomial $X(x)$ has three zeros $x_1\leq x_-$ and $x_2,x_3\geq x_+$.  $X(x)$ is positive for $x\in(x_1, x_2)$ and $x > x_3$. MBOs and EOs exist.
\item Region (IIIb): The polynomial $X(x)$ has two zeros $x_1\leq x_-$ and $x_2\geq x_+$.  $X(x)$ is positive for $x\in(x_1, x_2)$ and therefore MBOs crossing both horizons can be found in this region.
\item Region (IVb): The polynomial $X(x)$ has three zeros $x_1,x_2,x_3\leq x_-$.  $X(x)$ is positive for $x\in(x_1, x_2)$ and  $x > x_3$. There are BOs hidden behind the inner horizon and TWEOs.
\end{enumerate}

\begin{figure}[h]
	\centering
	\subfigure[$\delta=0$, $K=4$, $g=0.2$,  $a=0.5$, $b=0.4$, $L=0.5$, $J=0.3$]{
		\includegraphics[width=0.31\textwidth]{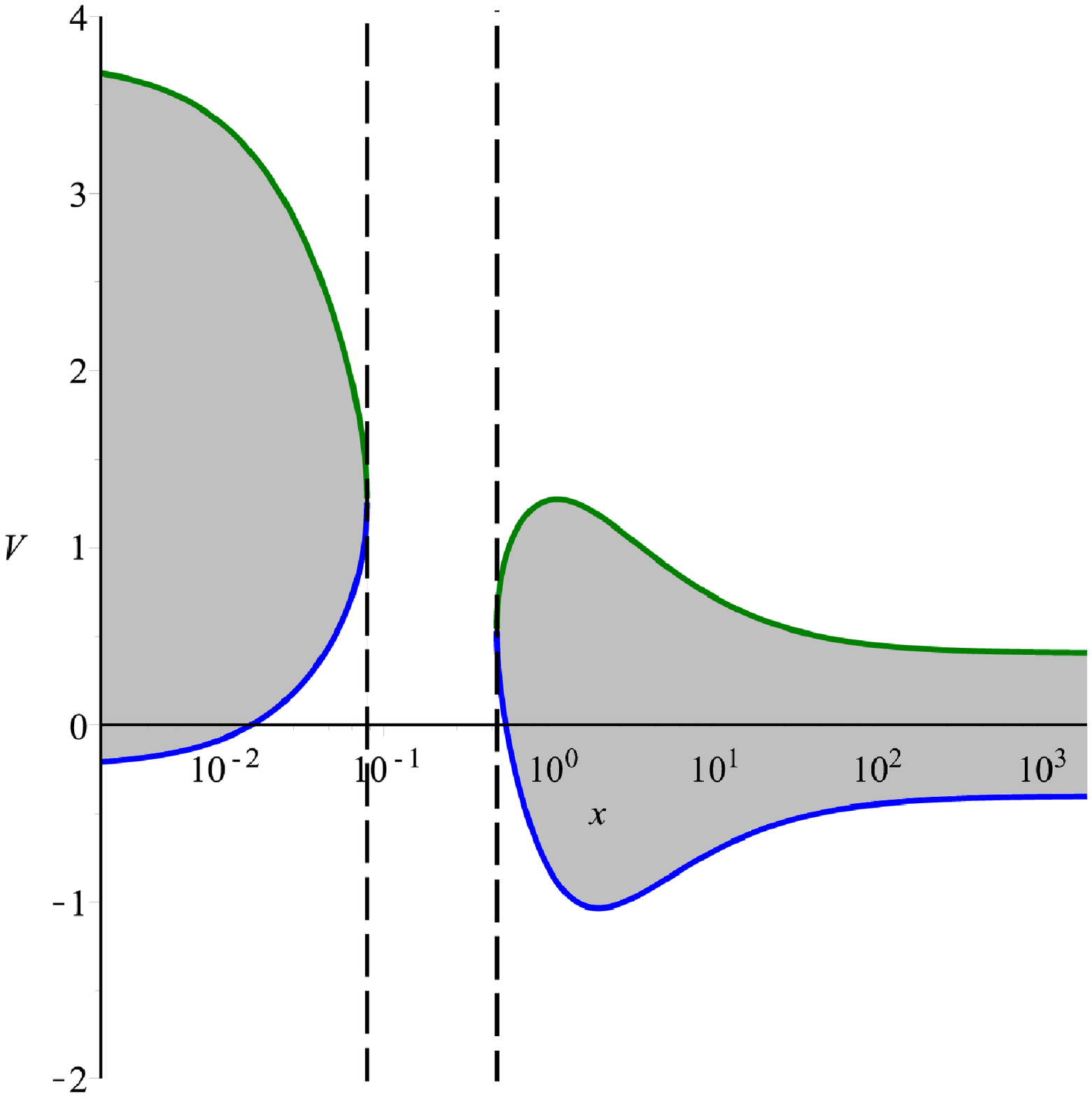}
		\label{pic:potential-lightlikea}
	}
	\subfigure[[$\delta=0$, $K=4$, $g=0.2$,  $a=0.5$, $b=0.4$, $L=0.5$, $J=0.3$]{
		\includegraphics[width=0.31\textwidth]{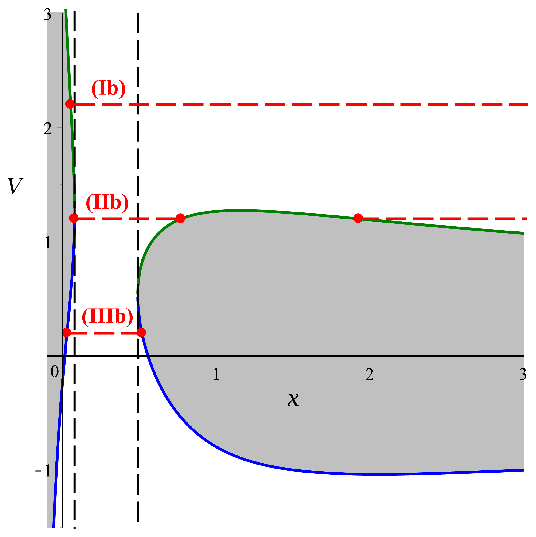}
		\label{pic:potential-lightlikeb}
	}
	\subfigure[$\delta=0$, $K=1.106$, $g=0.3$,  $a=0.55$, $b=0.4$, $L=0.75$, $J=0.5$]{
		\includegraphics[width=0.31\textwidth]{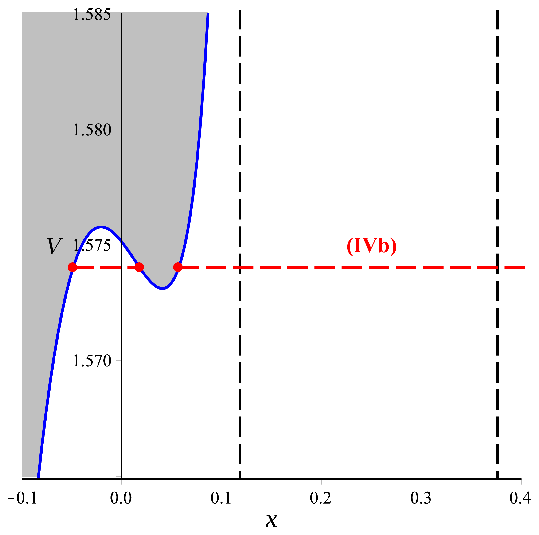}
		\label{pic:potential-lightlikec}
	}
	\caption{Plots of the effective potential for lightlike geodesics in a semilogarithmic overview plot \subref{pic:potential-lightlikea} and two detailed plots including orbit types \subref{pic:potential-lightlikeb}-\subref{pic:potential-lightlikec}. The green and blue curves are the two parts of the effective potential. The grey areas are forbidden by the $x$ equation and the hatched areas are forbidden by the $\theta$ equation. Geodesic motion is possible in the white areas only. The vertical black dashed lines indicate the position of the horizons. The red dashed lines are example energies and the red dots mark the turning points. The red numbers refer to the orbit types of Table \ref{tab:orbits-lightlike} and Fig. \ref{pic:parameterplot-lightlike}}
	\label{pic:potential-lightlike}
\end{figure}

 The corresponding orbit types can be visualized by effective potentials. Proceeding in the same way as for timelike geodesics, we define the effective potential according to Eq. \eqref{eq:radialpotential} with $\delta = 0$. In the case of null geodesics, the asymptotic behavior of the effective potential is given by
\begin{equation}
	\lim\limits_{r \rightarrow \infty} V_\pm = \pm \frac{g \sqrt{K}}{\sqrt{1 - \left(a^2 + b^2\right)g^2}}.
\end{equation}
This behavior is quite similar to the Kerr-Anti-de Sitter spacetime, but contrary to the timelike case, where the effective potentials diverge for radial infinity. Furthermore, the same divergences as discussed for the timelike case occur, due to the denominator $\alpha$ in Eq. \eqref{eq:radialpotential} defining the effective potential. Since $\alpha$ is independent of $\delta$, there is no difference to the timelike case for this special conditions. Figure \ref{pic:potential-lightlike} shows some typical effective potentials presenting the possible orbit types for null geodesics. The complete classification of lightlike orbits in the Myers-Perry-AdS spacetime is given in Table \ref{tab:orbits-lightlike}.

\begin{table}[h]
\begin{center}
\begin{tabular}{|cccc|}\hline
Region & Zeros & Range of $x$ & Orbit \\
\hline\hline
Ib & 1 &
\begin{pspicture}(0,-0.2)(6,0.2)
\psline[linewidth=0.5pt]{->}(0,0)(6,0)
\psline[linewidth=0.5pt](0,-0.2)(0,0.2)
\psline[linewidth=0.5pt,doubleline=true](2,-0.2)(2,0.2)
\psline[linewidth=0.5pt,doubleline=true](3.5,-0.2)(3.5,0.2)
\psline[linewidth=1.2pt]{*-}(1.5,0)(6,0)
\end{pspicture}
& TWEO
\\ \hline
IIb & 3 &
\begin{pspicture}(0,-0.2)(6,0.2)
\psline[linewidth=0.5pt]{->}(0,0)(6,0)
\psline[linewidth=0.5pt](0,-0.2)(0,0.2)
\psline[linewidth=0.5pt,doubleline=true](2,-0.2)(2,0.2)
\psline[linewidth=0.5pt,doubleline=true](3.5,-0.2)(3.5,0.2)
\psline[linewidth=1.2pt]{*-*}(1.5,0)(4,0)
\psline[linewidth=1.2pt]{*-}(4.5,0)(6,0)
\end{pspicture}
& MBO, EO
\\ \hline
IIIb & 2 &
\begin{pspicture}(0,-0.2)(6,0.2)
\psline[linewidth=0.5pt]{->}(0,0)(6,0)
\psline[linewidth=0.5pt](0,-0.2)(0,0.2)
\psline[linewidth=0.5pt,doubleline=true](2,-0.2)(2,0.2)
\psline[linewidth=0.5pt,doubleline=true](3.5,-0.2)(3.5,0.2)
\psline[linewidth=1.2pt]{*-*}(1.5,0)(4,0)
\end{pspicture}
& MBO
\\ \hline
IVb & 3 &
\begin{pspicture}(0,-0.2)(6,0.2)
\psline[linewidth=0.5pt]{->}(0,0)(6,0)
\psline[linewidth=0.5pt](0,-0.2)(0,0.2)
\psline[linewidth=0.5pt,doubleline=true](2,-0.2)(2,0.2)
\psline[linewidth=0.5pt,doubleline=true](3.5,-0.2)(3.5,0.2)
\psline[linewidth=1.2pt]{*-*}(0.5,0)(1,0)
\psline[linewidth=1.2pt]{*-}(1.5,0)(6,0)
\end{pspicture}
& BO, TWEO
\\ \hline\hline
\end{tabular}
\caption{Types of orbits for lightlike geodesics in the Myers-Perry-AdS spacetime. The range of the orbits is represented by thick lines. The turning points are marked by thick dots. The two vertical double lines indicate the position of the horizons and the single vertical line corresponds to the singularity.}
\label{tab:orbits-lightlike}
\end{center}
\end{table}	

\FloatBarrier

\subsubsection{The photon region}

If an observer points a telescope at a black hole, he or she will notice a region in the sky which stays dark. This is the shadow of a black hole. There are two kinds of light rays in the surroundings of a black hole: those escaping the black hole and those falling beyond the horizon. The two kinds of orbits are separated by unstable spherical light orbits, which can be found in the so-called photon region.
The shadow of the black hole is an image of the photon region as seen by a fixed observer and can be obtained via a coordinate transformation; see e.g., \cite{Grenzebach:2014fha}-\cite{Grenzebach:Springer}.

To obtain the photon region for the five-dimensional Myers-Perry-AdS black holes, we consider null geodesics on unstable spherical orbits. For $\delta=0$ these obey the conditions
\begin{equation}
	\frac{\dd x}{\dd \gamma}=0 \quad \text{and} \quad \frac{\dd^2 x}{\dd \gamma^2}=0 \, .
\label{eqn:pr-conditions}
\end{equation}
The conditions depend on the parameters $a$, $b$ and $g$ of the black hole and on the constants of motion $K$, $L$, $J$ and $E$. We define the impact parameters
\begin{equation}
	K_E=\frac{K}{E^2} \ , \quad  L_E=\frac{L}{E}\quad \text{and} \quad J_E=\frac{J}{E} \, .
\end{equation}
The conditions \eqref{eqn:pr-conditions} are solved for two of the impact parameters (here $K_E$ and $L_E$). However, the expressions are too long to be displayed here.

The $\theta$ equation of motion \eqref{eqn:theta-equation} yields
\begin{equation}
 \Theta(\theta)\geq 0 \,.
\end{equation}
By inserting $K_E$ and $L_E$ which we obtained before, we get a relation for the photon region depending on the coordinates $x$ and $\theta$ and the parameters $a$, $b$, $g$ and $J_E$. 

\begin{figure}[h]
	\centering
	\subfigure[$J_E=0$]{
		\includegraphics[width=0.31\textwidth]{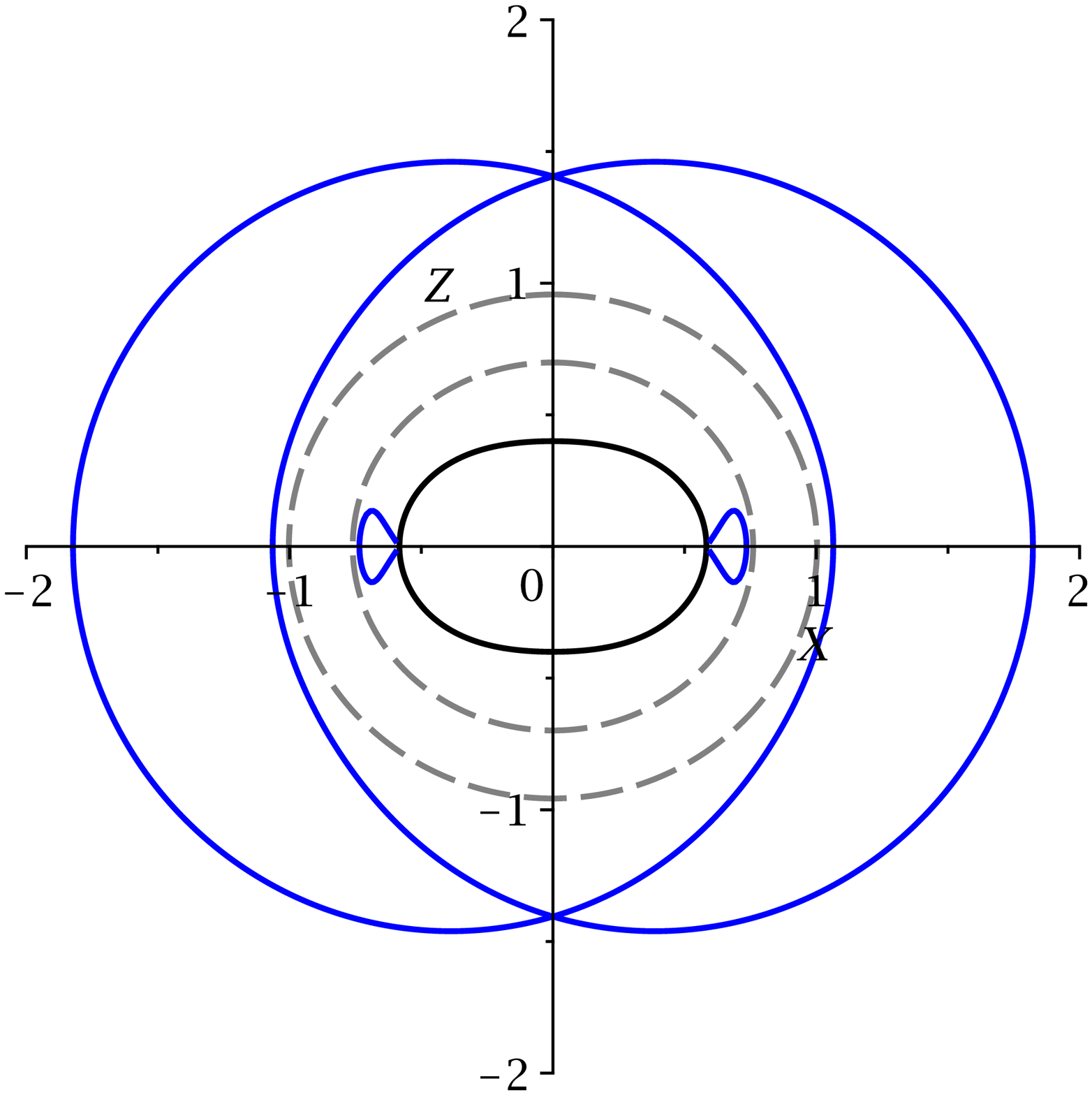}
	}
	\subfigure[$J_E=0.1$]{
		\includegraphics[width=0.31\textwidth]{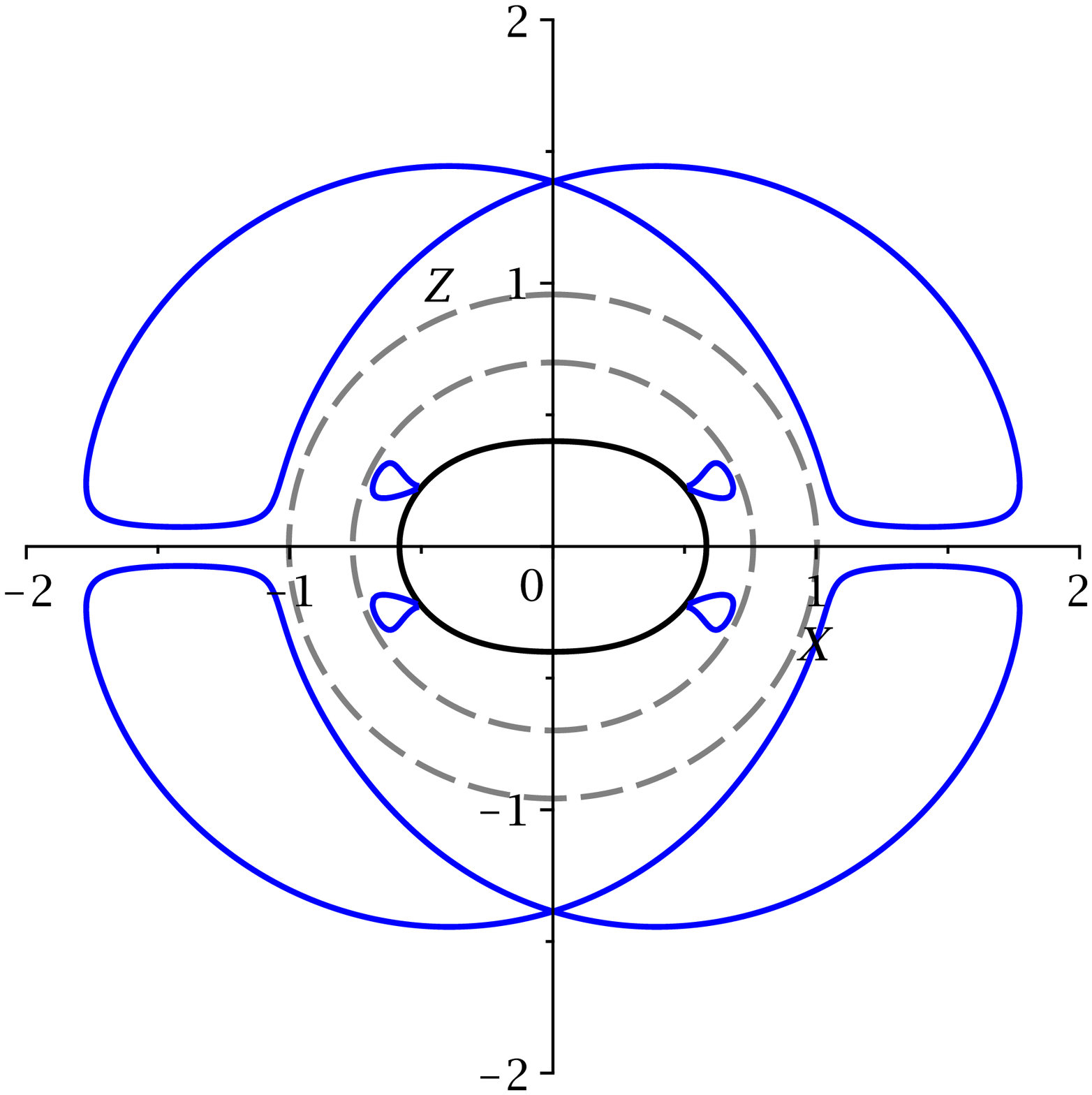}
	}
	\subfigure[$J_E=0.6$]{
		\includegraphics[width=0.31\textwidth]{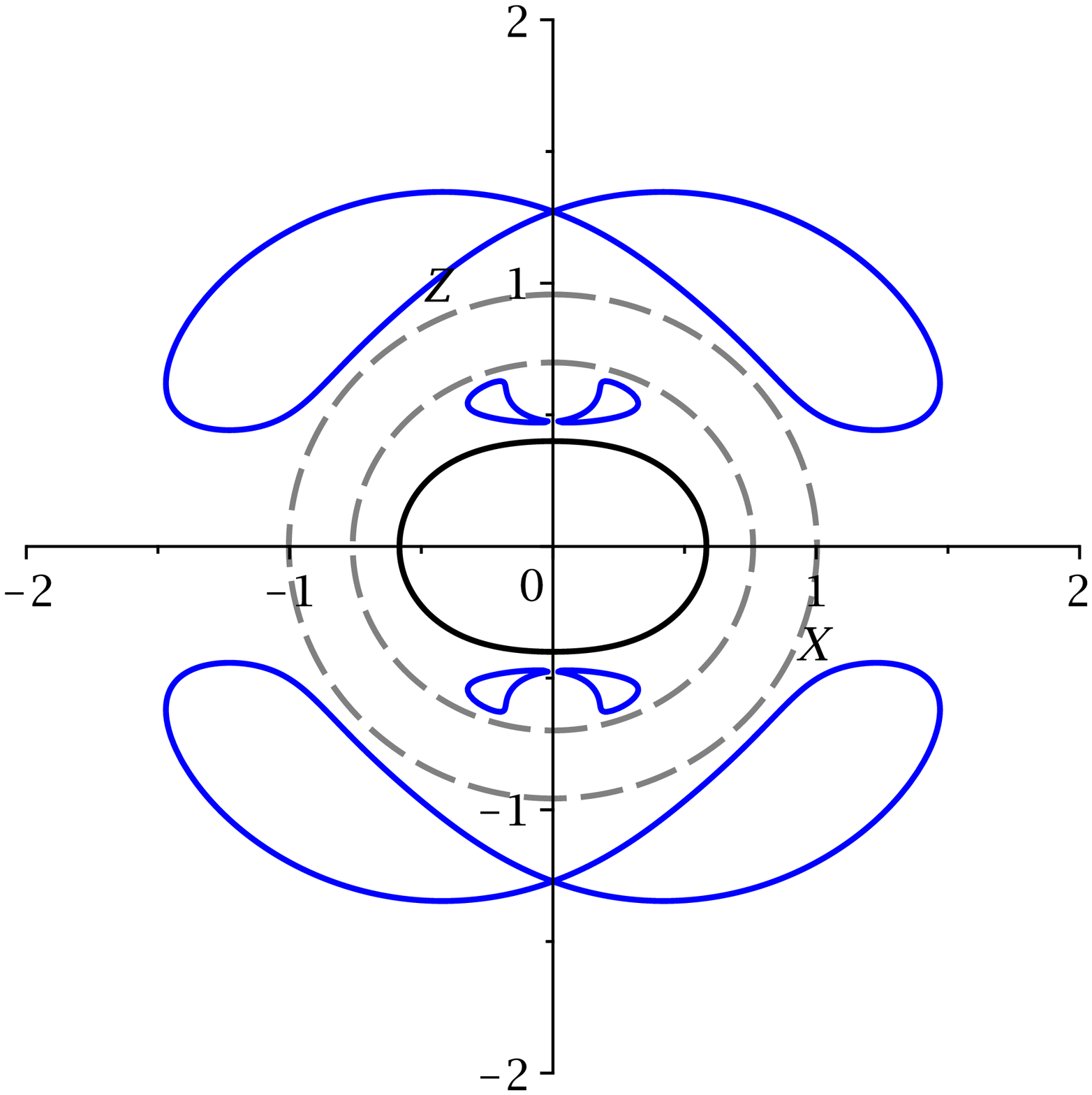}
	}
	\caption{Plots of the photon region in the Myers-Perrs-AdS spacetime for $a=0.5$, $b=0.4$, $g=0.1$ and different $J_E$ in the $X$-$Z$ plane. The black curve is the singularity and the dashed grey curves indicate the positions of the horizons. The blue curve is the boundary of the photon region.}
	\label{pic:photonregion}
\end{figure}

\begin{figure}[h]
	\centering
	\subfigure[$a=0.5$, $b=0$]{
		\includegraphics[width=0.43\textwidth]{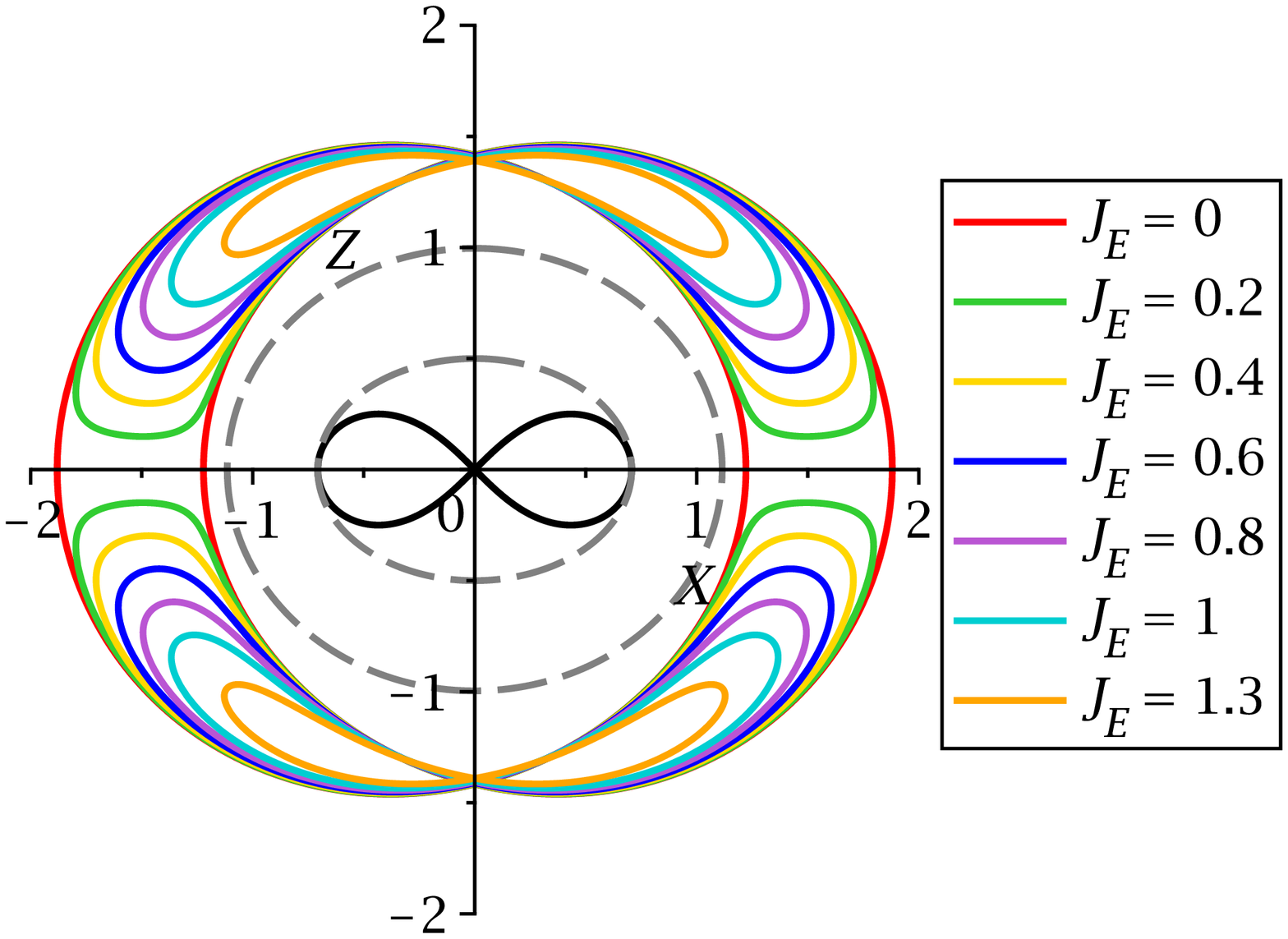}
	}
	\subfigure[$a=0.5$, $b=0.2$]{
		\includegraphics[width=0.43\textwidth]{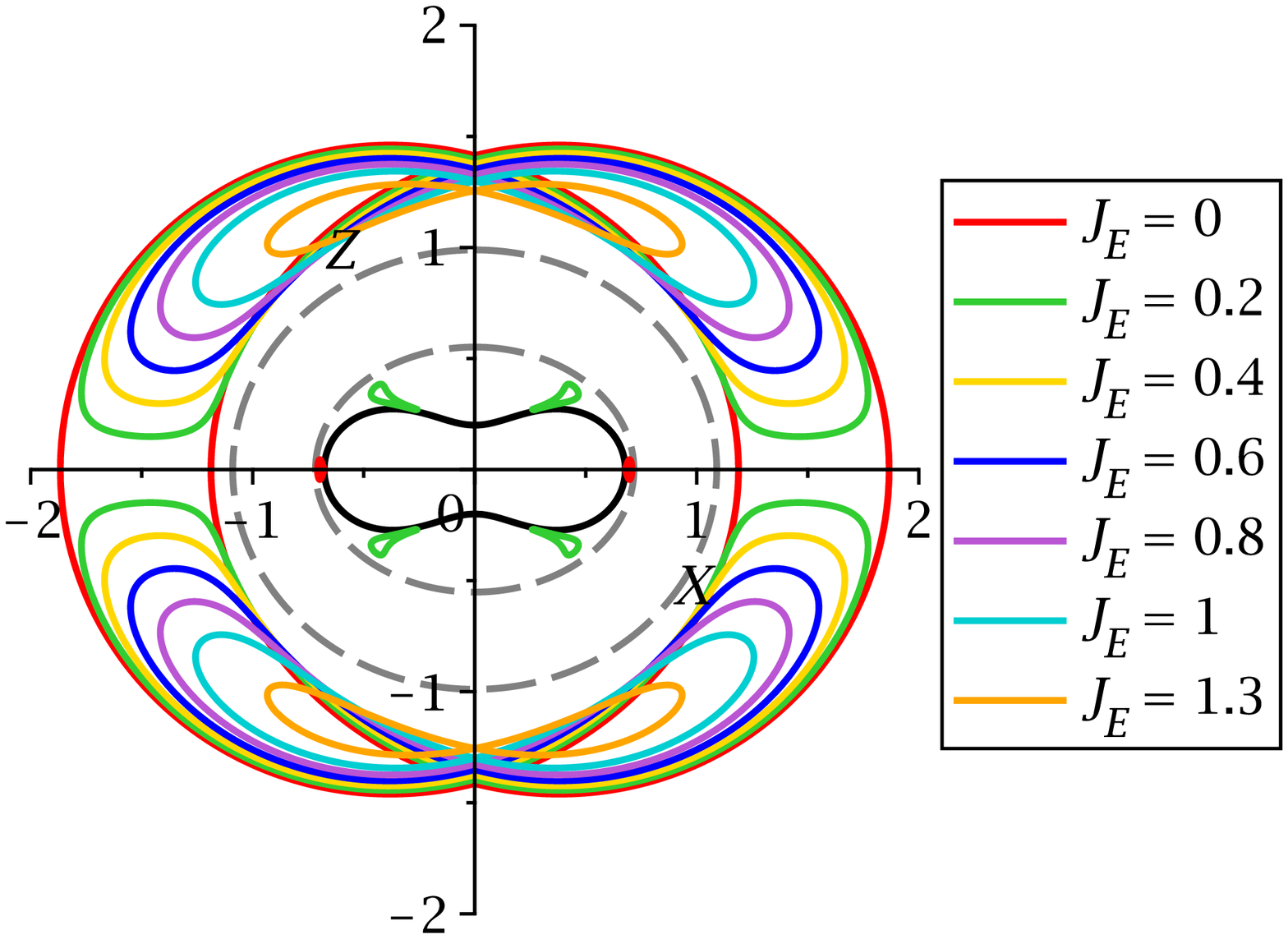}
	}
	\subfigure[$a=0.5$, $b=0.4$]{
		\includegraphics[width=0.43\textwidth]{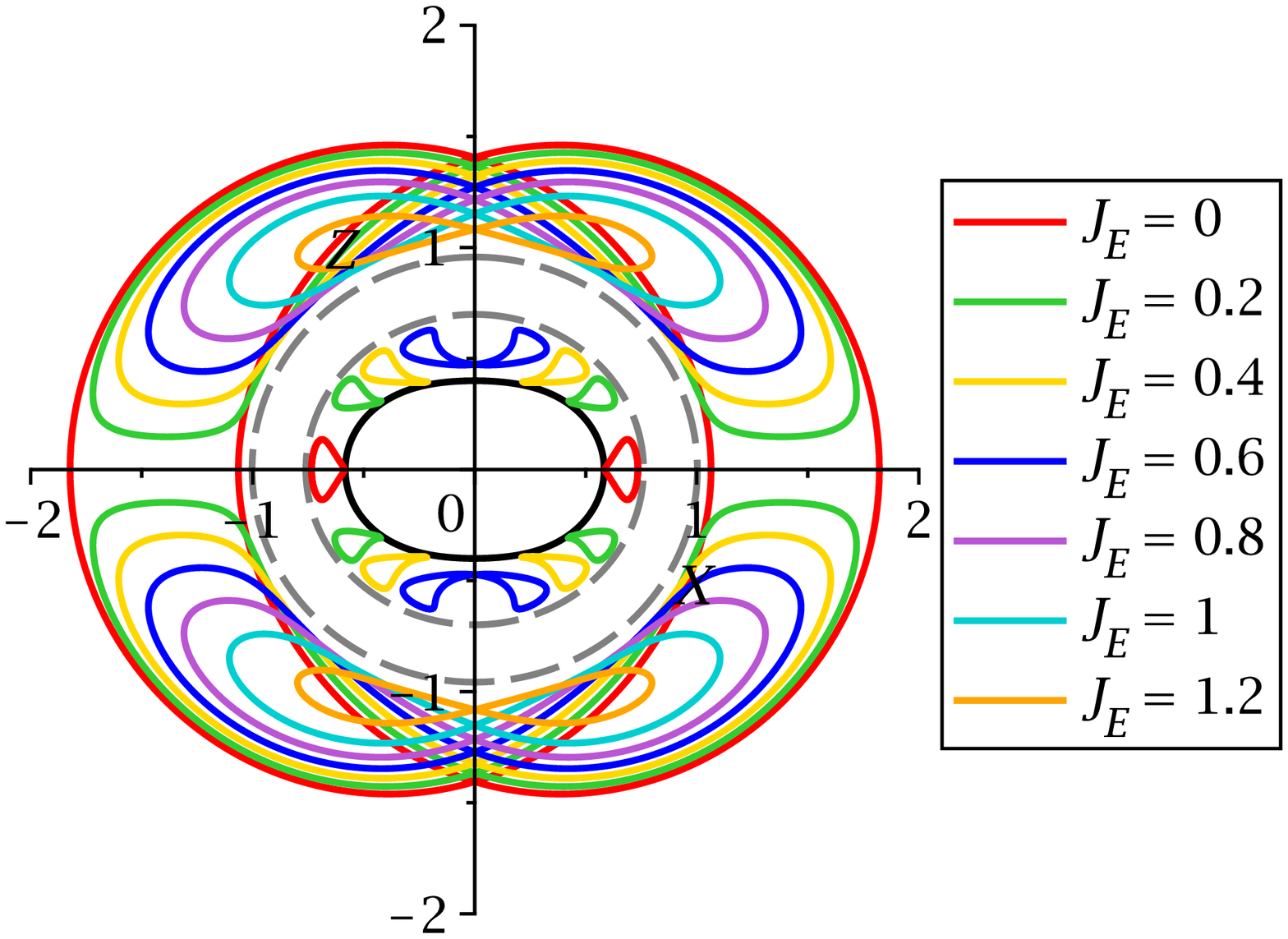}
	}
	\subfigure[$a=0.4$, $b=0.5$]{
		\includegraphics[width=0.43\textwidth]{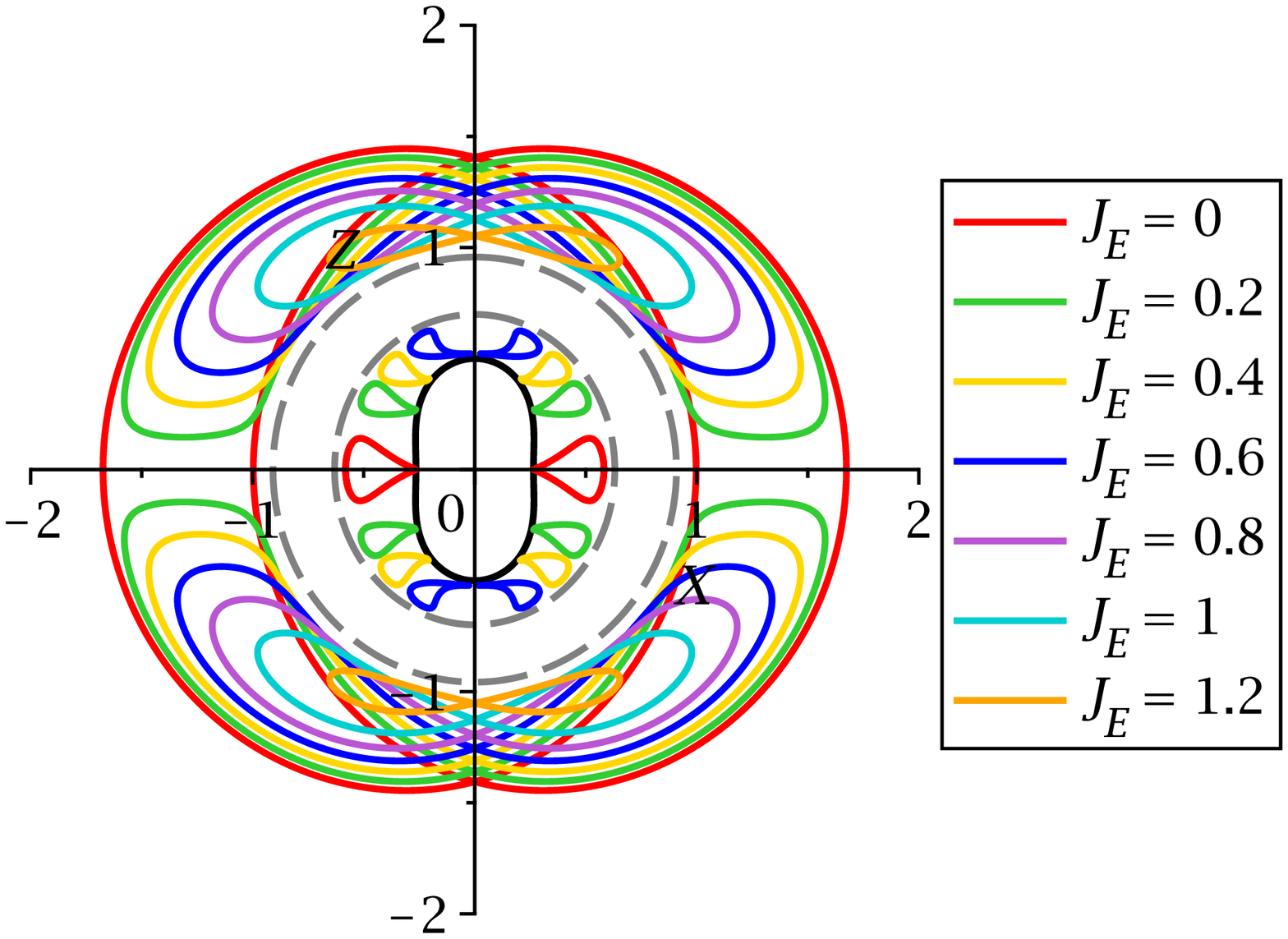}
	}
	\caption{Plots of the photon region in the Myers-Perr-AdS spacetime for $g=0.1$ and different $a$,$b$ in the $X$-$Z$-plane. Here we show plots for several $J_E$ in one picture to give an impression on what the full photon region would look like. The black curve is the singularity and the dashed grey curves indicate the positions of the horizons.}
	\label{pic:photonregion2}
\end{figure}

Figure \ref{pic:photonregion} shows examples of the photon region in the Myers-Perry-AdS spacetime. Usually, in four-dimensional spacetimes a photon region does not depend on the parameters of the light rays and is fully determined by the parameters of the black hole. However, in this case we do not have enough constraints to eliminate the angular momentum $J$ from the equation for the photon region. To give an impression on what the full photon region looks like, we display plots for several $J_E$ combined in one picture, see Fig. \ref{pic:photonregion2}. The photon region consists of crescent shaped areas for $J_E=0$ similar to the four-dimensional Kerr-(AdS) spacetime. In contrast to four-dimensional spacetimes (compare \cite{Grenzebach:2014fha, Grenzebach:2015oea, Grenzebach:Springer}) for $J_E\neq 0$ there is a gap in the photon region centered around the equatorial plane. The gap grows if $J_E$ increases.

Behind the horizons there are additional smaller parts of the photon region. Here stable and unstable spherical photon orbits can be found, as seen in the effective potential (Fig. \ref{pic:potential-lightlike}).


\subsection{Terminating orbits}

In this section we study the conditions for so-called terminating orbits which end at the singularity. As we have seen before the singularity in the Meyers-Perry-AdS spacetime is a complex structure which extends both in the $x$ direction and in the $\theta$ direction. The two rotation parameters $a$ and $b$ determine the shape of the closed surface $\rho(x,\theta)^2=0$, see Fig. \ref{pic:singularity}. To check if a geodesic hits the singularity, that is, a terminating orbit exists, one has to consider both the $x$ equation \eqref{eqn:x-equation} and the $\theta$ equation \eqref{eqn:theta-equation}. If light or a particle hits the singularity then  $\rho^2=0$ and therefore
\begin{equation}
	x=-\left( a^2\cos^2\theta + b^2\sin^2\theta \right) \quad\text{or}\quad  x=-\left( a^2\nu + b^2(1-\nu) \right) \, .
\end{equation}
Now we can write the effective potential $U(\nu)$ of the $\theta$ motion with $\nu=\frac{b^2+x}{b^2-a^2}$ as
\begin{equation}
 U_\pm(x) = \pm \sqrt{  \frac{\Delta_\nu}{\Xi_a a^2\frac{b^2+x}{b^2-a^2} + \Xi_b b^2\left(1-\frac{b^2+x}{b^2-a^2}\right)} \left[  \delta \left( a^2 \frac{b^2+x}{b^2-a^2} + b^2\left( 1 - \frac{b^2+x}{b^2-a^2} \right) \right) -K  +\frac{L^2\Xi_a}{1-\frac{b^2+x}{b^2-a^2}} +\frac{J^2\Xi_b}{\frac{b^2+x}{b^2-a^2}}  \right]   } 
\end{equation}
with $\Delta_\nu= 1-a^2g^2\frac{b^2+x}{b^2-a^2}-b^2g^2\left(1-\frac{b^2+x}{b^2-a^2}\right)$. Since $\nu\in[0,1]$ we only consider $U(x)$ in the range $x\in[-a^2,-b^2]$, which is also the $x$-range of $\rho^2=0$. The potential $U(\nu)$ diverges for $\nu\rightarrow 0$ and  $\nu\rightarrow 1$ [see Eq. \eqref{eqn:theta-potential}]. Similarly, the potential $U(x)$ diverges if $x$ approaches the closed surface of the singularity (from each side), i.e. if $x\rightarrow -a^2$ or  $x\rightarrow -b^2$.

\begin{figure}[h]
	\centering
	\subfigure[$K=1.25$]{
		\includegraphics[width=0.31\textwidth]{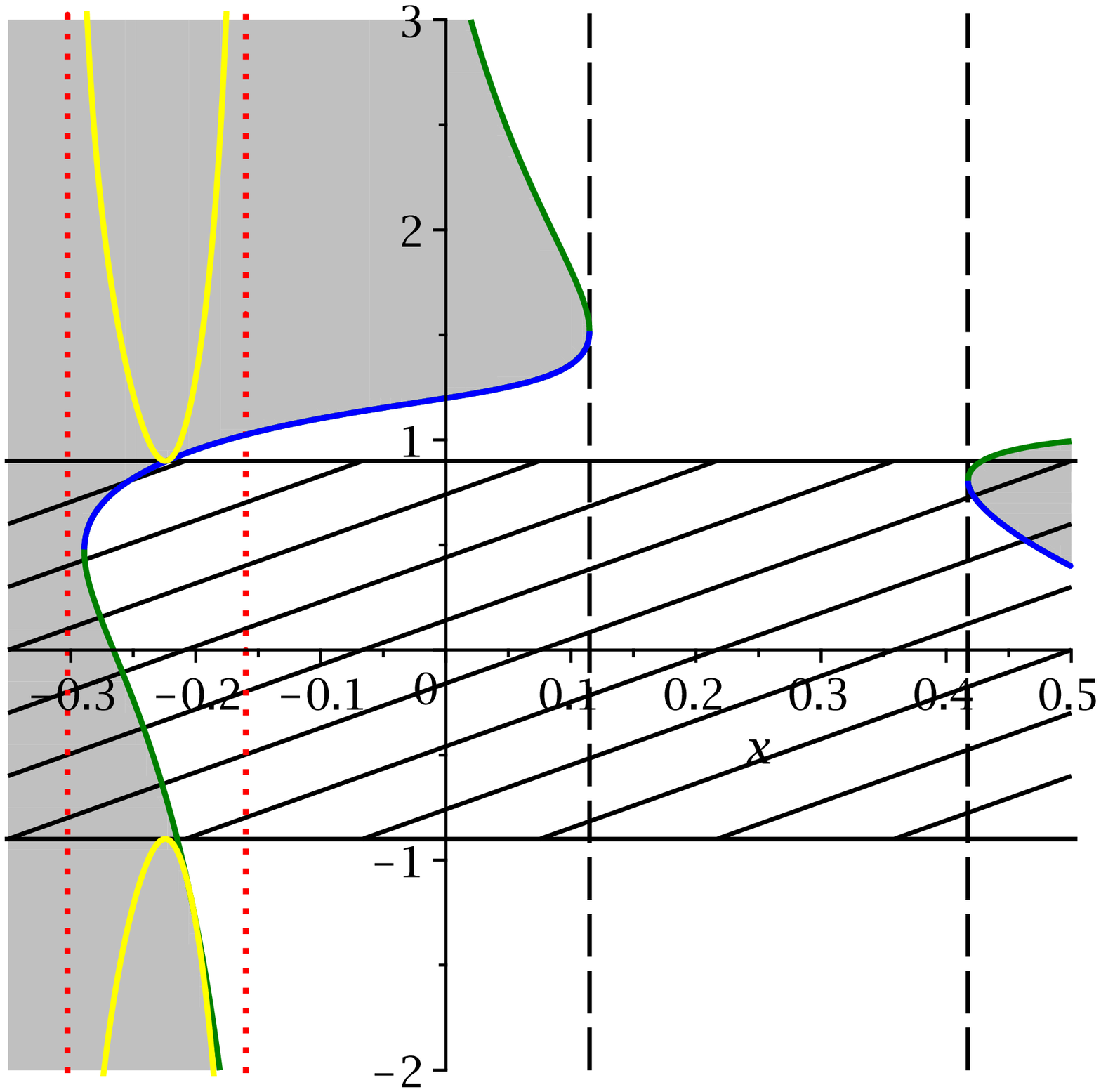}
	}
	\subfigure[$K=1.75$]{
		\includegraphics[width=0.31\textwidth]{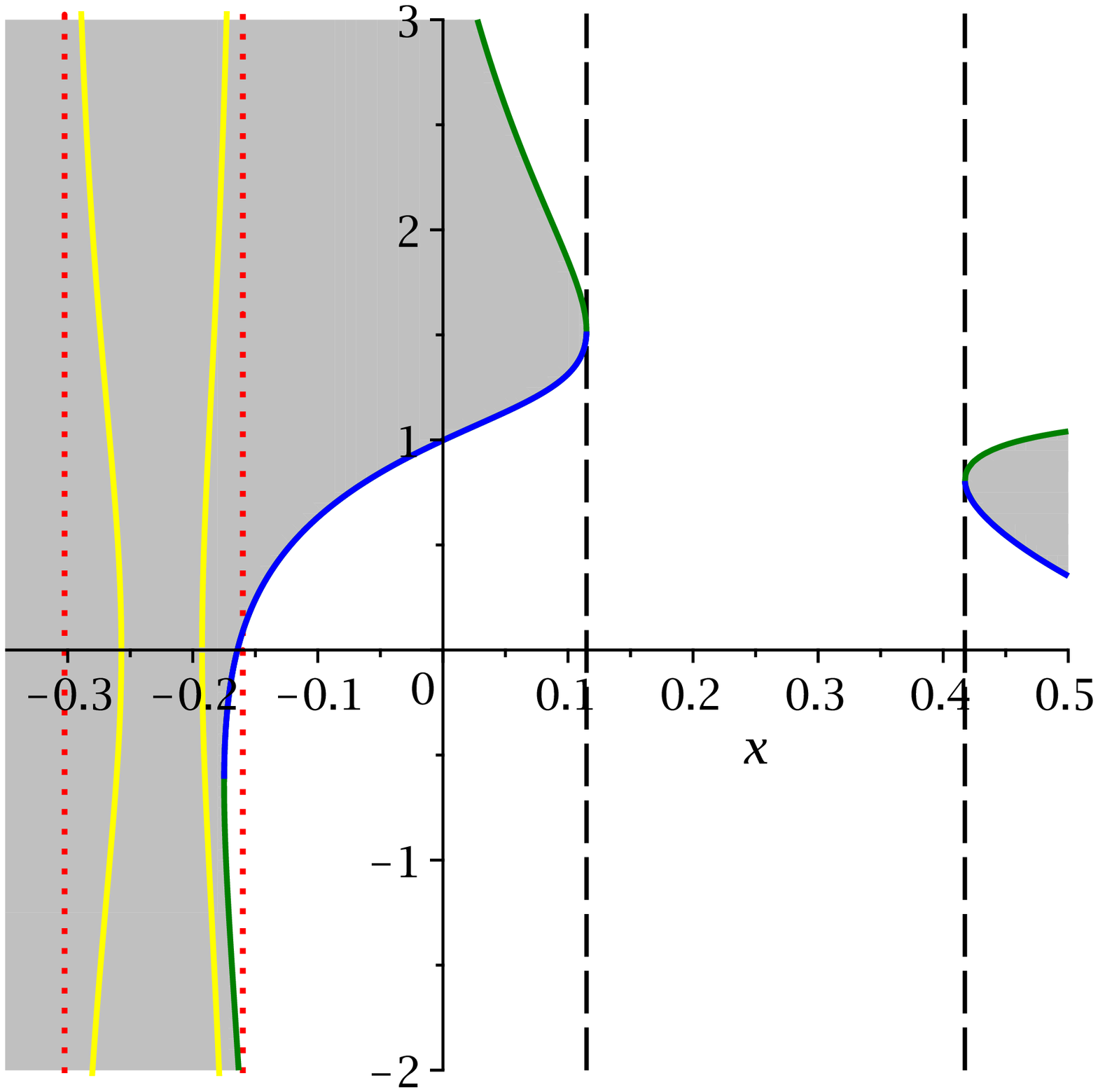}
	}
	\caption{Plots of the effective potential for $\delta=1$, $g=0.1$, $a=0.55$, $b=0.4$, $L=0.6$, $J=0.5$ and different values of $K$.The black dashed lines indicate the position of the horizons. The red dotted lines represent the range of the singularity $x\in[-a^2,-b^2]$. The blue and green curves are the effective potential $V(x)$ of the $x$ motion. Geodesic motion is allowed in the white regions but not in the grey regions. The yellow curve is the potential $U(x)$ for a particle hitting the singularity. In the dashed area motion is forbidden by the $\theta$ equation.}
	\label{pic:potential-terminating}
\end{figure}

$U(x)$ is the $\theta$ potential if a geodesic hits the singularity and can be plotted together with the effective potential $V(x)$ for the $x$ motion. Some plots are depicted in Fig. \ref{pic:potential-terminating}. Here the black dashed lines indicate the positions of the horizons and the red dotted lines represent the range of the singularity $x\in[-a^2,-b^2]$. A test particle does not automatically hit the singularity once it enters the region $x\in[-a^2,-b^2]$. It falls into the singularity if $\theta$ and $x$ fulfill the condition $\rho^2=0$. The blue and green curves are the effective potential $V(x)$ of the $x$ motion. Geodesic motion is allowed in the white regions but not in the grey regions. In the dashed area motion is forbidden by the $\theta$ equation. The potential $U(x)$ is shown as a yellow curve.

The test particle meets the singularity at the intersection point of $U(x)$ and $V(x)$. We find that there are two intersection points at the extrema of the potential $U$, which are at the same time the boundaries of the forbidden energy regime for the $\theta$ motion. Therefore, terminating orbits have constant $\theta$ and only appear for these specific energies. This behavior is already present in the five-dimensional Myers-Perry spacetime \cite{Kagramanova:2012hw, Diemer:2014lba}. 

In Fig. \ref{pic:potential-terminating}(a) the parameters are chosen such that certain energies are forbidden by the $\theta$ equation (dashed area). If a test particle has precisely the boundary energy of the forbidden region (that means this energy is the extremum of $U$), a terminating orbit is possible. However, in Fig. \ref{pic:potential-terminating}(b) all energies are allowed by the $\theta$ equation. The potential $U(x)$ cannot be reached by the test particles, because it lies entirely in the region forbidden by the $x$ equation. In this case it is not possible to find a value for the energy of the test particle so that it hits the singularity.

\FloatBarrier

\section{Spacelike geodesics and AdS/CFT}
\label{sec:spacelike}

Spacelike geodesics refer to test particles with imaginary mass ($\delta =-1$) and therefore they are usually not considered in the analysis of geodesic motion. However, there are applications in AdS/CFT. 
The observables on the asymptotic boundary of the AdS spacetime are described by CFT correlators or Feynman propagators. The correlation functions of operators in CFT on the boundary are dual to the correlation functions of fields in the bulk. The correlator of two operators corresponds to the Green function
\begin{equation}
	\langle \mathcal{O}(t,\boldsymbol{x}) \, \mathcal{O}(t',\boldsymbol{x}') \rangle = \int\! \exp [i\Delta L(\mathcal{P})]\mathcal{DP}
\end{equation}
where $L(\mathcal{P})$ is the proper length of the path $\mathcal{P}$ between the boundary points $(t,\boldsymbol{x})$ and $(t,\boldsymbol{x}')$. $L(\mathcal{P})$ is imaginary for spacelike trajectories. The mass $m$ of the bulk field is related to the conformal dimension $\Delta$ of the operator $\mathcal{O}$ by  $\Delta=1+\sqrt{1+m^2}$. For large masses (so that $\Delta\approx m$) the correlator can be computed in the WKB approximation
\begin{equation}
	\langle \mathcal{O}(t,\vec{x}) \, \mathcal{O}(t',\vec{x}') \rangle = \sum_{g} \exp (-\Delta L_{{g}}) \, .
\end{equation}
The correlator is now described by the sum over all spacelike geodesics between the boundary points. $L_g$ is the real proper length of a geodesic. Since the length diverges due to contributions near the AdS boundary, we have to renormalize it by removing the divergent part in pure AdS. It turns out that the sum is dominated by the shortest spacelike geodesic between the boundary points (see, e.g., \cite{Balasubramanian:1999zv}, \cite{Balasubramanian:2011ur} and \cite{Louko:2000tp}).

In this formalism one can study two-point correlators to calculate for instance the thermalization time  \cite{Balasubramanian:2011ur} or entanglement entropy \cite{Hubeny:2007xt}, \cite{AbajoArrastia:2010yt}. 

Geodesics relevant for AdS/CFT must have endpoints on the boundary which is at $r\rightarrow\infty$. Therefore we are looking for escape orbits which have a single turning point and reach infinity. If the turning point of a geodesic lies outside the horizons, both endpoints are located on a single boundary. In contrast, if a geodesic crosses a horizon the endpoints will be on two disconnected boundaries. The corresponding correlators will then be sensitive to the physics behind the horizon, which could be used to study black hole formation \cite{Balasubramanian:1999zv} or the black hole singularity \cite{Fidkowski:2003nf}.\\

\begin{figure}[h]
	\centering
	\includegraphics[width=0.31\textwidth]{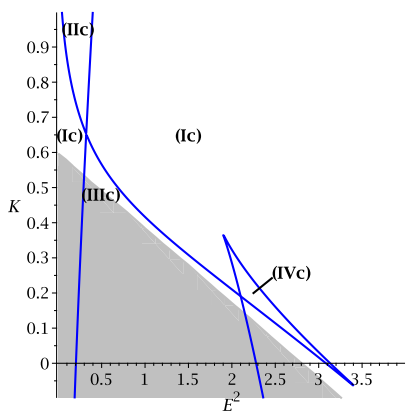}
	\caption{Parametric $K$-$E^2$ diagram for spacelike geodesics with $\delta=-1$, $g=0.1$, $a=0.5$, $b=0.4$, $L=0.4$, $J=0.5$. The blue lines separate regions with different numbers of zeros of $X(x)$ and different types of orbits. For $x>-a^2$ there is a single zero in region (Ic) and three zeros in regions (IIc)--(IVc). In the grey region geodesics cannot exist due to constraints coming from the $\vartheta$ equation.}
\label{pic:parameterplot-spacelike}
\end{figure}

We will now analyze the behavior of spacelike geodesics in the Myers-Perry-AdS spacetime. The $\theta$ motion is qualitatively the same as for timelike and null geodesics. Considering the $x$ motion we will perform the analysis as before and start with parametric $K$-$E^2$ diagrams which show the number of zeros of the polynomial $X(x)$ in different parameter regions. Figure \ref{pic:parameterplot-spacelike} depicts a parametric $K$-$E^2$ diagram for spacelike geodesics. In the grey part of the diagram $\Theta(\nu)$ is negative and therefore geodesic motion is not possible, whereas in the white part $\Theta(\nu)$ has two zeros. The blue curves correspond to double zeros of $X(x)$ and separate parameter regions with a different number of zeros. For $x>-a^2$ there is a single zero in region (Ic) and three zeros in the regions (IIc)--(IVc).\\

\begin{figure}[h]
	\centering
	\subfigure[$\delta=-1$, $K=2$ $g=0.1$, $a=0.5$, $b=0.4$, $L=0.4$, $J=0.5$]{
		 \includegraphics[width=0.31\textwidth]{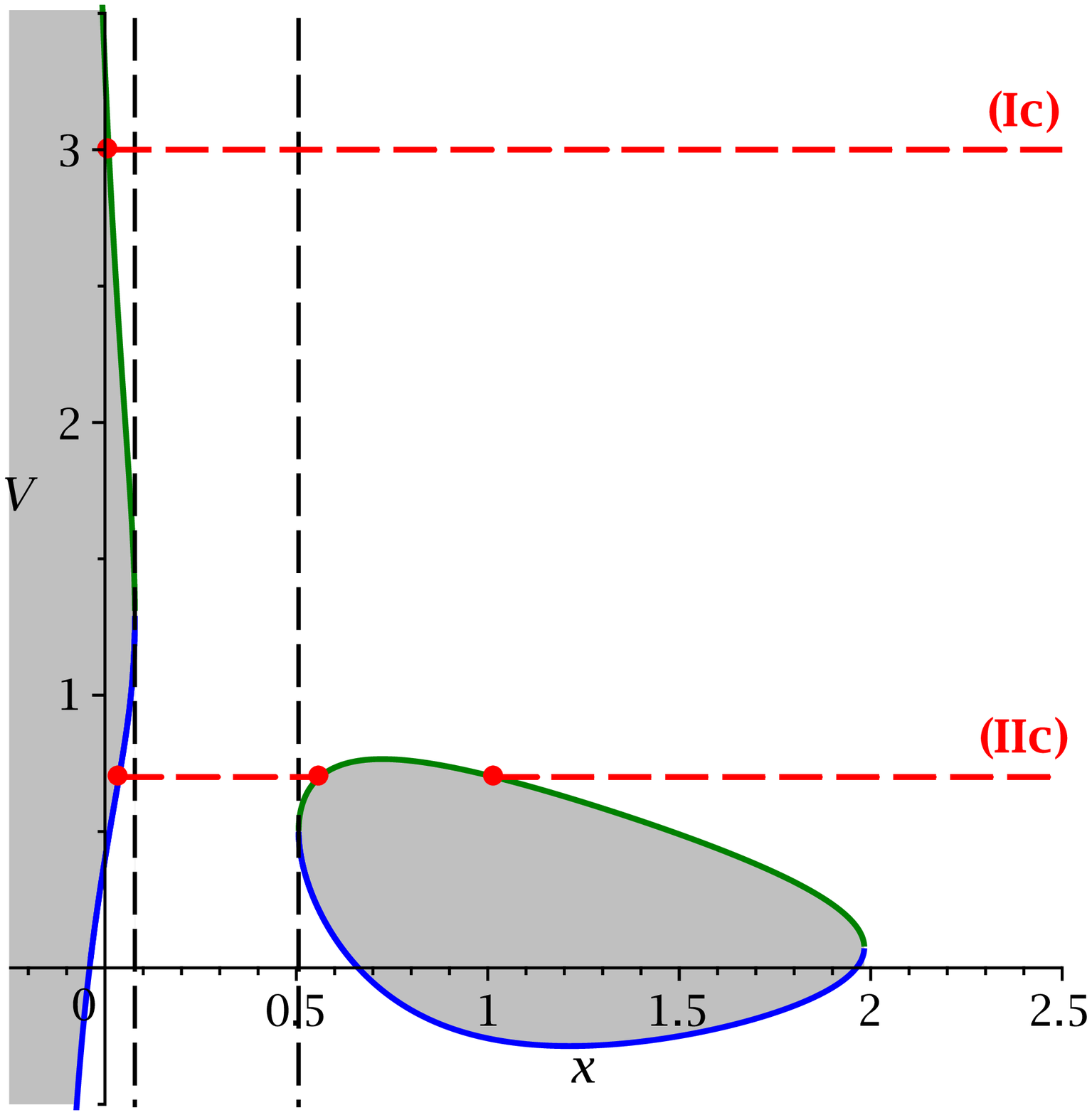}
	}
	\subfigure[$\delta=-1$, $K=0.4$, $g=0.1$, $a=0.4$, $b=0.3$, $L=0.4$, $J=0.5$]{
		 \includegraphics[width=0.31\textwidth]{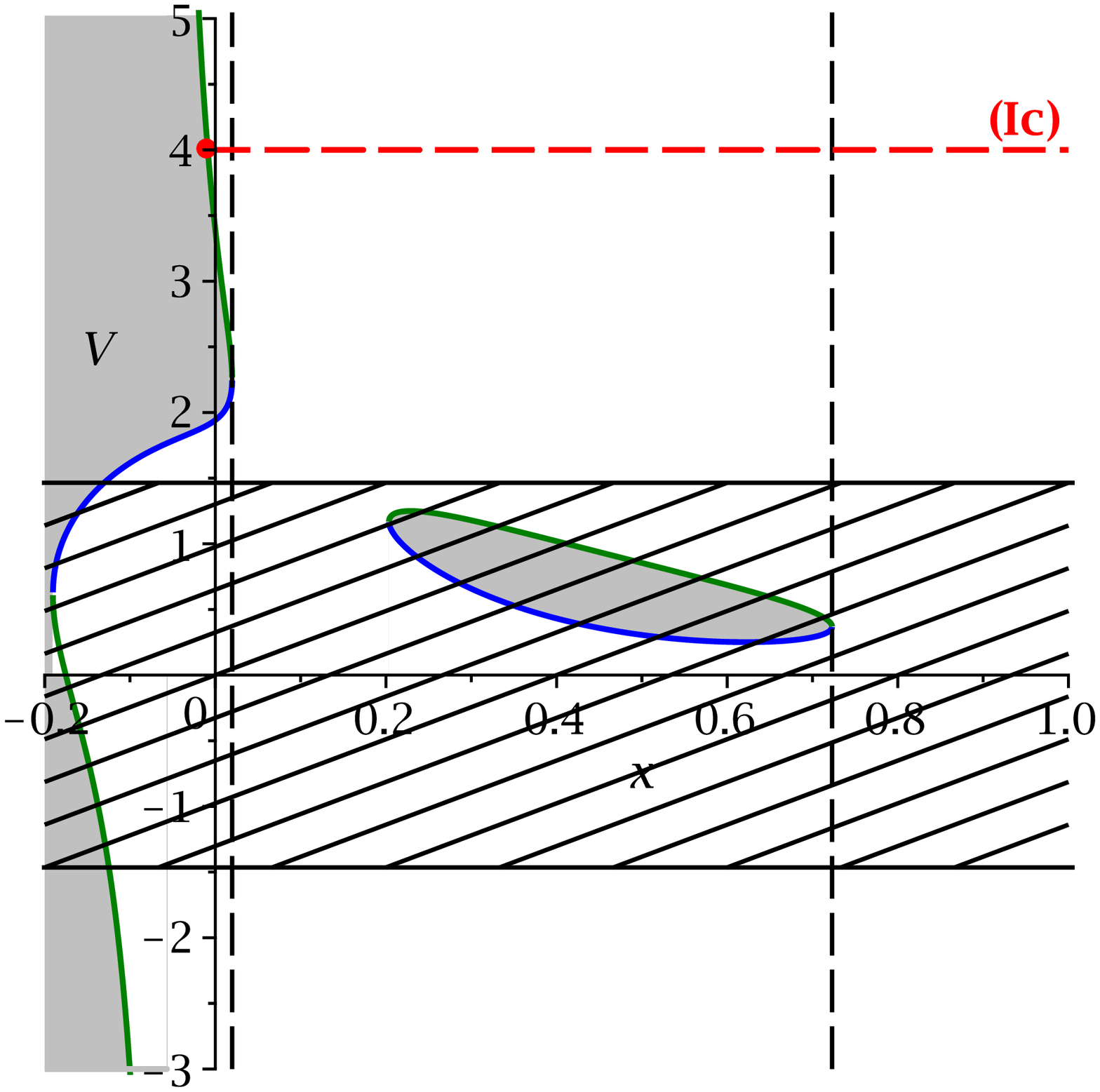}
	}
	\subfigure[$\delta=-1$, $K=0.19$, $g=0.1$, $a=0.5$, $b=0.4$, $L=0.3$, $J=0.5$]{
		 \includegraphics[width=0.31\textwidth]{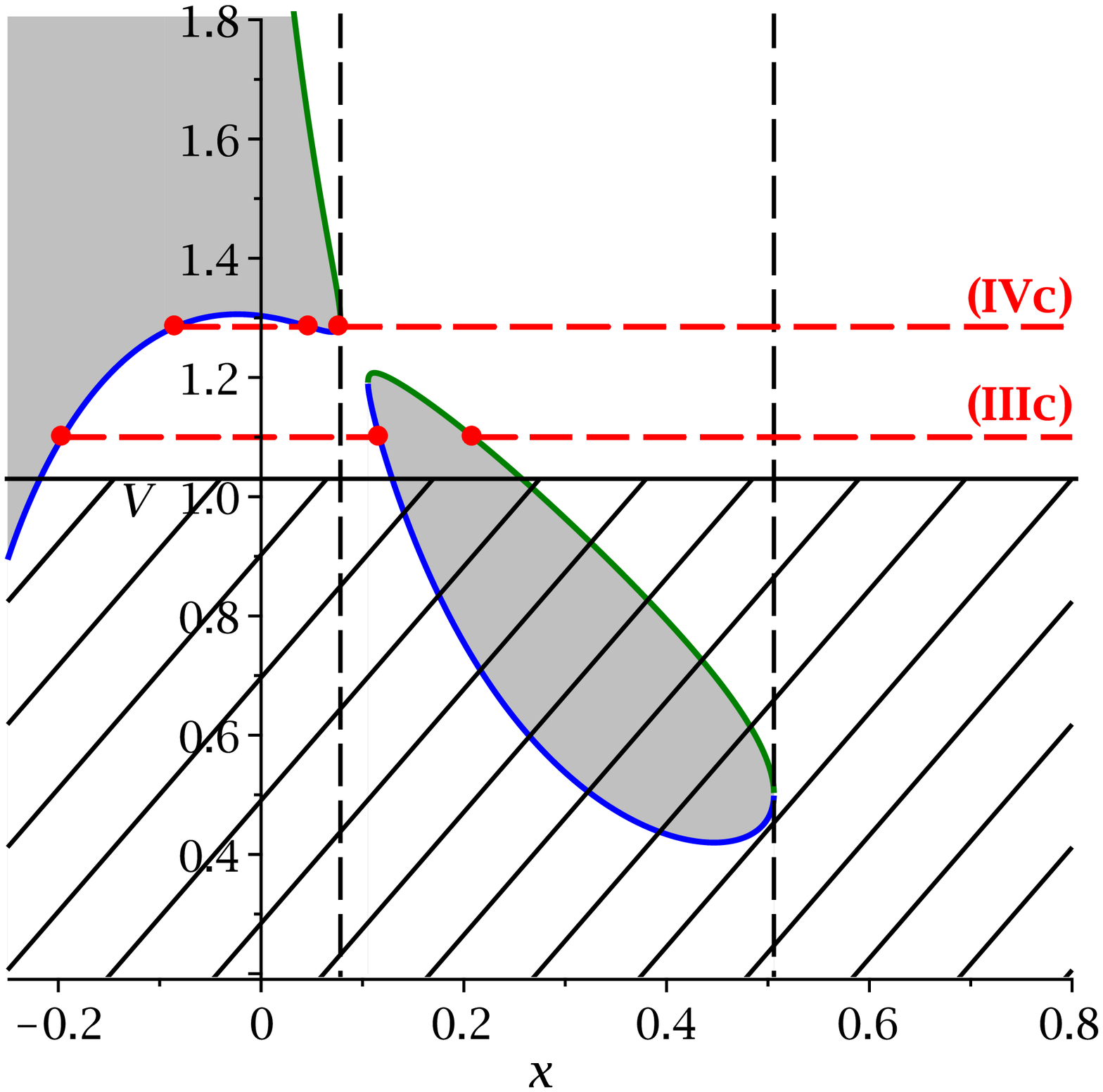}
	}
	\caption{Plots of the effective potential for spacelike geodesics. The green and blue curves are the two parts of the effective potential. The grey areas are forbidden by the $x$ equation and the hatched areas are forbidden by the $\theta$ equation. Geodesic motion is possible in the white areas only. The vertical black dashed lines indicate the position of the horizons. The red dashed lines are example energies and the red dots mark the turning points. The numbers refer to the orbit types of Table \ref{tab:orbits-spacelike} and Fig. \ref{pic:parameterplot-spacelike}}
	\label{pic:potential-spacelike}
\end{figure}

If we take the effective potential [see Eq. \eqref{eq:radialpotential}] into account, we can determine the orbit types in each region. Fig. \ref{pic:potential-spacelike} shows examples of the effective potential for spacelike geodesics. The green and blue curves are the two parts of the effective potential. The grey areas are forbidden by the $x$ equation and the hatched areas are forbidden by the $\theta$ equation. Geodesic motion is possible in the white areas only. The vertical black dashed lines indicate the position of the horizons. The red dashed lines are example energies for different orbits and the red dots mark the turning points.

We find the following orbit types in regions (Ic)-(IVc) (let $x_i<x_{i+1}$):
\begin{enumerate}
	\item Region (Ic): The polynomial $X(x)$ has a single zeros $x_1\leq x_-$. $X(x)$ is positive for $x\in[x_1, \infty)$ and therefore TWEOs crossing both horizons can be found in this region.
	\item Region (IIc): The polynomial $X(x)$ has three zeros $x_1\leq x_-$ and $x_2,x_3\geq x_+$.  $X(x)$ is positive for $x\in[x_1, x_2]$ and  $x\in[x_3, \infty)$. MBOs and EOs exist.
	\item Region (IIIc): The polynomial $X(x)$ has three zeros $x_1\leq x_-$ and $x_-\leq x_2, x_3\leq x_+$. $X(x)$ is positive for $x\in[x_1, x_2]$ and  $x\in[x_3, \infty)$. Here special kinds of MBOs and TWEOs exist, which cross just one of the horizons.
	\item Region (IVc): The polynomial $X(x)$ has three zeros $x_1,x_2,x_3\leq x_-$.  $X(x)$ is positive for $x\in[x_1, x_2]$ and  $x\in[x_3, \infty)$. There are BOs hidden behind the inner horizon and TWEOs crossing both horizons.
\end{enumerate}
An overview of the possible orbits types for timelike geodesics is shown in Table \ref{tab:orbits-spacelike}. Geodesics relevant for AdS/CFT with endpoints on the boundary exist in all four regions. However, only the EOs in region (IIc) return to the same boundary where they started. In regions (Ic), (IIIc) and (IVc) we find TWEOs which cross the horizons and go to another universe. Escaping geodesics in regions (Ic) and (IVc) reach past the inner horizon $x_-$ while escaping geodesics in region (IIIc) can probe behind $x_+$ but do not cross $x_-$.

Like EOs, the TWEOs can be considered as propagators in the AdS/CFT context \cite{Balasubramanian:1999zv, Louko:2000tp}. Boundary correlators can be used to probe physics behind the horizons and especially physics of the singularity, although there are limitations \cite{Kraus:2002iv}. Geodesics crossing the horizons and therefore connecting two different boundaries represent a pure state, an entangled state of the two field theories. In \cite{Levi:2003cx} it was argued that the region behind an event horizon is encoded in the ``hologram;'' however, the region behind a Cauchy horizon is not. 

Furthermore, geodesics crossing the horizon can also be applied to questions regarding the information paradox \cite{Papadodimas:2012aq}.

\begin{table}[h]
\begin{center}
\begin{tabular}{|cccc|}\hline
Region & Zeros & Range of $x$ & Orbit \\
\hline\hline
Ic & 1 &
\begin{pspicture}(0,-0.2)(6,0.2)
\psline[linewidth=0.5pt]{->}(0,0)(6,0)
\psline[linewidth=0.5pt](0,-0.2)(0,0.2)
\psline[linewidth=0.5pt,doubleline=true](2,-0.2)(2,0.2)
\psline[linewidth=0.5pt,doubleline=true](3.5,-0.2)(3.5,0.2)
\psline[linewidth=1.2pt]{*-}(1.5,0)(6,0)
\end{pspicture}
  & TWEO
\\ \hline
IIc & 3 &
\begin{pspicture}(0,-0.2)(6,0.2)
\psline[linewidth=0.5pt]{->}(0,0)(6,0)
\psline[linewidth=0.5pt](0,-0.2)(0,0.2)
\psline[linewidth=0.5pt,doubleline=true](2,-0.2)(2,0.2)
\psline[linewidth=0.5pt,doubleline=true](3.5,-0.2)(3.5,0.2)
\psline[linewidth=1.2pt]{*-*}(1.5,0)(4,0)
\psline[linewidth=1.2pt]{*-}(4.5,0)(6,0)
\end{pspicture}
  & MBO, EO
\\ \hline
IIIc & 3 &
\begin{pspicture}(0,-0.2)(6,0.2)
\psline[linewidth=0.5pt]{->}(0,0)(6,0)
\psline[linewidth=0.5pt](0,-0.2)(0,0.2)
\psline[linewidth=0.5pt,doubleline=true](2,-0.2)(2,0.2)
\psline[linewidth=0.5pt,doubleline=true](3.5,-0.2)(3.5,0.2)
\psline[linewidth=1.2pt]{*-*}(1.5,0)(2.5,0)
\psline[linewidth=1.2pt]{*-}(3,0)(6,0)
\end{pspicture}
  & MBO, TWEO
\\ \hline
IVc & 3 &
\begin{pspicture}(0,-0.2)(6,0.2)
\psline[linewidth=0.5pt]{->}(0,0)(6,0)
\psline[linewidth=0.5pt](0,-0.2)(0,0.2)
\psline[linewidth=0.5pt,doubleline=true](2,-0.2)(2,0.2)
\psline[linewidth=0.5pt,doubleline=true](3.5,-0.2)(3.5,0.2)
\psline[linewidth=1.2pt]{*-*}(0.5,0)(1,0)
\psline[linewidth=1.2pt]{*-}(1.5,0)(6,0)
\end{pspicture}
  & BO, TWEO
\\ \hline\hline
\end{tabular}
\caption{Types of orbits for spacelike geodesics in the Myers-Perry-AdS spacetime. The range of the orbits is represented by thick lines. The turning points are marked by thick dots. The two vertical double lines indicate the position of the horizons and the single vertical line corresponds to the singularity.}
\label{tab:orbits-spacelike}
\end{center}
\end{table}

\FloatBarrier

\section{Solution of the geodesic equations}
\label{sec:solutions}

In this section we solve the equations of motion \eqref{eqn:x-equation}--\eqref{eqn:t-equation}. The analytical solutions are given in terms of the  Weierstra{\ss} $\wp$, $\sigma$ and $\zeta$ functions.

\subsection{The $x$ equation}

On the right-hand side of the $x$ equation 
\begin{equation}
	\left(\frac{\dd x}{\dd \gamma}\right) ^2 = X(x) = \sum _{i=1}^4 a_i x^i \, .
\end{equation}
we find a polynomial of order four, which can be reduced to third order by the substitution  $x=\pm\frac{1}{y}+x_0$. Here $x_0$ is a zero of $X$. The equation 
\begin{equation}
	\left(\frac{\dd y}{\dd \gamma}\right) ^2 = \sum _{i=1}^3 b_i y^i \, .
\end{equation}
can then be transformed into the Weierstra{\ss} form by setting $y=\frac{1}{b_3}\left( 4z-\frac{b_2}{3}\right)$
\begin{equation}
	\left(\frac{\dd z}{\dd \gamma}\right)^2=4z^3-g_2^{x}z-g_3^{x}= P_3^{x} (z) \, .
	\label{eqn:weierstrass-form}
\end{equation}
The coefficients are
\begin{equation}
	g_2^{x}=\frac{b_2^2}{12} - \frac{b_1b_3}{4} \, , \qquad  g_3^{x}=\frac{b_1b_2b_3}{48} - \frac{b_0b_3^2}{16}-\frac{b_2^3}{216} \ .
\end{equation}
The solution of the $x$ equation \eqref{eqn:x-equation} is now trivial, since it is known that Eq. \eqref{eqn:weierstrass-form} is solved by the elliptic Weierstra{\ss} $\wp$ function (see e.g. \cite{Markushevich:1967})
\begin{equation}
	z(\gamma) = \wp\left(\gamma - \gamma'_{\rm in}; g_2^{x}, g_3^{x}\right) \ ,
\end{equation}
with the initial values $\gamma'_{\rm in}=\gamma_{\rm in}+\int^\infty_{z_{\rm in}}{\frac{\dd z}{\sqrt{4z^3-g_2^{x}z-g_3^{x}}}}$ and $z_{\rm in}=\pm\frac{b_3}{4(x_{\rm in}-x_0)} + \frac{b_2}{12}$ and $x_{\rm in}=r^2_{\rm in}$. 
Via resubstitution we obtain the solution of Eq. \eqref{eqn:x-equation}
\begin{equation}
	x(\gamma)=\pm \frac{b_3}{4 \wp\left(\gamma - \gamma'_{\rm in}; g_2^{x}, g_3^{x}\right) - \frac{b_2}{3}} +x_0\ ,
\end{equation}
and in terms of the $r$ coordinate we get
\begin{equation}
	r(\gamma)=\sqrt{ \pm \frac{b_3}{4 \wp\left(\gamma - \gamma'_{\rm in}; g_2^{x}, g_3^{x}\right) - \frac{b_2}{3}} +x_0} \ .
\end{equation}

\subsection{The $\theta$ equation}

The $\theta$ equation \eqref{eqn:theta-equation} can be simplified with $\nu=\cos\theta^2$
\begin{equation}
	\left( \frac{\dd \nu}{\dd\gamma} \right)^2 = \Theta_\nu =  \sum _{i=1}^4 c_i \nu^i \, ,
\end{equation}
so that $\Theta_\nu$ is a fourth order polynomial. To transform this into a polynomial of order three $\sum _{i=1}^3 d_i y^i$ we substitute $\nu=\pm\frac{1}{y}+\nu_0$, where $\nu_0$ is a zero of $\Theta_\nu$. Analogous to the previous section, another substitution $y=\frac{1}{d_3}\left( 4z-\frac{d_2}{3}\right)$ gives the Weierstra{\ss} form 
\begin{equation}
	\left(\frac{\dd z}{\dd \gamma}\right)^2=4z^3-g_2^{\nu}z-g_3^{\nu}= P_3^{\nu} (z) \, ,
	\label{eqn:weierstrass-form2}
\end{equation}
where the coefficients are
\begin{equation}
	g_2^{\nu}=\frac{d_2^2}{12} - \frac{d_1d_3}{4} \, , \qquad  g_3^{\nu}=\frac{d_1d_2d_3}{48} - \frac{d_0d_3^2}{16}-\frac{d_2^3}{216} \ .
\end{equation}
Again, the Eq. \eqref{eqn:weierstrass-form2} is solved by Weierstra{\ss} $\wp$ function. Therefore the solution of the $\theta$ equation \eqref{eqn:theta-equation} is
\begin{equation}
	\theta(\gamma)=\arccos \left( \sqrt{ \pm \frac{d_3}{4 \wp\left(\gamma - \gamma''_{\rm in}; g_2^{\nu}, g_3^{\nu}\right) - \frac{d_2}{3}} +\nu_0} \right) \ .
\end{equation}
where the initial values are $\gamma''_{\rm in}=\gamma_{\rm in}+\int^\infty_{z'_{\rm in}}{\frac{\dd z}{\sqrt{4z^3-g_2^{\nu}z-g_3^{\nu}}}}$ and $z'_{\rm in}=\pm\frac{d_3}{4(\nu_{\rm in}-\nu_0)} + \frac{d_2}{12}$ and $\nu_{\rm in}=\cos^2\theta_{\rm in}$.

\subsection{The $\phi$ equation}
\label{sec:phisol}
To solve the $\phi$ equation \eqref{eqn:phi-equation} we rewrite it using the $x$ equation \eqref{eqn:x-equation} and the $\theta$ equation in the form of Eq. \eqref{eqn:nu-equation}
\begin{align}
	\dd \phi =& \left\lbrace L\Xi_a (1+g^2x)(b^2+x)(b^2-a^2) + 2M \left[ Ea(b^2+x) -L(b^2+a^2g^2x) -Jab(1+g^2x) \right] \right\rbrace  \frac{\dd x}{\Delta_x \sqrt{X}}\nonumber\\ 
	&+ \frac{L\Xi_a}{1-\nu} \frac{\dd \nu}{\sqrt{\Theta_\nu}}
\end{align}
Integrating this equation yields two integrals $I_x(x)$ and $I_{\nu}(\nu)$ which can be solved separately
\begin{align}
	\phi -\phi_{\rm in} =& \int_{x_{\rm in}}^x\!\!\!\!\! \left\lbrace L\Xi_a (1+g^2x)(b^2+x)(b^2-a^2) + 2M \left[ Ea(b^2+x) -L(b^2+a^2g^2x) -Jab(1+g^2x) \right] \right\rbrace  \frac{\dd x}{\Delta_x \sqrt{X}}\nonumber\\ 
	&+ \int_{\nu_{\rm in}}^\nu \frac{L\Xi_a}{1-\nu} \frac{\dd \nu}{\sqrt{\Theta_\nu}} = I_x(x) +I_{\nu} (\nu) \, .
\end{align}
In the integral $I_x(x)$ we substitute ${x=\pm\frac{b_3}{4z-\frac{b_2}{3}}+x_0}$ to transform the polynomial $X(x)$ into the Weierstra{\ss} form $P_3^{x} (z)$. Then we apply a partial fraction decomposition
\begin{equation}
	I_{x}= \int_{z_{\rm in}}^{z} \! \left( A_0 + \sum^3_{i=1}\frac{A_i}{z-p_i} \right) \frac{\dd z}{\sqrt{P^{x}_3(z)}} \, .
\end{equation}
The constants $A_i$ which arise from the partial fraction decomposition depend on the parameters of the metric and the test particle. The poles $p_i$ correspond to the zeros of $\Delta_x$. Furthermore, with $z=\wp\left(\gamma - \gamma'_{\rm in}; g_2^{x}, g_3^{x}\right)=:\wp_{x}(v)$ and $v=\gamma-\gamma_{\rm in}'$ the integral $I_x$ acquires the form
\begin{equation}
	I_{x}= \int_{v_{\rm in}}^{v} \! \left( A_0 + \sum^3_{i=1}\frac{A_i}{\wp_{x}(v)-p_i} \right) \dd v \, .
\label{eqn:ell3rdkind1}
\end{equation}
The integral $I_{\nu}(\nu)$ can be treated in the same way by substituting first $\nu=\pm\frac{d_3}{4z-\frac{d_2}{3}}+\nu_0$ and then $z=\wp(\gamma-\gamma''_{\rm in};g_2^\nu,g_3^\nu)=:\wp_{\nu}(u)$ with $u=\gamma-\gamma_{\rm in}''$,
\begin{equation}
	I_{\nu}= \int_{u_{\rm in}}^{u} \! \left( B_0 + \frac{B_1}{\wp_{\nu}(u)-q_\phi} \right) \dd u \, .
\label{eqn:ell3rdkind2}
\end{equation}
The constants $B_0=\frac{L\Xi_a}{1-\nu_0}$, $B_1=\pm\frac{L\Xi_a d_3}{(1-\nu_0)^2}$ and the pole $q_\phi=\frac{d_2(1-\nu_0)\pm 3d_3}{12(1-\nu_0)}$ arise in a partial fraction decomposition. 

Equations \eqref{eqn:ell3rdkind1} and \eqref{eqn:ell3rdkind2} show that $I_x$ and $I_{\nu}$ are elliptic integrals of the third kind. Those can be solved in terms of the $\wp$, $\sigma$ and $\zeta$ functions as shown in, e.g., \cite{Kagramanova:2010bk, Grunau:2010gd, Enolski:2011id}. Hence, we can write the solution of the $\phi$ equation \eqref{eqn:phi-equation} as
\begin{equation}
	\begin{split}
		\phi(\gamma) &= A_0(v-v_{\rm in}) + \sum^3_{i=1}\frac{A_i}{\wp'_{x}(v_i)}\left( 2\zeta_{x}(v_i)(v-v_{\rm in}) + \ln\frac{\sigma_{x}(v-v_i)}{\sigma_{x}(v_{\rm in}-v_i)} - \ln\frac{\sigma_{x}(v+v_i)}{\sigma_{x}(v_{\rm in}+v_i)}\right) \\
		&+B_0(u-u_{\rm in}) + \frac{B_1}{\wp'_{\nu}(u_\phi)}\left( 2\zeta_{\nu}(u_\phi)(u-u_{\rm in}) + \ln\frac{\sigma_{\nu}(u-u_\phi)}{\sigma_{\nu}(u_{\rm in}-u_\phi)} - \ln\frac{\sigma_{\nu}(u+u_\phi)}{\sigma_{\nu}(u_{\rm in}+u_\phi)}\right) \\
		&+ \phi_{\rm in} \, .
	 \end{split}
\end{equation}
Here $\wp'$ is the derivative of the Weierstra{\ss} $\wp$ function and $p_i=\wp_{x}(v_i)$, $q_\phi=\wp_{\nu}(u_\phi)$, $v=\gamma-\gamma_{\rm in}'$,  $u=\gamma-\gamma_{\rm in}''$. We also defined
\begin{eqnarray}
	\wp_{x}(v) &= \wp (v, g_2^{x}, g_3^{x})\, , \qquad \wp_\nu (u)&= \wp (u, g_2^{\nu}, g_3^{\nu}) \, ,\nonumber\\
	\zeta_{x}(v) &= \zeta (v, g_2^{x}, g_3^{x})\, , \qquad \zeta_\nu (u)&= \zeta (u, g_2^{\nu}, g_3^{\nu}) \, ,\\
	\sigma_{x}(v) &= \sigma (v, g_2^{x}, g_3^{x})\, , \qquad \sigma_\nu (u)&= \sigma (u, g_2^{\nu}, g_3^{\nu}) \, .\nonumber
\end{eqnarray}

\subsection{The $\psi$ equation}
The $\psi$ equation \eqref{eqn:psi-equation} can be rewritten by employing the same substitutions as in Sec. \ref{sec:phisol}
\begin{equation}
	\psi-\psi_{\rm in} = \int_{v_{\rm in}}^{v} \! \left( C_0 + \sum^3_{i=1}\frac{C_i}{\wp_{x}(v)-p_i} \right) \dd v + \int_{u_{\rm in}}^{u} \! \left( D_0 + \frac{D_1}{\wp_{\nu}(u)-q_\psi} \right) \dd u \, .
\end{equation}
The constants $C_i$, $D_0=\frac{J\Xi_b}{\nu_0}$, and $D_1=\frac{J\Xi_b d_3}{4\nu_0^2}$ result from a partial fraction decomposition. The poles $p_i$ are the same as in Sec. \ref{sec:phisol}, but the pole $q_\psi$ is  $q_\psi=\frac{d_2\nu_0\mp 3d_3}{12\nu_0}$. Analogously to Sec. \ref{sec:phisol} the solution of the $\psi$ equation is
\begin{equation}
	\begin{split}
		\psi(\gamma) &= C_0(v-v_{\rm in}) + \sum^3_{i=1}\frac{C_i}{\wp'_{x}(v_i)}\left( 2\zeta_{x}(v_i)(v-v_{\rm in}) + \ln\frac{\sigma_{x}(v-v_i)}{\sigma_{x}(v_{\rm in}-v_i)} - \ln\frac{\sigma_{x}(v+v_i)}{\sigma_{x}(v_{\rm in}+v_i)}\right) \\
		&+D_0(u-u_{\rm in}) + \frac{D_1}{\wp'_{\nu}(u_\psi)}\left( 2\zeta_{\nu}(u_\psi)(u-u_{\rm in}) + \ln\frac{\sigma_{\nu}(u-u_\psi)}{\sigma_{\nu}(u_{\rm in}-u_\psi)} - \ln\frac{\sigma_{\nu}(u+u_\psi)}{\sigma_{\nu}(u_{\rm in}+u_\psi)}\right) \\
		&+ \psi_{\rm in} \, ,
	 \end{split}
\end{equation}
where $q_\psi=\wp_{\nu}(u_\psi)$, $v=\gamma-\gamma_{\rm in}'$,  $u=\gamma-\gamma_{\rm in}''$.

\subsection{The $t$ equation}
The $t$ equation \eqref{eqn:t-equation} can be rewritten by employing the same substitutions as in Sec. \ref{sec:phisol}
\begin{equation}
	t-t_{\rm in} =  \int_{v_{\rm in}}^{v} \! \left( F_0 + \sum^3_{i=1}\frac{F_i}{\wp_{x}(v)-p_i} \right) \dd v + \int_{u_{\rm in}}^{u} \! \left( G_0 + \frac{G_1}{\wp_{\nu}(u)-q_t} \right) \dd u \, .
\end{equation}
where the constants $F_i$,$G_i$ and the poles $p_i$, $q_t$ arise from a partial fraction decomposition. Note that the poles $p_i$ are the same as in Sec. \ref{sec:phisol}. Analogously to Sec. \ref{sec:phisol} the solution of the $t$ equation is
\begin{equation}
	\begin{split}
		t(\gamma) &= F_0(v-v_{\rm in}) + \sum^3_{i=1}\frac{F_i}{\wp'_{x}(v_i)}\left( 2\zeta_{x}(v_i)(v-v_{\rm in}) + \ln\frac{\sigma_{x}(v-v_i)}{\sigma_{x}(v_{\rm in}-v_i)} - \ln\frac{\sigma_{x}(v+v_i)}{\sigma_{x}(v_{\rm in}+v_i)}\right) \\
		&+G_0(u-u_{\rm in}) + \frac{G_1}{\wp'_{\nu}(u_t)}\left( 2\zeta_{\nu}(u_t)(u-u_{\rm in}) + \ln\frac{\sigma_{\nu}(u-u_t)}{\sigma_{\nu}(u_{\rm in}-u_t)} - \ln\frac{\sigma_{\nu}(u+u_t)}{\sigma_{\nu}(u_{\rm in}+u_t)}\right) \\
		&+ t_{\rm in} \, ,
	 \end{split}
\end{equation}
where $q_t=\wp_{\nu}(u_t)$, $v=\gamma-\gamma_{\rm in}'$,  $u=\gamma-\gamma_{\rm in}''$.

\section{The orbits}
\label{sec:orbits}
\begin{figure}[h]
	\centering
	\subfigure[BO with parameters $\delta=1$, $K=2$, $g=0.1$,  $a=0.5$, $b=0.4$, $L=0.2$, $J=0.8$ and $E=1.12$.]{
		 \includegraphics[width=0.28\textwidth]{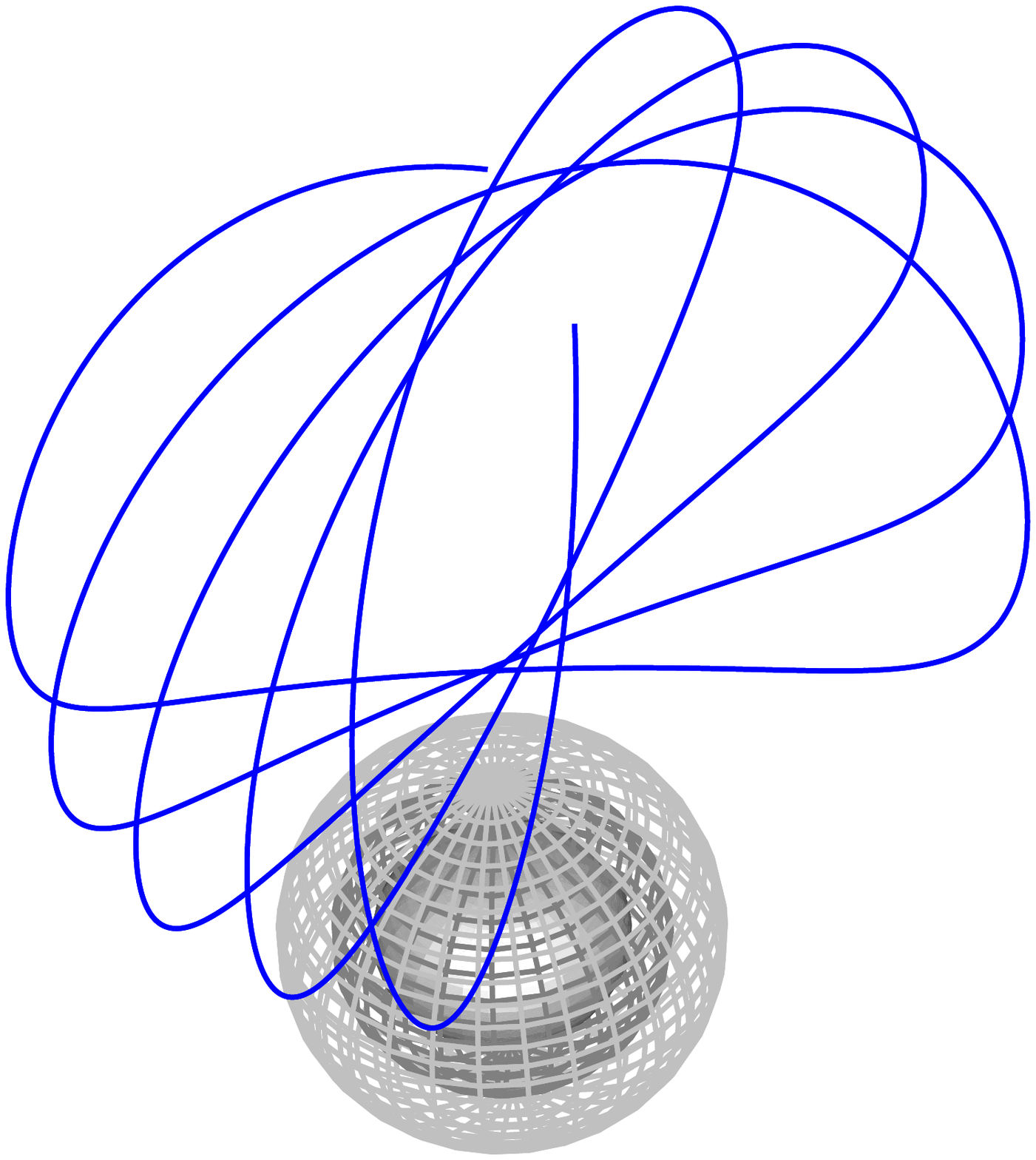}
	}
	\subfigure[BO hidden behind the horizons with parameters $\delta=1$, $K=5$, $g=0.1$,  $a=-0.4$, $b=-0.3$, $L=-2.45$, $J=-0.5$ and $E=6.122$.]{
		 \includegraphics[width=0.28\textwidth]{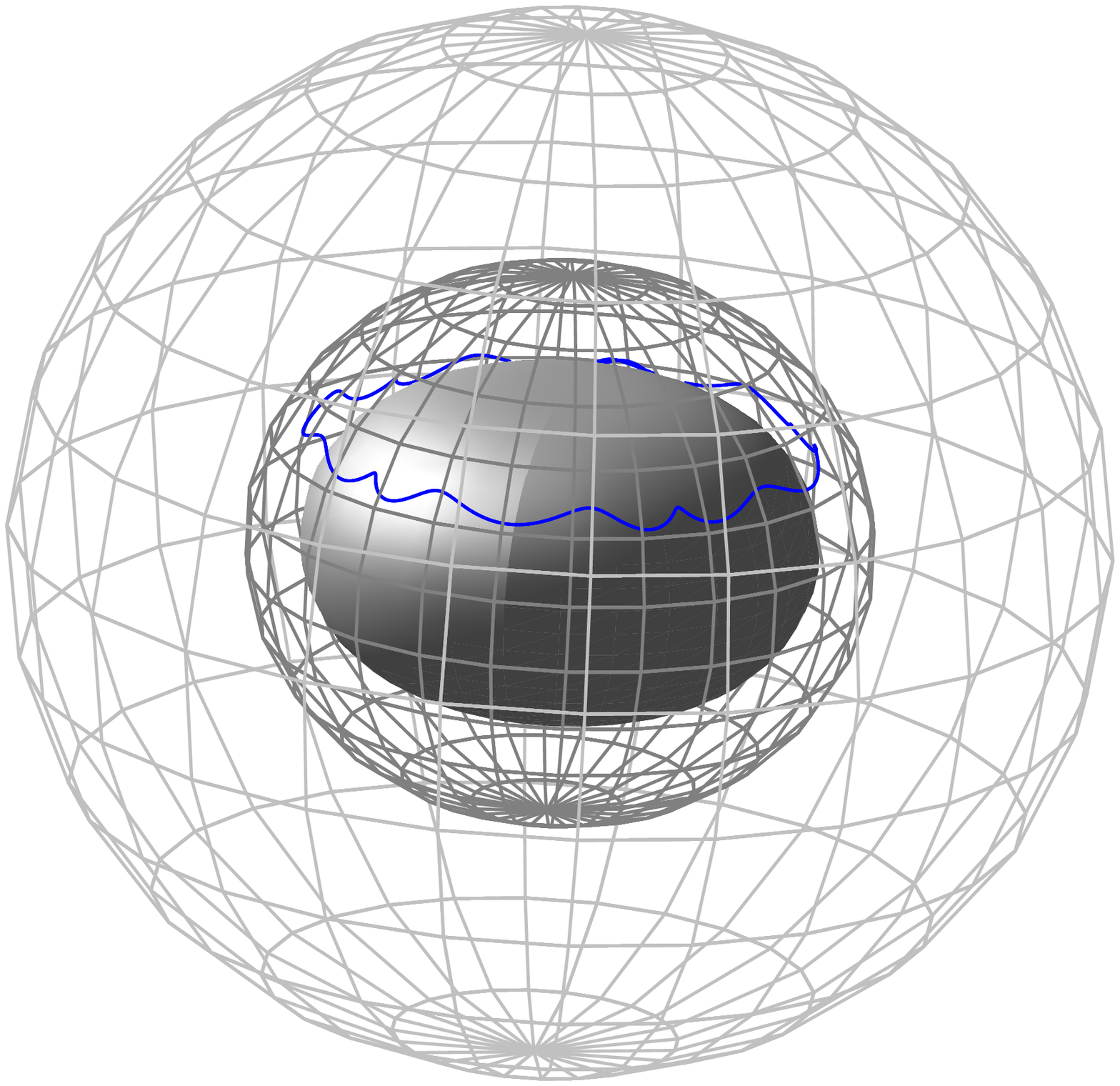}
	}
	\subfigure[MBO with parameters $\delta=1$, $K=0.9$, $g=0.1$,  $a=0.5$, $b=0.4$, $L=0.2$, $J=0.5$ and $E=0.7$.]{
		 \includegraphics[width=0.28\textwidth]{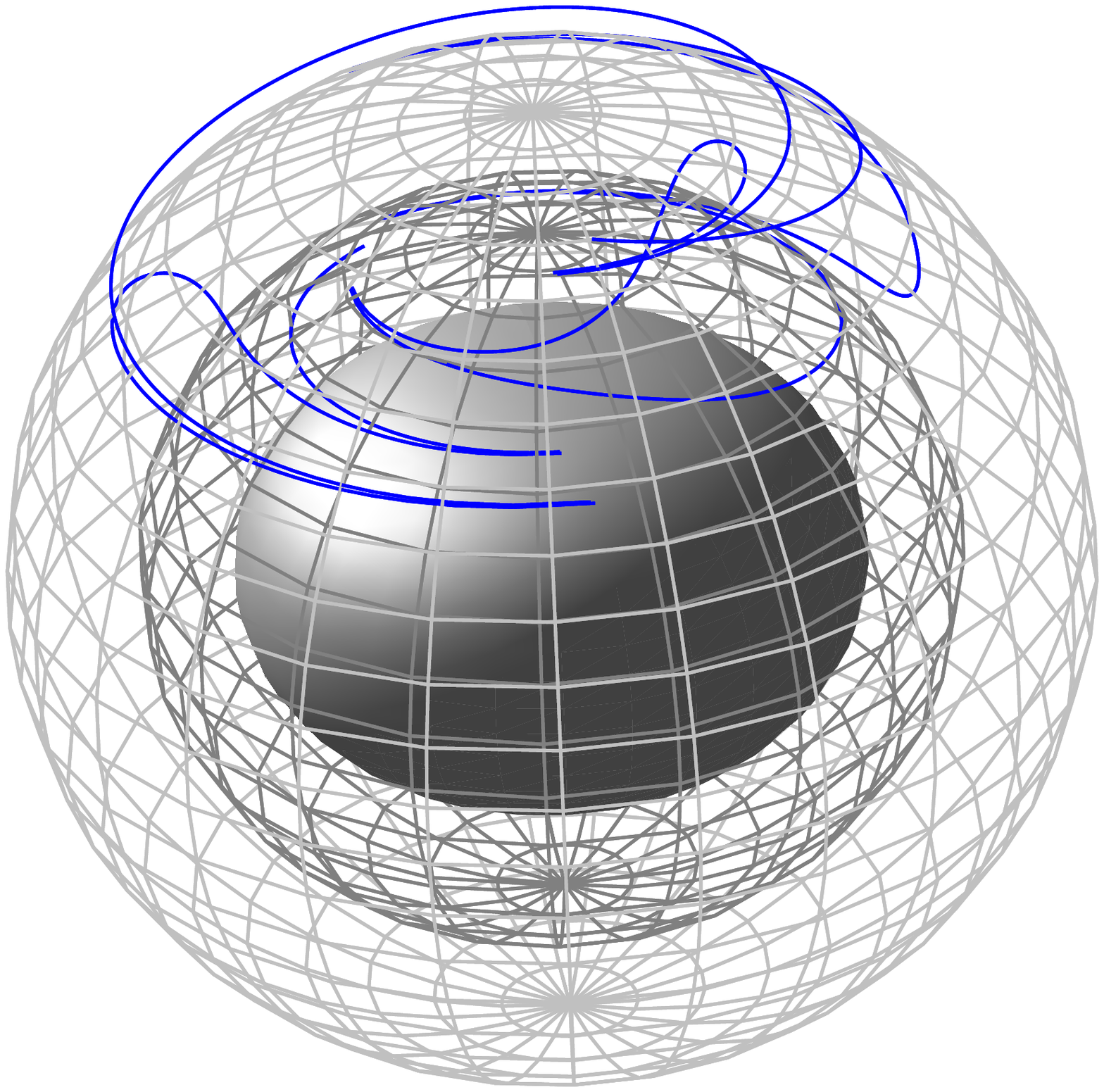}
	}
	\caption{Example plots of timelike geodesics in the Myers-Perry-AdS spacetime. The blue curve depicts the orbit and the wire frame spheroids are the inner and outer horizons. The grey solid spheroid is the singularity.}
	\label{pic:orbits-timelike}
\end{figure}

In this sections we plot examples of the orbits using the analytical solutions of the equations of motion. Figure \ref{pic:orbits-timelike} shows three timelike geodesics, a bound orbit (\ref{pic:orbits-timelike}(a)), a bound orbit hidden behind the horizons [\ref{pic:orbits-timelike}(b)] and a many-world bound orbit crossing both horizons (\ref{pic:orbits-timelike}(c)). Discontinuities in the orbit in Fig. \ref{pic:orbits-timelike}(c) can be observed when the geodesic crosses a horizon. These are caused by divergencies in the $\phi$ equation which occur on the horizons. The divergencies are due to the choice of coordinates and also appear in the $\psi$  and $t$ equation.

In contrast to timelike geodesics, which only move on bound orbits, null and spacelike geodesics can approach the black hole and then escape its gravity after a turning point. However, bound orbits with $r>r_+$ do not exist for null and spacelike geodesics.

\begin{figure}[h]
	\centering
	\subfigure[EO with parameters $\delta=0$, $K=5$, $g=0.2$,  $a=-0.4$, $b=-0.3$, $L=-0.7$, $J=-0.7$ and $E=1.4$.]{
		 \includegraphics[width=0.28\textwidth]{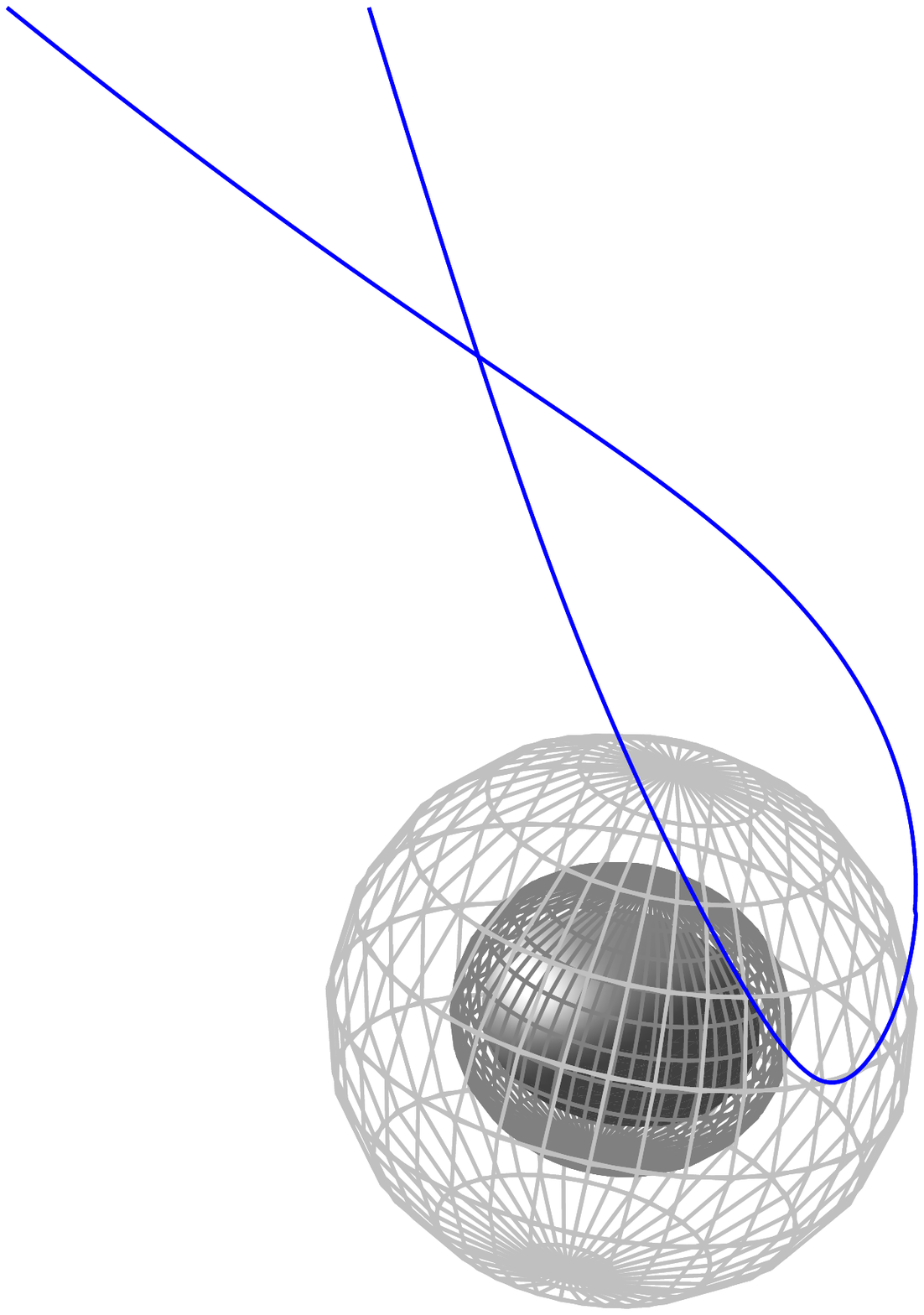}
	}
	\subfigure[TWEO with parameters $\delta=0$, $K=3$, $g=0.25$,  $a=-0.4$, $b=-0.3$, $L=-0.7$, $J=-0.5$ and $E=1.7$.]{
		 \includegraphics[width=0.28\textwidth]{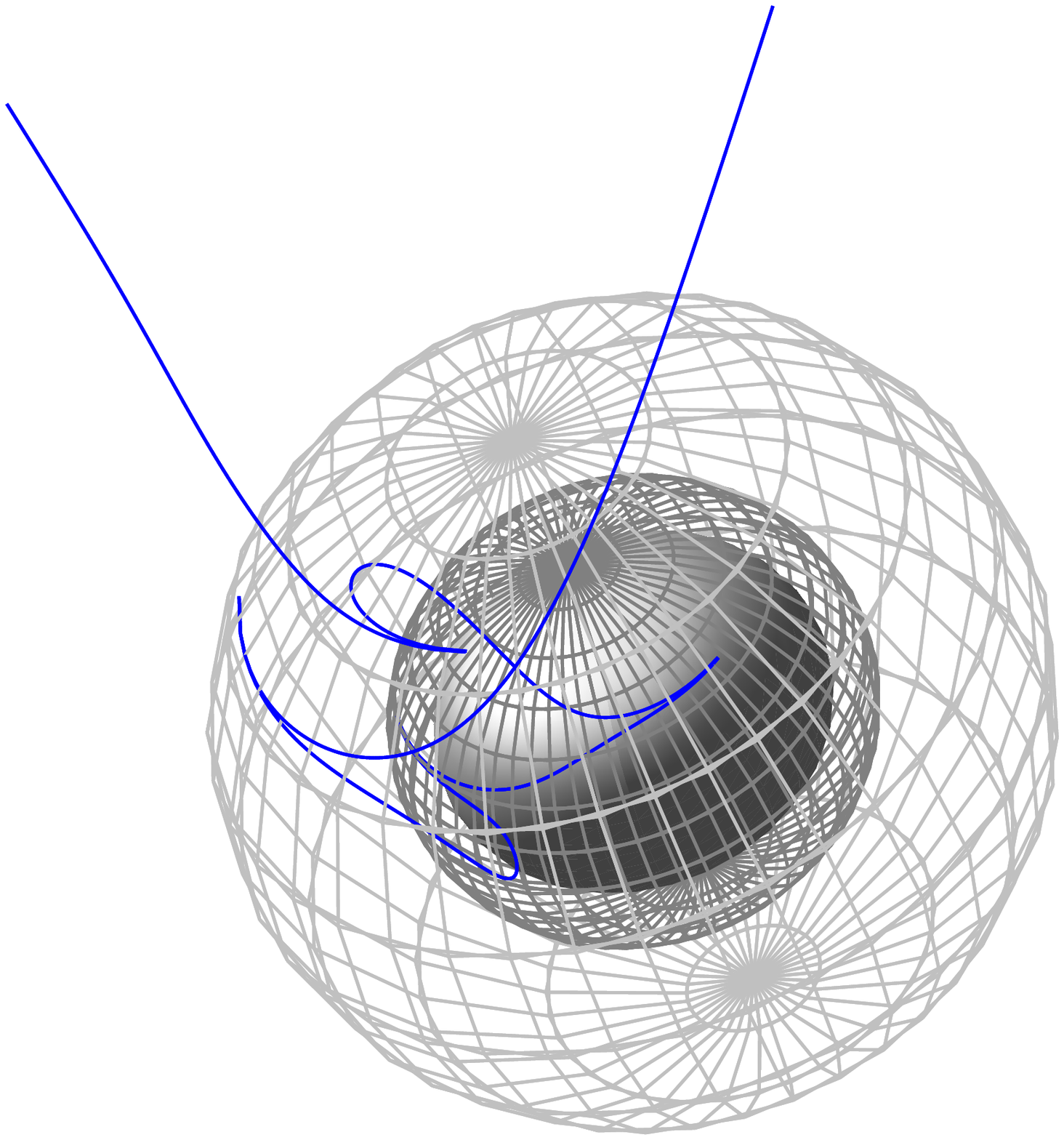}
	}
	\subfigure[TO with parameters $\delta=0$, $K=5$, $g=0.1$,  $a=-0.4$, $b=-0.3$, $L=-2.45$, $J=-0.5$ and $E\approx 5.9954$.]{
		 \includegraphics[width=0.28\textwidth]{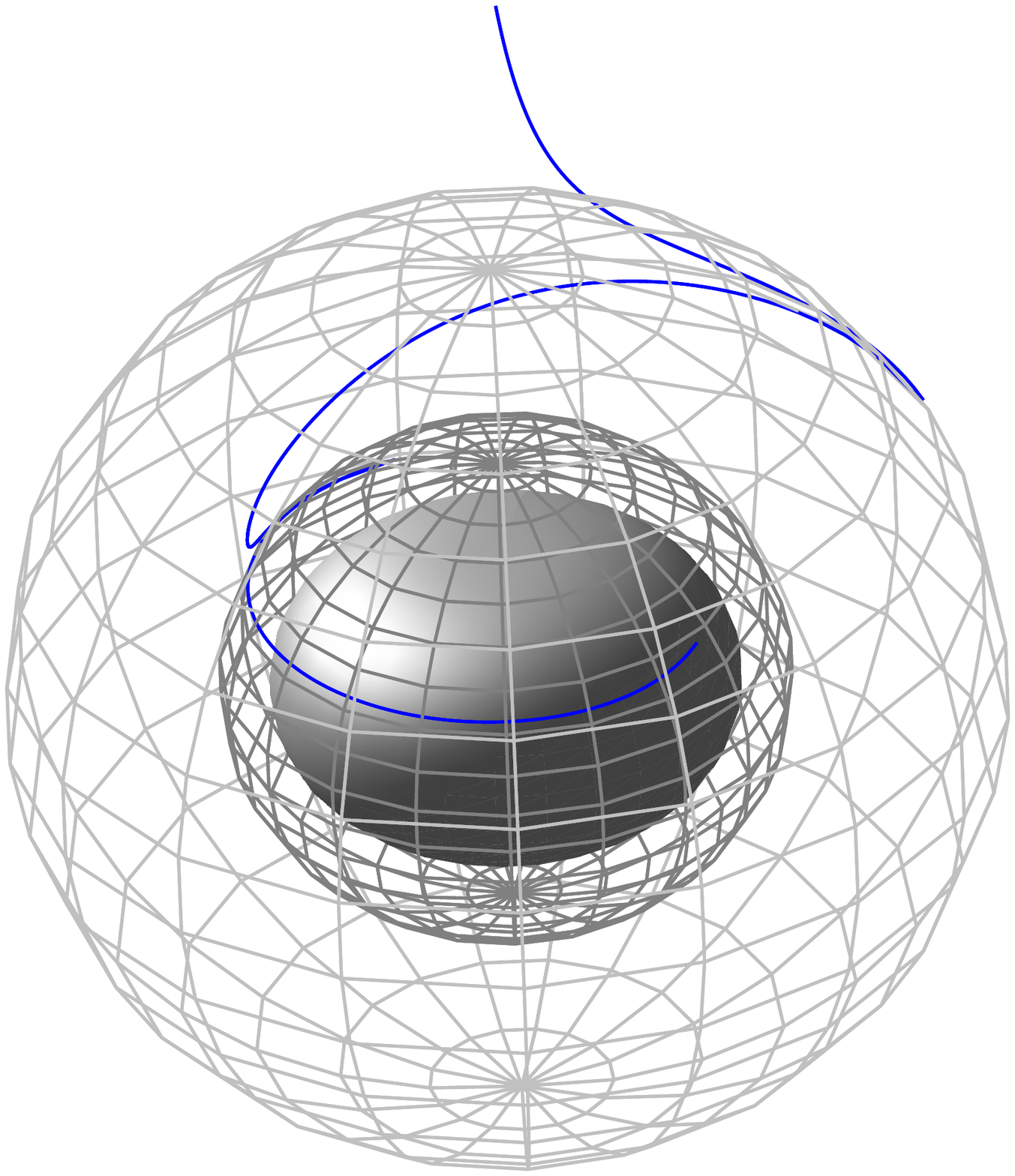}
	}
	\caption{Example plots of null geodesics in the Myers-Perry-AdS spacetime. The blue curve depicts the orbit and the wire frame spheroids are the inner and outer horizons. The grey solid spheroid is the singularity.}
	\label{pic:orbits-null}
\end{figure}

Examples of null geodesics are depicted in Fig. \ref{pic:orbits-null}. Here an escape orbit [ref{pic:orbits-null}(a)] and a two-world escape orbit crossing the horizons [\ref{pic:orbits-null}(b)] can be seen. Terminating orbits [\ref{pic:orbits-null}(c)], which end  in the singularity,  have a constant $\theta$ value and their energy an extremum of the effective $\theta$ potential.

\begin{figure}[h]
	\centering
	\subfigure[EO with parameters $\delta=-1$, $K=1$, $g=0.1$,  $a=0.5$, $b=0.4$, $L=0.3$, $J=0.5$ and $E=0.5782$.]{
		 \includegraphics[width=0.28\textwidth]{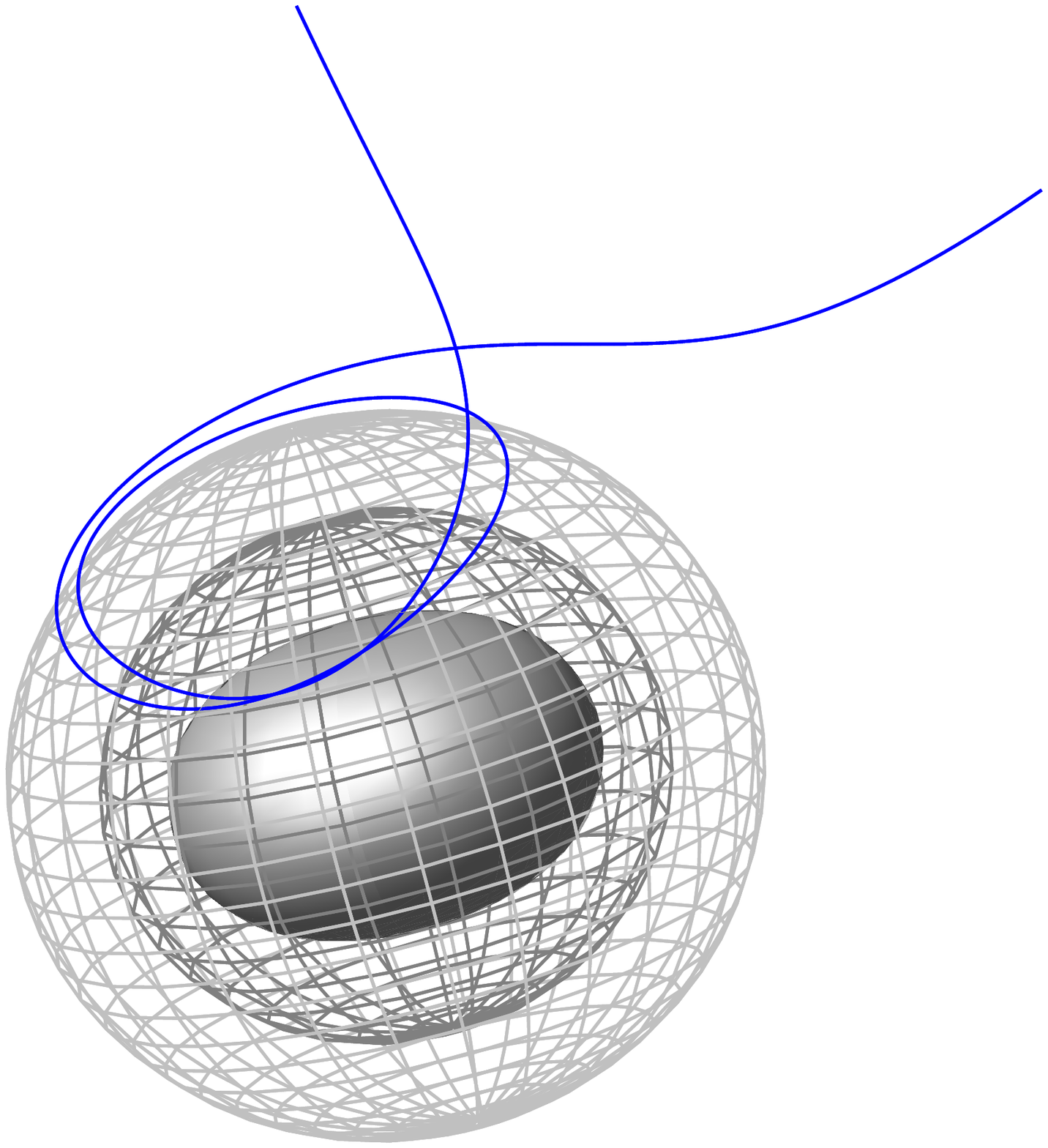}
	}
	\subfigure[TWEO crossing both horizons with parameters $\delta=-1$, $K=2$, $g=0.1$,  $a=0.5$, $b=0.45$, $L=1.2$, $J=0.4$ and $E=4$.]{
		 \includegraphics[width=0.28\textwidth]{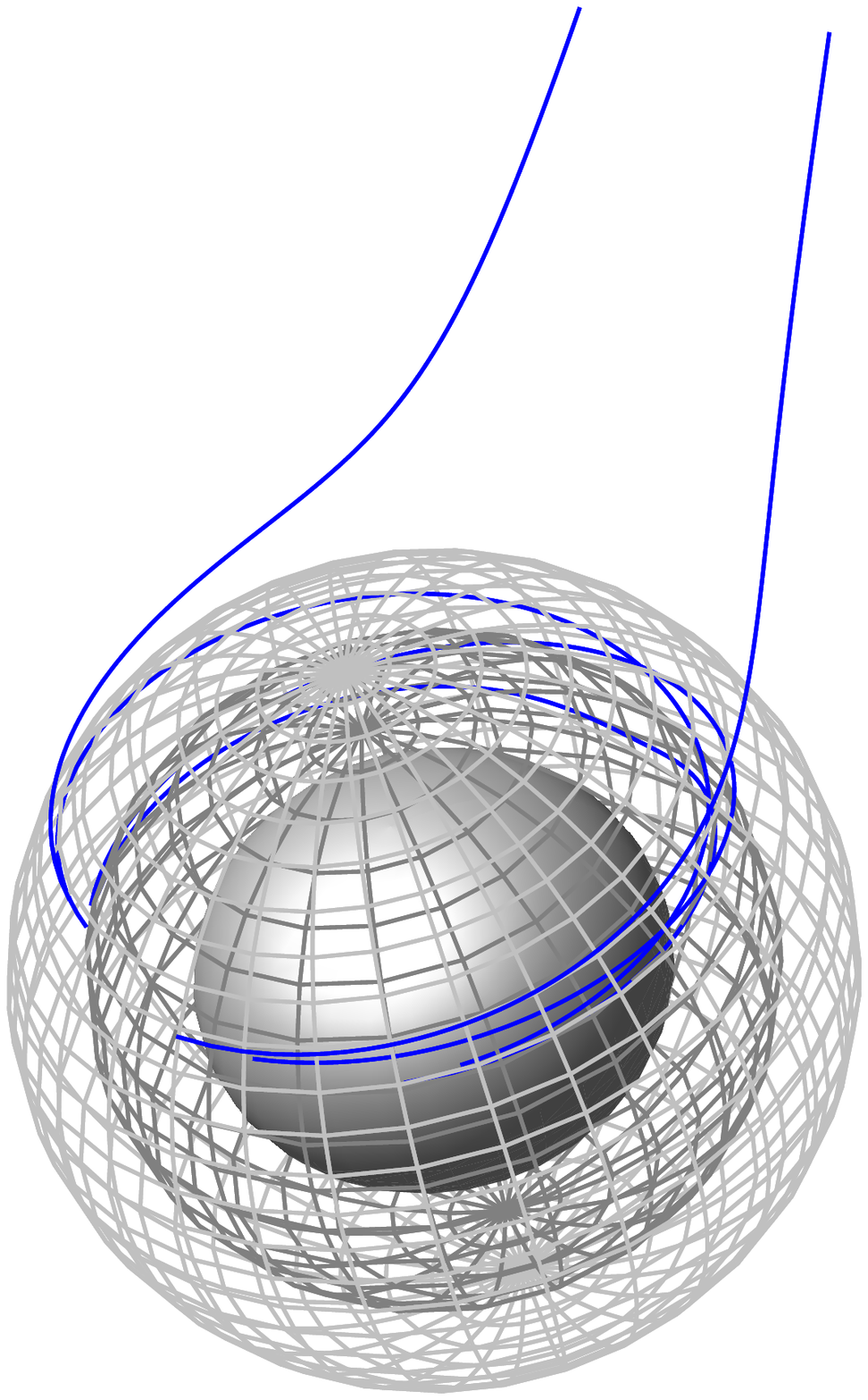}
	}
	\subfigure[TWEO crossing one horizon with parameters $\delta=-1$, $K=0.22$, $g=0.1$,  $a=0.5$, $b=0.4$, $L=0.3$, $J=0.5$ and $E=1.1$.]{
		 \includegraphics[width=0.28\textwidth]{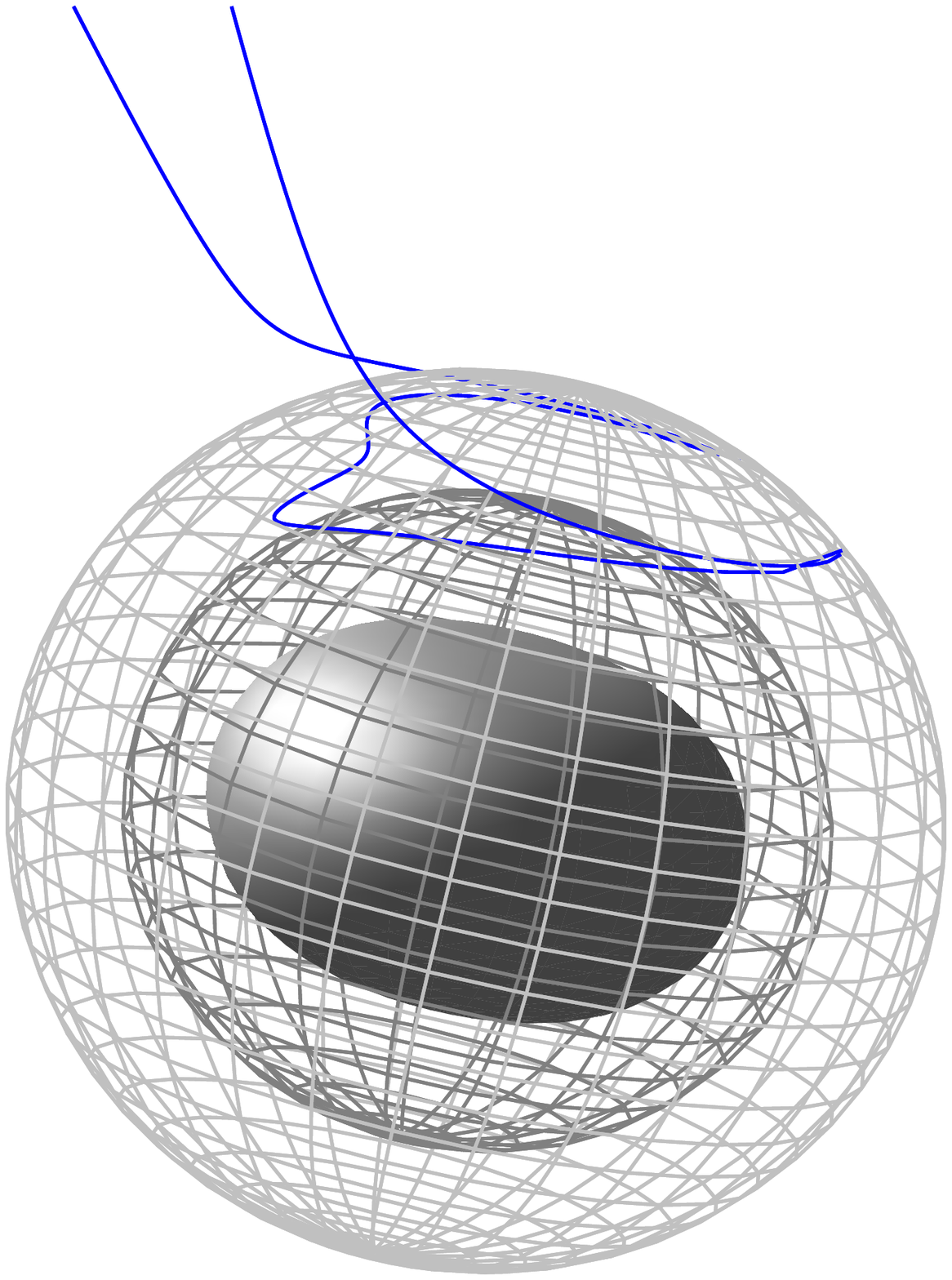}
	}
	\caption{Example plots of spacelike geodesics in the Myers-Perry-AdS spacetime. The blue curve depicts the orbit and the wire frame spheroids are the inner and outer horizons. The grey solid spheroid is the singularity.}
	\label{pic:orbits-spacelike}
\end{figure}

Spacelike geodesics are shown in Fig. \ref{pic:orbits-spacelike}. Here an escape orbit [\ref{pic:orbits-spacelike}(a)], and a two-world escape orbit crossing both horizons [\ref{pic:orbits-spacelike}(b)] are shown. Since spacelike geodesics move faster than light, there are orbits which cross just a single horizon, see Fig. \ref{pic:orbits-spacelike}(c) for a two-world escape orbit of this kind.

\FloatBarrier

\section{Conclusion}
\label{sec:conclusion}
In this article we studied the Myers-Perry-AdS black hole which is  characterized by its mass, two independent rotation parameters and a negative cosmological constant. We derived the equations of motion and solved them analytically in terms of the Weierstra{\ss} $\wp$, $\sigma$ and $\zeta$ functions. To analyze the geodesics we used parametric diagrams and effective potentials. 

Timelike geodesics move on bound orbits and many-world bound orbits which cross both horizons. They never reach the AdS boundary at infinity. For null geodesics we found many-world bound orbits and escape orbits which reach the boundary. Additionally, two-world escape orbits which cross both horizons, exist for null geodesics.

Using the analytical solutions one can calculate the exact orbits and their properties, such as the periastron shift of a bound orbit or the light deflection of an escape orbit. The observables can be computed with formulas analogous to those given in \cite{Hackmann:2010zz}. The analytical solutions of the equations of motion for null geodesics can be applied to calculate the shadow of the black hole.
\\

Since spacelike geodesics have applications in the AdS/CFT correspondence, we studied their properties too. The relevant geodesics, which correspond to correlation functions, must have endpoints on the boundary. For spacelike geodesics bound orbits hidden behind the inner horizon and  many-world bound orbits which cross both horizons exist. We found escape orbits with endpoints on a single boundary and also two-world escape orbits crossing both horizons with endpoints on two different boundaries. For certain parameter ranges there are many-world bound orbits and two-world escape orbits which just cross one of the horizons.
\\

For future work it would be very interesting to study the geodesic motion of charged particles in the charged Myers-Perry-AdS spacetime. In minimal five-dimensional gauged supergravity a charged generalization of the spacetime considered in the present article was found \cite{Chong:2005hr}. We are confident that the geodesic equations for charged particles in this spacetime can be solved analytically in terms of elliptic or hyperelliptic functions. In the AdS/CFT context, charged correlations functions are interesting to study, for example, the Schwinger pair production or induced emission \cite{Brecher:2004gn}.

\section{Acknowledgements}
We would like to thank  Anastasia Golubtsova and Jutta Kunz for interesting and fruitful discussions. 
S.G. gratefully acknowledges support by the DFG (Deutsche Forschungsgemeinschaft/ German Research Foundation) within the Research Training Group 1620 ``Models of Gravity.''

\bibliographystyle{unsrt}

\begin{thebibliography}{99}

%\cite{Maldacena:1997re}
\bibitem{Maldacena:1997re} 
  J.~M.~Maldacena,
  %``The Large N limit of superconformal field theories and supergravity,''
  Int.\ J.\ Theor.\ Phys.\  {\bf 38}, 1113 (1999)
  [Adv.\ Theor.\ Math.\ Phys.\  {\bf 2}, 231 (1998)]
  [hep-th/9711200].

%\cite{Carter:1968ks}
\bibitem{Carter:1968ks} 
  B.~Carter,
  %``Hamilton-Jacobi and Schr\"odinger separable solutions of Einstein's equations,''
  Commun.\ Math.\ Phys.\  {\bf 10}, 280 (1968).

%\cite{Kerr:1963ud}
\bibitem{Kerr:1963ud} 
  R.~P.~Kerr,
  %``Gravitational field of a spinning mass as an example of algebraically special metrics,''
  Phys.\ Rev.\ Lett.\  {\bf 11}, 237 (1963).
  %doi:10.1103/PhysRevLett.11.237

%\cite{Myers:1986un}
\bibitem{Myers:1986un} 
  R.~C.~Myers and M.~J.~Perry,
  %``Black Holes in Higher Dimensional Space-Times,''
  Annals Phys.\  {\bf 172}, 304 (1986).
  %doi:10.1016/0003-4916(86)90186-7

%\cite{Hawking:1998kw}
\bibitem{Hawking:1998kw} 
  S.~W.~Hawking, C.~J.~Hunter and M.M.~Taylor-Robinson,
  %``Rotation and the AdS / CFT correspondence,''
  Phys.\ Rev.\ D {\bf 59}, 064005 (1999)
  %doi:10.1103/PhysRevD.59.064005
  [hep-th/9811056].

%\cite{Gibbons:2004js}
\bibitem{Gibbons:2004js} 
  G.~W.~Gibbons, H.~Lu, D.~N.~Page and C.~N.~Pope,
  %``Rotating black holes in higher dimensions with a cosmological constant,''
  Phys.\ Rev.\ Lett.\  {\bf 93}, 171102 (2004)
  %doi:10.1103/PhysRevLett.93.171102
  [hep-th/0409155].

%\cite{Gibbons:2004uw}
\bibitem{Gibbons:2004uw} 
  G.~W.~Gibbons, H.~Lu, D.~N.~Page and C.~N.~Pope,
  %``The General Kerr-de Sitter metrics in all dimensions,''
  J.\ Geom.\ Phys.\  {\bf 53}, 49 (2005)
  %doi:10.1016/j.geomphys.2004.05.001
  [hep-th/0404008].

%\cite{Chen:2006xh}
\bibitem{Chen:2006xh} 
  W.~Chen, H.~Lu and C.~N.~Pope,
  %``General Kerr-NUT-AdS metrics in all dimensions,''
  Class.\ Quant.\ Grav.\  {\bf 23}, 5323 (2006)
  %doi:10.1088/0264-9381/23/17/013
  [hep-th/0604125].

%\cite{Chong:2005hr}
\bibitem{Chong:2005hr} 
  Z.-W.~Chong, M.~Cvetic, H.~Lu and C.~N.~Pope,
  %``General non-extremal rotating black holes in minimal five-dimensional gauged supergravity,''
  Phys.\ Rev.\ Lett.\  {\bf 95}, 161301 (2005)
  %doi:10.1103/PhysRevLett.95.161301
  [hep-th/0506029].

%\cite{Gauntlett:2002nw}
\bibitem{Gauntlett:2002nw} 
  J.~P.~Gauntlett, J.~B.~Gutowski, C.~M.~Hull, S.~Pakis and H.~S.~Reall,
  %``All supersymmetric solutions of minimal supergravity in five- dimensions,''
  Class.\ Quant.\ Grav.\  {\bf 20}, 4587 (2003)
  %doi:10.1088/0264-9381/20/21/005
  [hep-th/0209114].
  %%CITATION = doi:10.1088/0264-9381/20/21/005;%%

%\cite{Balasubramanian:1999zv}
\bibitem{Balasubramanian:1999zv} 
  V.~Balasubramanian and S.~F.~Ross,
  %``Holographic particle detection,''
  Phys.\ Rev.\ D {\bf 61}, 044007 (2000)
  %doi:10.1103/PhysRevD.61.044007
  [hep-th/9906226].

%\cite{Kraniotis:2003ig}
\bibitem{Kraniotis:2003ig} 
  G.~V.~Kraniotis and S.~B.~Whitehouse,
  %``Exact calculation of the perihelion precession of mercury in general relativity, the cosmological constant and jacobi's inversion problem,''
  Class.\ Quant.\ Grav.\  {\bf 20}, 4817 (2003)
  %doi:10.1088/0264-9381/20/22/007
  [astro-ph/0305181].

%\cite{Kraniotis:2004cz}
\bibitem{Kraniotis:2004cz} 
  G.~V.~Kraniotis,
  %``Precise relativistic orbits in Kerr space-time with a cosmological constant,''
  Class.\ Quant.\ Grav.\  {\bf 21}, 4743 (2004)
  %doi:10.1088/0264-9381/21/19/016
  [gr-qc/0405095].

%\cite{Kraniotis:2005zm}
\bibitem{Kraniotis:2005zm} 
  G.~V.~Kraniotis,
  %``Frame-dragging and bending of light in Kerr and Kerr-(anti) de Sitter spacetimes,''
  Class.\ Quant.\ Grav.\  {\bf 22}, 4391 (2005)
  %doi:10.1088/0264-9381/22/21/001
  [gr-qc/0507056].

%\cite{Kraniotis:2006ux}
\bibitem{Kraniotis:2006ux} 
  G.~V.~Kraniotis,
  %``Periapsis and gravitomagnetic precessions of stellar orbits in Kerr and Kerr-de Sitter black hole spacetimes,''
  Class.\ Quant.\ Grav.\  {\bf 24}, 1775 (2007)
  %doi:10.1088/0264-9381/24/7/007
  [gr-qc/0602056].

%\cite{Fujita:2009bp}
\bibitem{Fujita:2009bp} 
  R.~Fujita and W.~Hikida,
  %``Analytical solutions of bound timelike geodesic orbits in Kerr spacetime,''
  Class.\ Quant.\ Grav.\  {\bf 26}, 135002 (2009)
  %doi:10.1088/0264-9381/26/13/135002
  [arXiv:0906.1420 [gr-qc]].

%\cite{Hackmann:2009nh}
\bibitem{Hackmann:2009nh} 
  E.~Hackmann, V.~Kagramanova, J.~Kunz and C.~L\"ammerzahl,
  %``Analytic solutions of the geodesic equation in axially symmetric space-times,''
  Europhys.\ Lett.\  {\bf 88}, 30008 (2009)
  %doi:10.1209/0295-5075/88/30008
  [arXiv:0911.1634 [gr-qc]].

%\cite{Hackmann:2010zz}
\bibitem{Hackmann:2010zz} 
  E.~Hackmann, C.~L\"ammerzahl, V.~Kagramanova and J.~Kunz,
  %``Analytical solution of the geodesic equation in Kerr-(anti) de Sitter space-times,''
  Phys.\ Rev.\ D {\bf 81}, 044020 (2010)
  %doi:10.1103/PhysRevD.81.044020
  [arXiv:1009.6117 [gr-qc]].

%\cite{Frolov:2003en}
\bibitem{Frolov:2003en} 
  V.~P.~Frolov and D.~Stojkovic,
  %``Particle and light motion in a space-time of a five-dimensional rotating black hole,''
  Phys.\ Rev.\ D {\bf 68}, 064011 (2003)
  %doi:10.1103/PhysRevD.68.064011
  [gr-qc/0301016].

%\cite{Vasudevan:2004ca}
\bibitem{Vasudevan:2004ca} 
  M.~Vasudevan, K.~A.~Stevens and D.~N.~Page,
  %``Separability of the Hamilton-Jacobi and Klein-Gordon equations in Kerr-de Sitter metrics,''
  Class.\ Quant.\ Grav.\  {\bf 22}, 339 (2005)
  %doi:10.1088/0264-9381/22/2/007
  [gr-qc/0405125].

%\cite{Vasudevan:2004mr}
\bibitem{Vasudevan:2004mr} 
  M.~Vasudevan, K.~A.~Stevens and D.~N.~Page,
  %``Particle motion and scalar field propagation in Myers-Perry black hole spacetimes in all dimensions,''
  Class.\ Quant.\ Grav.\  {\bf 22}, 1469 (2005)
  %doi:10.1088/0264-9381/22/7/017
  [gr-qc/0407030].

%\cite{Krtous:2008tb}
\bibitem{Krtous:2008tb} 
  P.~Krtous, V.~P.~Frolov and D.~Kubiznak,
  %``Hidden Symmetries of Higher Dimensional Black Holes and Uniqueness of the Kerr-NUT-(A)dS spacetime,''
  Phys.\ Rev.\ D {\bf 78}, 064022 (2008)
  %doi:10.1103/PhysRevD.78.064022
  [arXiv:0804.4705 [hep-th]].

%\cite{Kubiznak:2006kt}
\bibitem{Kubiznak:2006kt} 
  D.~Kubiznak and V.~P.~Frolov,
  %``Hidden Symmetry of Higher Dimensional Kerr-NUT-AdS Spacetimes,''
  Class.\ Quant.\ Grav.\  {\bf 24}, no. 3, F1 (2007)
  %doi:10.1088/0264-9381/24/3/F01
  [gr-qc/0610144].

%\cite{Frolov:2006pe}
\bibitem{Frolov:2006pe} 
  V.~P.~Frolov, P.~Krtous and D.~Kubiznak,
  %``Separability of Hamilton-Jacobi and Klein-Gordon Equations in General Kerr-NUT-AdS Spacetimes,''
  JHEP {\bf 0702}, 005 (2007)
  %doi:10.1088/1126-6708/2007/02/005
  [hep-th/0611245].

%\cite{Frolov:2008jr}
\bibitem{Frolov:2008jr} 
  V.~P.~Frolov and D.~Kubiznak,
  %``Higher-Dimensional Black Holes: Hidden Symmetries and Separation of Variables,''
  Class.\ Quant.\ Grav.\  {\bf 25}, 154005 (2008)
  %doi:10.1088/0264-9381/25/15/154005
  [arXiv:0802.0322 [hep-th]].

%\cite{Enolski:2010if}
\bibitem{Enolski:2010if} 
  V.~Z.~Enolski, E.~Hackmann, V.~Kagramanova, J.~Kunz and C.~L\"ammerzahl,
  %``Inversion of hyperelliptic integrals of arbitrary genus with application to particle motion in General Relativity,''
  J.\ Geom.\ Phys.\  {\bf 61}, 899 (2011)
  %doi:10.1016/j.geomphys.2011.01.001
  [arXiv:1011.6459 [gr-qc]].

%\cite{Chervonyi:2015ima}
\bibitem{Chervonyi:2015ima} 
  Y.~Chervonyi and O.~Lunin,
  %``Killing(-Yano) Tensors in String Theory,''
  JHEP {\bf 1509}, 182 (2015)
  %doi:10.1007/JHEP09(2015)182
  [arXiv:1505.06154 [hep-th]].

%\cite{Kagramanova:2012hw}
\bibitem{Kagramanova:2012hw} 
  V.~Kagramanova and S.~Reimers,
  %``Analytic treatment of geodesics in five-dimensional Myers-Perry space--times,''
  Phys.\ Rev.\ D {\bf 86}, 084029 (2012)
  %doi:10.1103/PhysRevD.86.084029
  [arXiv:1208.3686 [gr-qc]].

%\cite{Diemer:2014lba}
\bibitem{Diemer:2014lba} 
  V.~Diemer, J.~Kunz, C.~L\"ammerzahl and S.~Reimers,
  %``Dynamics of test particles in the general five-dimensional Myers-Perry spacetime,''
  Phys.\ Rev.\ D {\bf 89}, no. 12, 124026 (2014)
  %doi:10.1103/PhysRevD.89.124026
  [arXiv:1404.3865 [gr-qc]].
  
  %\cite{Reimers:2016czc}
  \bibitem{Reimers:2016czc} 
  S.~Paranjape and S.~Reimers,
  %``Dynamics of test particles in the five-dimensional, charged, rotating Einstein-Maxwell-Chern-Simons spacetime,''
  Phys.\ Rev.\ D {\bf 94}, no. 12, 124003 (2016)
  %doi:10.1103/PhysRevD.94.124003
  [arXiv:1609.03557 [gr-qc]].

%\cite{Delsate:2015ina}
\bibitem{Delsate:2015ina} 
  T.~Delsate, J.~V.~Rocha and R.~Santarelli,
  %``Geodesic motion in equal angular momenta Myers-Perry-AdS spacetimes,''
  Phys.\ Rev.\ D {\bf 92}, no. 8, 084028 (2015)
  %doi:10.1103/PhysRevD.92.084028
  [arXiv:1507.03602 [gr-qc]].

%\cite{Gibbons:1999uv}
\bibitem{Gibbons:1999uv}
  G.~W.~Gibbons and C.~A.~R.~Herdeiro,
  %``Supersymmetric rotating black holes and causality violation,''
  Class.\ Quant.\ Grav.\  {\bf 16} (1999) 3619
  %doi:10.1088/0264-9381/16/11/311
  [hep-th/9906098].

%\cite{Diemer:2013fza}
\bibitem{Diemer:2013fza} 
  V.~Diemer and J.~Kunz,
  %``Supersymmetric rotating black hole spacetime tested by geodesics,''
  Phys.\ Rev.\ D {\bf 89}, no. 8, 084001 (2014)
  %doi:10.1103/PhysRevD.89.084001
  [arXiv:1312.6540 [gr-qc]].

%\cite{Herdeiro:2000ap}
\bibitem{Herdeiro:2000ap}
  C.~A.~R.~Herdeiro,
  %``Special properties of five-dimensional BPS rotating black holes,''
  Nucl.\ Phys.\ B {\bf 582} (2000) 363
  %doi:10.1016/S0550-3213(00)00335-7
  [hep-th/0003063].

%\cite{Gibbons:2009um}
\bibitem{Gibbons:2009um} 
  G.~Gibbons and H.~Kodama,
  %``Repulsons in the Myers-Perry Family,''
  Prog.\ Theor.\ Phys.\  {\bf 121}, 1361 (2009)
  %doi:10.1143/PTP.121.1361
  [arXiv:0901.1203 [hep-th]].

%\cite{Carter:1968rr}
\bibitem{Carter:1968rr} 
  B.~Carter,
  %``Global structure of the Kerr family of gravitational fields,''
  Phys.\ Rev.\  {\bf 174}, 1559 (1968).

%\cite{Mino:2003yg}
\bibitem{Mino:2003yg} 
  Y.~Mino,
  %``Perturbative approach to an orbital evolution around a supermassive black hole,''
  Phys.\ Rev.\ D {\bf 67}, 084027 (2003)
  [gr-qc/0302075].

%\cite{Grenzebach:2014fha}
\bibitem{Grenzebach:2014fha} 
  A.~Grenzebach, V.~Perlick and C.~L\"ammerzahl,
  %``Photon Regions and Shadows of Kerr-Newman-NUT Black Holes with a Cosmological Constant,''
  Phys.\ Rev.\ D {\bf 89}, no. 12, 124004 (2014)
  %doi:10.1103/PhysRevD.89.124004
  [arXiv:1403.5234 [gr-qc]].

%\cite{Grenzebach:2015uva}
\bibitem{Grenzebach:2015uva} 
  A.~Grenzebach,
  %``Aberrational Effects for Shadows of Black Holes,''
  Fund.\ Theor.\ Phys.\  {\bf 179}, 823 (2015)
  %doi:10.1007/978-3-319-18335-0_25
  [arXiv:1502.02861 [gr-qc]].

%\cite{Grenzebach:2015oea}
\bibitem{Grenzebach:2015oea} 
  A.~Grenzebach, V.~Perlick and C.~L\"ammerzahl,
  %``Photon Regions and Shadows of Accelerated Black Holes,''
  Int.\ J.\ Mod.\ Phys.\ D {\bf 24}, no. 09, 1542024 (2015)
  %doi:10.1142/S0218271815420249
  [arXiv:1503.03036 [gr-qc]].

%\cite{Grenzebach:Springer}
\bibitem{Grenzebach:Springer} 
A.~Grenzebach,
``The Shadow of Black Holes -- An Analytic Description'', 
SpringerBriefs in Physics (Springer, Heidelberg, 2016)

%\cite{Balasubramanian:2011ur}
\bibitem{Balasubramanian:2011ur} 
  V.~Balasubramanian, A.~Bernamonti, J.~ deBoer, N.~Copland, B.~Craps, E.~Keski-Vakkuri, B.~Muller, A.~Schafer, M.~Shigemori, W.~Staessens
  %``Holographic Thermalization,''
  Phys.\ Rev.\ D {\bf 84}, 026010 (2011)
  %doi:10.1103/PhysRevD.84.026010
  [arXiv:1103.2683 [hep-th]].

%\cite{Louko:2000tp}
\bibitem{Louko:2000tp} 
  J.~Louko, D.~Marolf and S.~F.~Ross,
  %``On geodesic propagators and black hole holography,''
  Phys.\ Rev.\ D {\bf 62}, 044041 (2000)
  %doi:10.1103/PhysRevD.62.044041
  [hep-th/0002111].

%\cite{Hubeny:2007xt}
\bibitem{Hubeny:2007xt} 
  V.~E.~Hubeny, M.~Rangamani and T.~Takayanagi,
  %``A Covariant holographic entanglement entropy proposal,''
  JHEP {\bf 0707}, 062 (2007)
  %doi:10.1088/1126-6708/2007/07/062
  [arXiv:0705.0016 [hep-th]].

%\cite{AbajoArrastia:2010yt}
\bibitem{AbajoArrastia:2010yt} 
  J.~Abajo-Arrastia, J.~Aparicio and E.~Lopez,
  %``Holographic Evolution of Entanglement Entropy,''
  JHEP {\bf 1011}, 149 (2010)
  doi:10.1007/JHEP11(2010)149
  [arXiv:1006.4090 [hep-th]].

%\cite{Fidkowski:2003nf}
\bibitem{Fidkowski:2003nf} 
  L.~Fidkowski, V.~Hubeny, M.~Kleban and S.~Shenker,
  %``The Black hole singularity in AdS / CFT,''
  JHEP {\bf 0402}, 014 (2004)
  %doi:10.1088/1126-6708/2004/02/014
  [hep-th/0306170].

%\cite{Kraus:2002iv}
\bibitem{Kraus:2002iv} 
  P.~Kraus, H.~Ooguri and S.~Shenker,
  %``Inside the horizon with AdS / CFT,''
  Phys.\ Rev.\ D {\bf 67}, 124022 (2003)
  doi:10.1103/PhysRevD.67.124022
  [hep-th/0212277].

%\cite{Levi:2003cx}
\bibitem{Levi:2003cx} 
  T.~S.~Levi and S.~F.~Ross,
  %``Holography beyond the horizon and cosmic censorship,''
  Phys.\ Rev.\ D {\bf 68}, 044005 (2003)
  doi:10.1103/PhysRevD.68.044005
  [hep-th/0304150].

%\cite{Papadodimas:2012aq}
\bibitem{Papadodimas:2012aq} 
  K.~Papadodimas and S.~Raju,
  %``An Infalling Observer in AdS/CFT,''
  JHEP {\bf 1310}, 212 (2013)
  doi:10.1007/JHEP10(2013)212
  [arXiv:1211.6767 [hep-th]].

%\cite{Markushevich:1967}
\bibitem{Markushevich:1967}
  A.~I.~Markushevich, {\em Theory of Functions of a Complex Variable} (Prentice-Hall, Englewood Cliffs, NJ, 1967), Vol. III.

%\cite{Kagramanova:2010bk}
\bibitem{Kagramanova:2010bk} 
  V.~Kagramanova, J.~Kunz, E.~Hackmann and C.~L\"ammerzahl,
  %``Analytic treatment of complete and incomplete geodesics in Taub-NUT space-times,''
  Phys.\ Rev.\ D {\bf 81}, 124044 (2010)
  [arXiv:1002.4342 [gr-qc]].

%\cite{Grunau:2010gd}
\bibitem{Grunau:2010gd} 
  S.~Grunau and V.~Kagramanova,
  %``Geodesics of electrically and magnetically charged test particles in the Reissner-Nordstr\"om space-time: analytical solutions,''
  Phys.\ Rev.\ D {\bf 83}, 044009 (2011)
  [arXiv:1011.5399 [gr-qc]].

%\cite{Enolski:2011id}
\bibitem{Enolski:2011id} 
  V.~Enolski, B.~Hartmann, V.~Kagramanova, J.~Kunz, C.~L\"ammerzahl and P.~Sirimachan,
  %``Hyperelliptic integrals and Ho\v{r}ava-Lifshitz black hole space-times,''
  Journal of mathematical physics {\bf{53}}, 012504 (2012)  
  [arXiv:1106.2408 [gr-qc]].

%\cite{Brecher:2004gn}
\bibitem{Brecher:2004gn} 
  D.~Brecher, J.~He and M.~Rozali,
  %``On charged black holes in anti-de Sitter space,''
  JHEP {\bf 0504}, 004 (2005)
  %doi:10.1088/1126-6708/2005/04/004
  [hep-th/0410214].

\end{thebibliography}

\end{document}